\documentclass[twocolumn]{aastex701} 
\usepackage{placeins}
\usepackage{amsmath}

\newcommand*{\teff}{T_{\mathrm{eff}}}
\newcommand*{\ang}{\mbox{\AA}}
\newcommand*{\water}{H$_{2}$O}
\begin{document}

\title{The Composition of M-dwarf Exoplanet Hosts: Atmospheric Parameters and O, Mg, Si, and Fe Abundances from IR Spectroscopy}

\author[0000-0002-0475-3662,sname=Maas]{Zachary G. Maas} 
\affil{Department of Astronomy, Indiana University, Bloomington, IN 47405, USA}
\email[show]{zmaas@iu.edu}  
\author[0000-0003-0595-5132,sname=Hinkel]{Natalie R.\ Hinkel}
\affiliation{Louisiana State University, Department of Physics and Astronomy, 202 Nicholson Hall, Baton Rouge, LA 70803, USA}
\email{natalie.hinkel@gmail.com}  
\author[0000-0001-7875-6391,sname=Mace]{Gregory N. Mace} 
\affil{Department of Astronomy and McDonald Observatory, The University of Texas, Austin, TX 78712, USA}
\email{gmace@astro.as.utexas.edu}  
\author[0000-0001-8991-3110,sname=Unterborn]{Cayman T. Unterborn} 
\affil{Southwest Research Institute, 6220 Culebra Road, San Antonio, TX }
\email{kmanunterborn@gmail.com}  
\author[0000-0002-3007-206X,sname=Pilachowski]{Catherine Pilachowski} 
\affil{Department of Astronomy, Indiana University, Bloomington, IN 47405, USA}
\email{cpilacho@iu.edu}  
\author[0000-0002-3456-5929,sname=Sneden]{Christopher Sneden} 
\affil{Department of Astronomy and McDonald Observatory, The University of Texas, Austin, TX 78712, USA}
\email{chris@astro.as.utexas.edu}  
\author[0009-0001-5775-9945,sname=Flores]{Isabella Flores} 
\affil{Department of Astronomy, Indiana University, Bloomington, IN 47405, USA}
\email{isabella@isabellaﬂores.me}  
\author[0000-0003-3570-422X,sname=Schulze]{Joseph G. Schulze} 
\affiliation{Louisiana State University, Department of Physics and Astronomy, 202 Nicholson Hall, Baton Rouge, LA 70803, USA}
\email{jschulze@lsu.edu}  

\begin{abstract}

Determining M-dwarf abundances is an important step to discerning the chemical evolution of the Milky Way and characterizing exoplanetary systems, especially since these objects commonly host small exoplanets. However, a dearth of M-dwarf abundance measurements exist compared to Sun-like stars, partially due to their complicated optical spectra dominated by molecular absorption lines and low luminosities. To overcome these issues, we have obtained high resolution H-band spectra of 40 planet-hosting M-dwarf stars and 2 M-dwarfs in wide binaries with FGK primaries with IGRINS on Gemini South and on the McDonald Observatory 2.7m telescope, and NIRSPEC on Keck. Abundances were derived by comparing each observed spectrum to synthetic spectral grids generated with MOOG and MARCS atmospheric models. Effective temperatures were derived using OH and \water{} lines and [Fe/H] from Fe I and FeH lines. Both Mg and Si abundances were determined using atomic features. Overall, we find that our abundances are broadly consistent with previous literature M-dwarf abundances. We complemented our abundances analysis with UVW velocities and Milky Way stellar population membership probabilities and identified 2 potential thick disk stars from our kinematic and chemical analysis: GJ 1252 and L181-1 which was a previously known thick disk star. Kinematic stellar population probabilities and a low [Fe/H] abundance makes Kepler 42 a likely thick disk star, although uncertain Mg and Si abundances make classification difficult. Lastly, we have determined Si/Mg and Fe/Mg molar ratios for M-dwarf exoplanet hosts and find the abundance distributions leads to exoplanet core-mass fractions between 20 - 40$\%$.

\end{abstract}


\section{Introduction}
\label{sec::intro}

M-dwarf stars are cool, low-mass, low-luminosity stars with long main-sequence lifetimes. These objects are the most common type of star found in the solar neighborhood \citep{reyle21} but are often excluded from large-scale abundance surveys due to their relatively complicated photospheres. In contrast to FGK-stars, M-dwarf optical spectra are dominated by strong TiO and VO molecular features while near-IR wavelengths are impacted by a large number of \water{} lines \citep{allard00} due to their cool effective temperatures ($\teff{}$s). A variety of approaches have been developed to understand the fundamental parameters and chemical inventories of M-dwarfs. For example, photometry and parallaxes of these objects have been successfully used to calculate M-dwarf atmospheric parameters, especially when results are benchmarked to interferometric radii \citep{casagrande08,rojasayala12,mann13,hejazi15,mann15}. Low-resolution spectroscopy has also been used to identify $\teff{}$- and metallicity-sensitive spectral features in order to measure bulk metallicities using optical \citep{hejazi22} and infrared spectra \citep{newton15}.

Challenges defining the continuum for M-dwarf spectra, identifying absorption features that yield accurate/precise abundances, and modeling their complicated atmospheres have limited the number of elemental abundances measured in M-dwarfs. Multiple successful parameter estimations have utilized moderate-to-high resolution spectra in the near-IR between 0.8--1.6 $\mu$m (the z, J, and H IR bands). Atmospheric parameters have been determined by analyzing molecular features instead of Fe line ionization balance typically used in FGK-star abundances (e.g., \citealt{passegger19,houdebine19,souto20}) while linelists have been adapted to account for the large numbers of molecular line transitions, especially \water{} \citep{smith21}. Studies that target specific regions free from large blends have been effective in measuring [Fe/H] ratios from FeH and Fe I lines from moderate to high resolution spectra \citep{bean06,wende10,neves14,lindgren16,lindgren17,rajpurohit18,houdebine19,lopezvaldivia19,kuznestov19}. Similarly, \citet{woolf20} measured Fe and Ti abundances in regions with minimal TiO contamination using high resolution optical spectra.

Recent efforts have more fully accessed the chemical inventory of M-dwarfs. As a result, successful abundance measurements for elements such as C, O, Mg, Si, etc., have been determined from abundance analysis using a combination of atomic and molecular lines \citep{tsuji15,souto17,abia20,souto22,ishikawa22,tabernero24,hejazi24,melo24}. Abundances have been tested by measuring the composition of stars in M-dwarf and FGK binary pairs. Studies have found that M-dwarf secondaries have consistent abundances derived from near-IR spectra when compared to their FGK-star companion \citep{lindgren16,souto22}. While it is often assumed that M-dwarfs in general follow similar abundance trends as FGK-type stars, there is currently not enough direct evidence to make statistical comparisons. To-date, only $\sim$1500 total abundances have been measured for 17 non-Fe elements in $\sim$400 M-dwarfs, or 3-4 elements per M-dwarf, as opposed to the millions of abundances that have been measured across 100+ elements for FGK-type stars \citep{hinkel14, majewski17}. Expanding the sample of M-dwarf abundances will allow the different stellar types to be more equivalently and robustly analyzed with respect to another while also filling in crucial gaps in our understanding of the chemical evolution of the Milky Way, especially those aspects/elements not accessible in larger main sequence stars.

Catalogs of reliable M-dwarf abundances, such as those studied in FGK star and M-dwarf binary pairs, have also been used to determine M-dwarf abundances via machine learning techniques \citep{birky20,maldonado20,marfil21,zhang25,olander25,wanderley25}. A data-driven abundance analysis is a promising direction for identifying usable features from complicated M-dwarf spectra and analyzing parameters from large quantities of spectra \citep{behmard25}. These studies rely on thorough sets of M-dwarf training data that span a variety of possible atmospheric parameters and compositions. Unfortunately, the level to which we understand M-dwarf composition is a small compared to FGK-type stars. Expanding the number of precise M-dwarf abundances across different metallicities and spectral sub-types may provide additional objects for training data to incorporate into machine learning abundance techniques.

Additional abundance studies have been motivated by the finding that M-dwarfs are likely common hosts to small exoplanets \citep{dressing15}. In recent years, hundreds of exoplanets around M-dwarfs have been detected with radii that span from terrestrial to Neptunian in size \citep[$\sim$0.5-4 R$_{\oplus}$,][]{henry24,christiansen25}. Many additional small planet detections are anticipated for M-dwarfs since the star-planet mass and radius ratios make it more conducive to find an Earth-sized planet around a smaller star, as compared to higher mass/radius FGK-stars. The large number of small M-dwarf planets enable a more complete census of their properties and a more holistic search for habitable worlds. 

\begin{deluxetable*}{ c c c c c c c c c}
\tablewidth{0pt} 
\tabletypesize{\footnotesize}
\tablecaption{Sample and Observation Summary \label{table::obslog}} 
 \tablehead{\colhead{Star Name} & \colhead{Date Obs.} & \colhead{H} & \colhead{K$_{s}$} & \colhead{$\pi$} & \colhead{$\sigma$ $\pi$} & \colhead{RV} & \colhead{$\sigma$RV} & \colhead{Observatory} \\
 \colhead{} & \colhead{} & \colhead{(Mag.)} & \colhead{(Mag.)} & \colhead{(mas)}& \colhead{(mas)} & \colhead{km s$^{-1}$} & \colhead{km s$^{-1}$} & \colhead{}  }
\startdata
GJ4102	&	2023-03-26T09:30:45.239	&	8.00	&	7.69	&	80.1134	&	0.0206	&	--10.70	&	0.34	&	(1)	\\
GJ436	&	2023-04-01T02:16:46.909	&	6.32	&	6.07	&	102.3014	&	0.0302	&	9.49	&	0.33	&	(1)	\\
GJ436	&	2016-02-29-10:23:22.932	&	6.32	&	6.07	&	102.3014	&	0.0302	&	9.44	&	0.33	&	(2)	\\
Gliese486 &	2021-05-25T00:13:09.871	&	6.67	&	6.36	&	123.7756	&	0.0329	&	18.31	&	0.30	&	(1)	\\
Kepler 138	& 2022-09-14T08:14:09.090 &   9.68	&  9.51 & 14.9019  & 0.0097 & --37.12 & 0.35 & (3) \\
\enddata
\tablecomments{H and K$_{s}$ from 2MASS \citep{skrutskie06}; $\pi$ and $\sigma \pi$ from Gaia DR3 \citep{gaia23}}
\tablecomments{Observatory Sources: (1) IGRINS on Gemini South, (2) IGRINS on the 2.7 McDonald Observatory, (3) NIRSPEC on Keck}
\tablecomments{The full table is available in machine readable form.}
\end{deluxetable*}

While the interiors of small rocky planets have a direct influence over a planet's ability to support life and the identification of potential biosignatures \citep[e.g.,][]{foley16, hinkel20, lisse20, krissansentotton25}, it is not currently feasible to directly observe planet's interior or surface composition. Instead, when combined with a planet's mass and radius, stellar abundances of the rock-building elements (Fe, Mg, Si) can enable studies of planetary interior structure and mineralogy \citep[e.g.,][]{schulze21,unterborn23,schulze26}. These abundances, when converted to molar ratios (Si/Mg and Fe/Mg), are noted to be roughly equivalent between Earth, Mars, and the Sun \citep{mcdonough03,wang19} due to their (semi-)refractory nature. As a result, the stellar abundances of the rock-building elements may be used as a one-to-one proxy for an exoplanet's bulk composition within exoplanet models \citep{thiabaud15,souto18,adibekyan21,schulze21,unterborn23,adibekyan24,schulze26}. Most importantly, models that use planetary mass and radius measurements in conjunction with host star abundances can be used to deduce the classification of rocky Earth-like planets, super-Mercuries, and volatile rich mini-Neptune exoplanets \citep{schulze21,unterborn23,schulze26}, which are fundamentally different from a habitability perspective. Increasing the number of known M-dwarf planet-hosting abundances will enhance interior structure modeling especially as large scale Mg and Si abundances \citep{tabernero24} for targeted systems \citep{souto18,weisserman26} become more prevalent. 

In this paper, we have measured the atmospheric parameters and rock-building block elements (Fe, Mg, Si) and O for 40 exoplanet hosting M-dwarfs. Our target selection, high-resolution IR spectroscopic observations, and data reduction methodology are described in section \ref{sec:sample_and_observations}. The abundance analysis and uncertainty calculations are explained in section \ref{sec:atmospheric_analysis}. A discussion of our results, including comparisons to the literature, M-dwarf disk populations status, and molar abundance ratios, are disseminated in section \ref{sec:discussion}. Finally, our conclusions are listed in section \ref{sec:conclusions}. 

\section{Sample Selection and Data Reduction}
\label{sec:sample_and_observations}

\subsection{Sample Selection Criteria and Spectra Sources}

\begin{figure*}
\centering
	\includegraphics[width=6.5in]{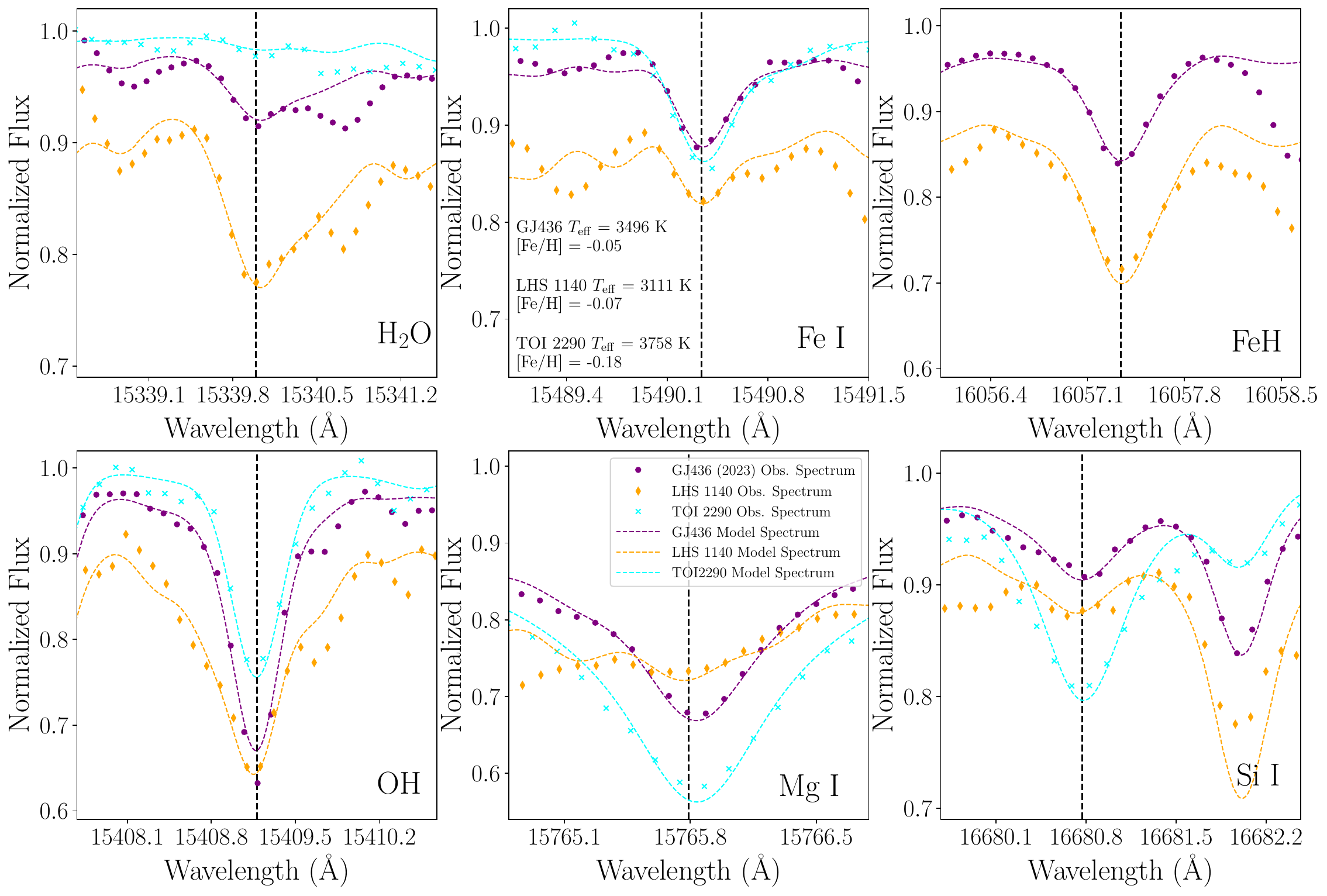}
    \caption{Example fits to the observed spectrum of GJ 436 ($\teff{}$ = 3495 K, [Fe/H] = -0.05 dex, 2023 observation), LHS 1140 ($\teff{}$ = 3111 K, [Fe/H] = --0.07 dex), and TOI 2290 ($\teff{}$ = 3758 K, [Fe/H] = --0.18 dex) are shown as solid purple, gold, and cyan points, respectively. GJ 436 and LHS 1140 spectra were obtained with IGRINS while the TOI 2290 spectrum was obtained with NIRSPEC. The FeH line falls in an order gap for our NIRSPEC spectra and is not shown in this figure. Best fitting synthetic spectra are shown as dashed lines. Synthetic spectra creation and fitting methodology is described in section \ref{sec:atmospheric_analysis}.} 
    \label{fig:spectra_fits}
\end{figure*}

Our sample of planet-hosting M-dwarfs was chosen from the NASA Exoplanet Archive \citep{christiansen25}. Stars that hosted at least one exoplanet with a radius $<$ 1.5 R$_{\oplus}$ were initially selected for observation to maximize our likelihood that these planets were rocky and not mini-Neptunes. M-dwarfs with expected $\teff{}$ $\lesssim$ 3000 K, as estimated from photometry and spectral type, were removed from the sample due to increasingly complicated atmospheres (e.g., dust formation, \citealt{allard12}) and very weak atomic absorption features. Stars with K$_{s}$ $>$ 13 mags were not targeted due to the challenges obtaining high signal-to-noise ratios (SNRs) $\gtrsim$ 100 per pixel to model weak atomic absorption features in cooler or relatively metal-poor objects. Fits to features are shown in Fig. \ref{fig:spectra_fits} demonstrating representative absorption features in our spectra. The figure also shows how the continuum and absorption line strengths change for high temperature M-dwarfs (TOI 2290 with $\teff{}$ = 3758 K) to cooler stars (LHS 1140 at $\teff{}$ = 3111 K).

High resolution H-band IR spectra were directly obtained for this work or accessed from archived observations \citep{sawczynec25}. This bandpass was chosen as it was used in previous analysis to measure M-dwarf abundances (e.g., \citealt{souto17,souto22,tabernero24,hejazi24}) and contains relevant lines for rock-building elements for our study. We obtained spectra using IGRINS (1.45 - 2.45 $\mu$m at R $\sim$ 45,000) on Gemini South \citep{yuk10,wang10,gullysasntiago12,moon12,han12,park14,oh14,jeon14,mace16,mace18} and using NIRSPEC on Keck \citep{mclean98,martin18}. A total of 39 spectra were obtained from IGRINS on Gemini South, 2 spectra from IGRINS on the McDonald Observatory 2.7m telescope, and 13 spectra from NIRSPEC on Keck. For NIRSPEC, observations were obtained using the 0.288" slit to achieve a resolution of R$\sim$37,500, spanning 1.43 - 1.79 $\mu$m. In some cases, target stars were observed multiple times and had observations for different programs available in the IGRINS archive. The Keck and IGRINS spectra in Fig. \ref{fig:spectra_fits} illustrate how each instrument has sufficient resolution to detect critical atomic and molecular features (line selection criteria is discussed in section \ref{sec:atmospheric_analysis}). A list of our final stars, the telescope used to acquire the data, and radial velocities (RVs) are shown in Table \ref{table::obslog}. 

\subsection{IGRINS and Keck NIRSPEC Data Reduction}

For both facilities, each object was observed in an ABBA pattern, where the object was nodded between two `A' and `B' positions on the slit. During the data reduction process, the A and B images were subtracted from one another to remove sky background and bias voltage. A0V standards were observed after each M-dwarf and used for telluric correction. 

Both newly acquired IGRINS observations and archival data were processed using the \texttt{IGRINS PLP v3} data reduction pipeline \citep{kaplan24,sawczynec25}. The pipeline has several updates including improved telluric correction which especially reduces impact for \water{} lines near $\sim$15000 $\ang$ and for the Mg I lines near $\sim$15740 - 15770 $\ang{}$. The wavelength solution was derived from skylines and the line position precision was improved with a flexture correction in the v3 pipeline. 

Keck NIRSPEC Data reduction was completed using \texttt{IRAF} data reduction software \citep{tody86,tody93}. Cosmic rays were first removed from images using the \texttt{cosmicrays} task. The flat field images were combined then corrected using an averaged dark image. The images were corrected for scattered light using the \texttt{apscatter} task and extracted using \texttt{apall}. Wavelength corrections were determined using skylines extracted from the object frames. The positions of skylines varied between A and B positions by several pixels in the bluest order and by a single pixel in the reddest order. The wavelength solution was derived separately for both slit positions with the \texttt{ecidentify} task using OH lines positions from \citet{rousselot00,oliva15,brooke16} and O$_{3}$ lines from \citet{oliva15}. Some blended doublets from \citet{rousselot00} are resolved at the higher resolution of NIRSPEC and updated line positions were adopted from \citet{brooke16}. The spectra were combined and telluric corrected using the \texttt{telluric} task. 

\subsection{Radial Velocity Corrections and Initial Normalization}

The final reduced spectra were normalized using \texttt{iSpec} \citep{blancocuaresma14,blancocuaresma19} to determine RVs, provided in Table \ref{table::obslog}. The fitting process removes both the stellar continuum and the psuedo-continuum created from a large quantity of \water{} lines in the infrared H-band (e.g., \citealt{hejazi24}), with additional steps explained in section \ref{subsec::teff}. Setting the regions near prominent absorption lines to unity is sufficient for the initial radial velocity measurements. Once normalized, the spectra were cross-correlated with an M-dwarf template spectrum of a representative star at $\teff{}$ = 3500 K, logg(g) = 5, and [Fe/H] = 0. The template spectrum was generated using the methodology discussed in section \ref{sec:atmospheric_analysis}. Uncertainties on each RV measurement were determined by \texttt{iSpec} and typically ranged from 0.2 - 0.4 km s$^{-1}$. 

\begin{figure}
\centering
	\includegraphics[width=3.3in]{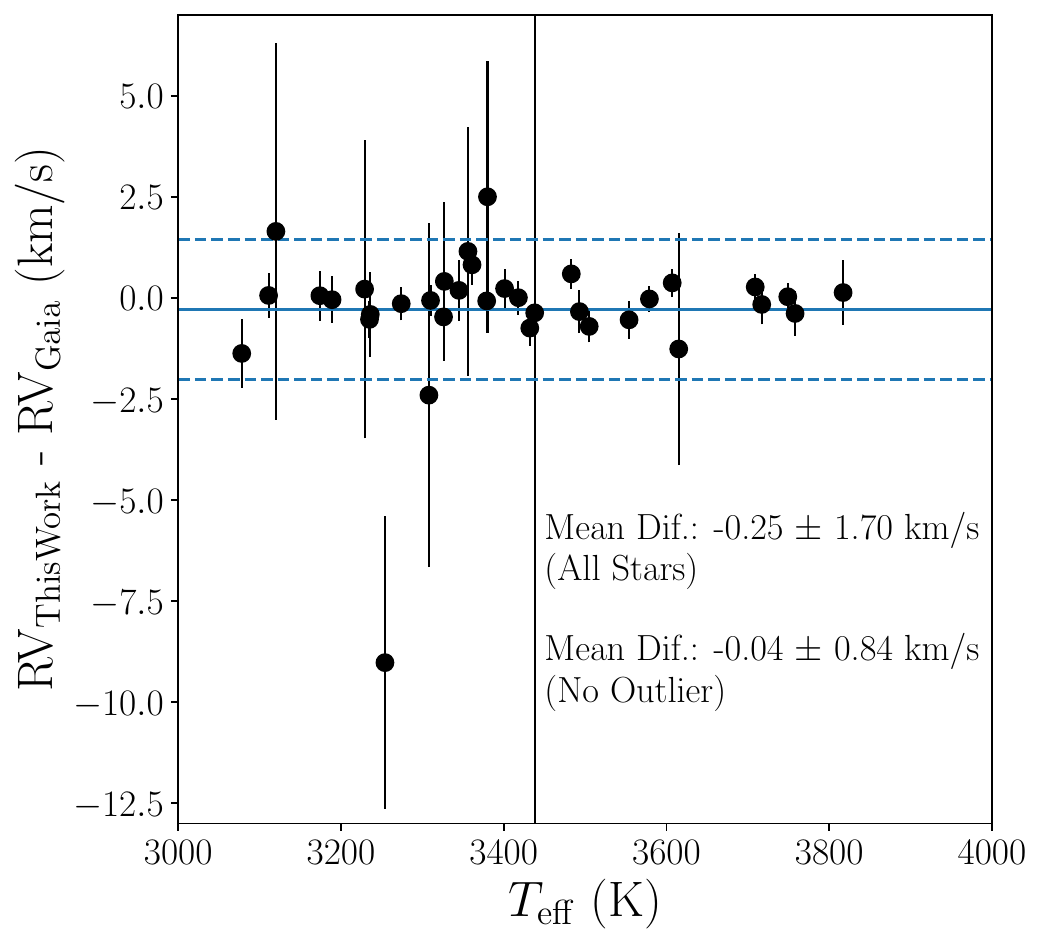}
    \caption{A RV comparison between this work and Gaia DR3 values \citep{gaia23} as a function of effective temperature. Uncertainties on the RV difference are propagated errors from both data sets. The blue solid line is the mean difference, the dashed lines are the 1$\sigma$ standard deviation using the total star sample. The mean difference listed without the outlier star only excludes TIC 439867639, which has an RV difference of --9.02 km s$^{-1}$.} 
    \label{fig:rvs}
\end{figure}

Our RVs were compared to Gaia DR3 values \citep{gaia23} in Fig. \ref{fig:rvs}, and we find good agreement for stars in common with both studies. An average difference of --0.25 $\pm$ 1.7 km s$^{-1}$ was determined between both data sets. One star, TIC 439867639, has a Gaia RV = 33.85 $\pm$ 3.59 km s$^{-1}$ and we found an RV = 24.83 $\pm$ 0.43 km s$^{-1}$. Our value matches a previous APOGEE RV measurement of 24.1970 $\pm$ 0.1136 km s$^{-1}$ \citep{apogeedr17}. Removing the outlier measurement leads to a mean difference of --0.04 $\pm$ 0.84 km s$^{-1}$, consistent with propagated average uncertainties from both studies. The lack of systematic error with $\teff{}$ suggests the synthetic spectra are appropriately modeling strong line locations across the spectral range. 

\section{M-dwarf Atmospheric Analysis}
\label{sec:atmospheric_analysis}
\subsection{Spectral Synthesis Tools and Linelist Construction}

Synthetic spectra for the atmospheric analysis was calculated using MOOG (v. 2019 \citealt{sneden73}). The 2019 version of MOOG was upgraded with the ability to model some triatomic molecules, most importantly \water. MARCS model atmospheres \citep{gustafsson08} were also used when generating synthetic spectra, which have been previously used for M-dwarf abundance analysis \citep{abia20,ishikawa22,souto22,melo24}. We did not interpolate the MARCS model atmosphere grid and instead adopted existing models that most closely matched our stars in $\teff{}$, log(g), and [Fe/H]. Common interpolation grids do not recommend extending past 3800 K (e.g., \texttt{INTERPOL MARCS} Fortran Code\footnote{\url{https://marcs.astro.uu.se/software.html}}) and previous work has found interpolation of cool atmospheres ($\teff{}$ $<$ 4000 K) are more likely to be subjected to systematic errors \citep{wheeler24}. We applied a microturbulence of 1 km s$^{-1}$ for each star, consistent with previous M-dwarf analysis in the literature (e.g., \citealt{souto17,souto22}). All abundances were normalized to the \citet{asplund09} scale. Lastly, we used photometric surface gravities for all spectral synthesis, with parameter derivation described in section \ref{subsec:phot}.

Atomic H-band lines were incorporated from the APOGEE linelists \citep{shetrone15,smith21}. Molecular line data was adopted from a variety of sources, with most data accessed from the ExoMol database, specifically \water{} \citep{polyansky18} and CO \citep{li15}. ExoMol data was translated to HITRAN format using available scripts\footnote{\url{https://www.exomol.com/software/}} then converted to MOOG compatible formats. Other molecular species, such as TiO \citep{mckemmish17} and VO \citep{mckemmish16} were synthesized in the IR H-band but absorption was less than 1$\%$ of the continuum in our bandpass. FeH data was adopted from two sources; the F$^{4}\Delta_{i}$-X$^{4}\Delta_{i}$ transitions information was from \citet{dulick03} while E$^{4}\Pi$-A$^{4}\Pi$ FeH line data was from \citet{hargreaves10}. We note only lines with known transition probabilities were added to the linelist from \citet{hargreaves10} and we did not include lines with oscillator strengths calculated from line intensities. OH \citep{brooke16} and CN \citep{brooke14} were added to the linelist although CN lines are much weaker in M-dwarfs compared to cool giant stars. Lastly, other weak lines from additional species have previously been tested and few detectable lines are expected to impact the spectra in the H-band \citep{flores24}.

\begin{figure*}
\centering
	\includegraphics[width=7.0in]{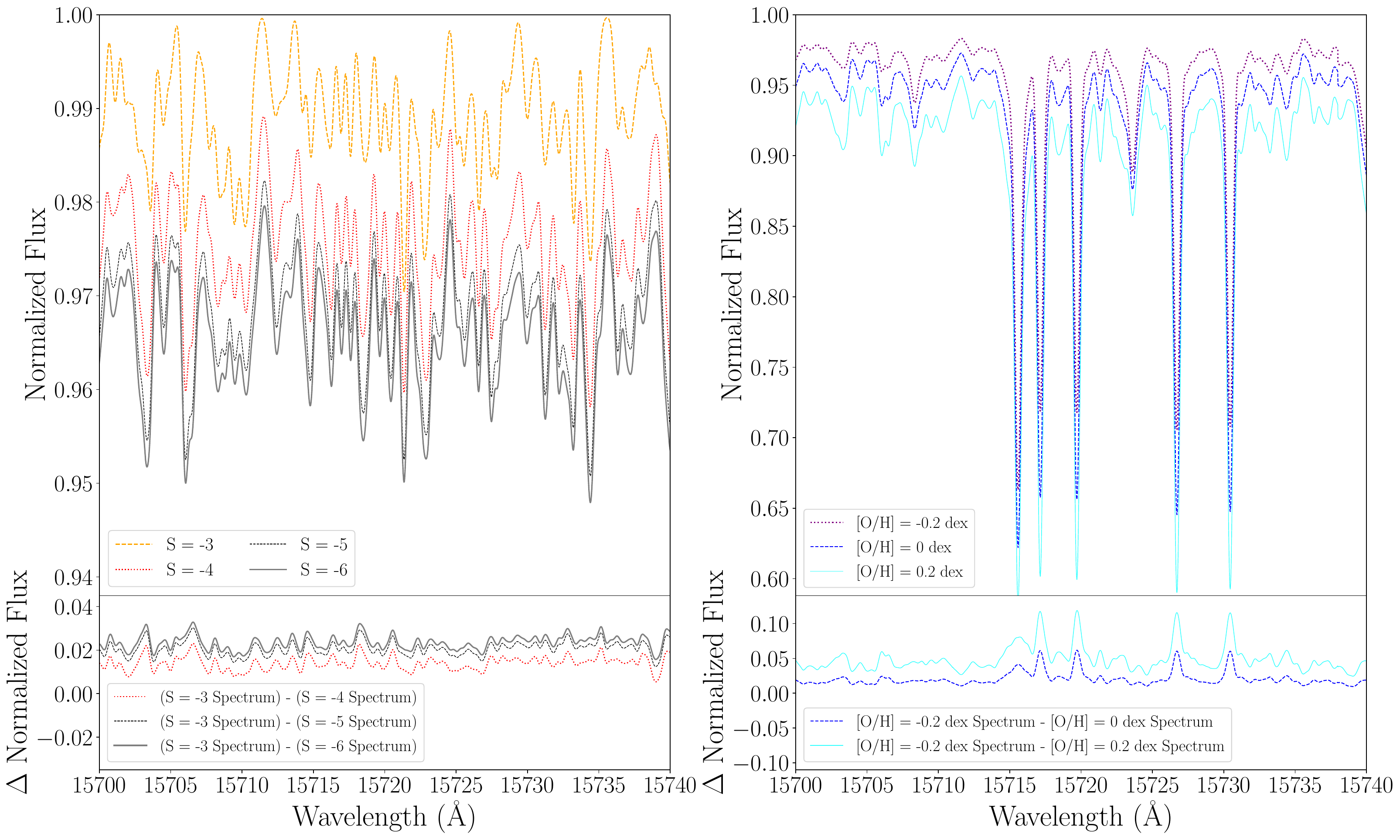}
    \caption{Left Panels: Spectral synthesis of \water{} linelists created using different \water{} line strength cut-off criteria (S = log(gf$\lambda$)-($\Theta$ * $\chi$)). Smaller values of S means a larger number of weak water lines were included in the linelists. Right Panels: Three synthetic spectra created using the full linelists with different [O/H] abundances, displayed over the same wavelength region as the top panel. Synthetic spectra for both plots were created using an atmosphere with $\teff{}$ = 3500 K, log(g) = 5.0 dex, and [Fe/H] = 0.0 dex. The two top panels show the generated synthetic spectra while the bottom panels display differences with the synthetic spectra.}
    \label{fig:water_line_strength}
\end{figure*}

The number of \water{} absorption features in the IR H-band is significant and potentially thousands of lines can impact each region in the synthetic spectrum. Strong \water{} absorption features are detectable and the weak lines combine to create a pseudo-continuum that impact abundance derivations of other lines. We followed the methodology of \citet{smith21} and characterized line strength with S = log(gf$\lambda$)-($\Theta$ * $\chi$), where $gf$ refers to the line oscillator strength multiplied by a statistical weight, $\Theta$ = 5040/$\teff{}$, and $\chi$ is the excitation potential of each line. We synthesized linelists that contained only \water{} lines with different line strength cut-offs, shown for wavelengths near the Mg I lines used for Mg abundance derivation, in Fig. \ref{fig:water_line_strength}. An atmospheric model with solar metallicity, $\teff$ = 3500 K, and a log(g) = 5, was used when creating the synthetic spectra. Each linelist only contained \water{} lines to isolate their effect on the continuum. We continued generating synthetic spectra until the continuum depth remained approximately constant at levels detectable in our IR spectra, at an strength of S = 6 as shown in Fig. \ref{fig:water_line_strength} (left panels). Therefore, we adopt a line strength cut of S $<$ --6. At this level, the associated continuum depth is lowered by $\sim$2 - 3 $\%$. We also found the going from S = --3 to the S = --4 model introduce weak blends, while additional weak lines primarily contribute to the continuum, as seen in Fig. \ref{fig:water_line_strength}.

\subsection{Photometric Parameters}
\label{subsec:phot}

The M-dwarf atmospheric parameters were initially estimated using \citet{mann15} photometric relationships. The V-band magnitude was calculated using Gaia DR3 photometry \citep{gaia16,gaia23} and a conversion in the Gaia documentation\footnote{\url{https://gea.esac.esa.int/archive/documentation/GEDR3/Data_processing/chap_cu5pho/cu5pho_sec_photSystem/cu5pho_ssec_photRelations.html}}. The absolute $\teff{}$s were calculated from the BP-RP and J-H colors, along with the relationships in \citet{mann15} using 2MASS photometry \citep{skrutskie06}. Reddening corrections were not applied to the photometry since the target stars are relatively close. Masses and radii were calculated using absolute V-band magnitude and formulae from \citet{mann15} for radii and masses from \citet{delfosse00}. The surface gravity was then calculated using the mass and radii measurements for each star. The parameters can be found in Table \ref{tab:star_final_params}. 

\subsection{Effective Temperature Measurements and Oxygen Abundances}
\label{subsec::teff}

We determined [O/H] abundances and $\teff$ for our M-dwarf stars following the methodology of \citet{souto17,souto20}. Both OH and \water{} line strengths have different behavior as a function of $\teff$, where \water{} decreases with increasing $\teff$ and OH line strengths are relatively constant for a given [O/H] abundance. We created initial synthetic spectra of only \water{} lines and OH lines. Then we identified features that met a minimum threshold of 3$\%$ of the continuum at the center of the lines at each $\teff{}$ to ensure that a common set of lines were used for all stars. The observed spectrum was scaled to match the pseudo-continuum for each model within the spectral synthesis grid. Abundances were derived for each line using a $\chi^{2}$ minimization routine that centered over a 1 \ang{} region around each line center to minimize influence from nearby blends. The $\chi^2$ reduced statistic was calculated for each initial fit and all line profiles were visually inspected for blends. Features that were either blended, had $\chi^2$ reduced values $>$ 10, or gave anomalous abundances $>$ 3$\sigma$ of the line average for $>$50\% of the spectra were removed from the final set of OH and \water{} lines used in the analysis. We identified 19 OH lines and 5 \water{} lines that typically gave consistent abundance results across our sample (example lines analyzed shown in Fig. \ref{fig:spectra_fits}). A linelist of our OH and \water{} lines is given in Appendix A. [O/H] abundances were derived before other elements, like Fe, since O abundances affect the psuedo-continuum contribution from \water{} lines. To demonstrate the impact on the continuum from different O abundances, we show synthetic spectra at different O abundances for a $\teff{}$ = 3500 K, log(g) = 5.0, [Fe/H] = 0.0 dex atmospheric model in the right panels of Fig. \ref{fig:water_line_strength}.

\begin{figure*}
\centering
	\includegraphics[width=7in]{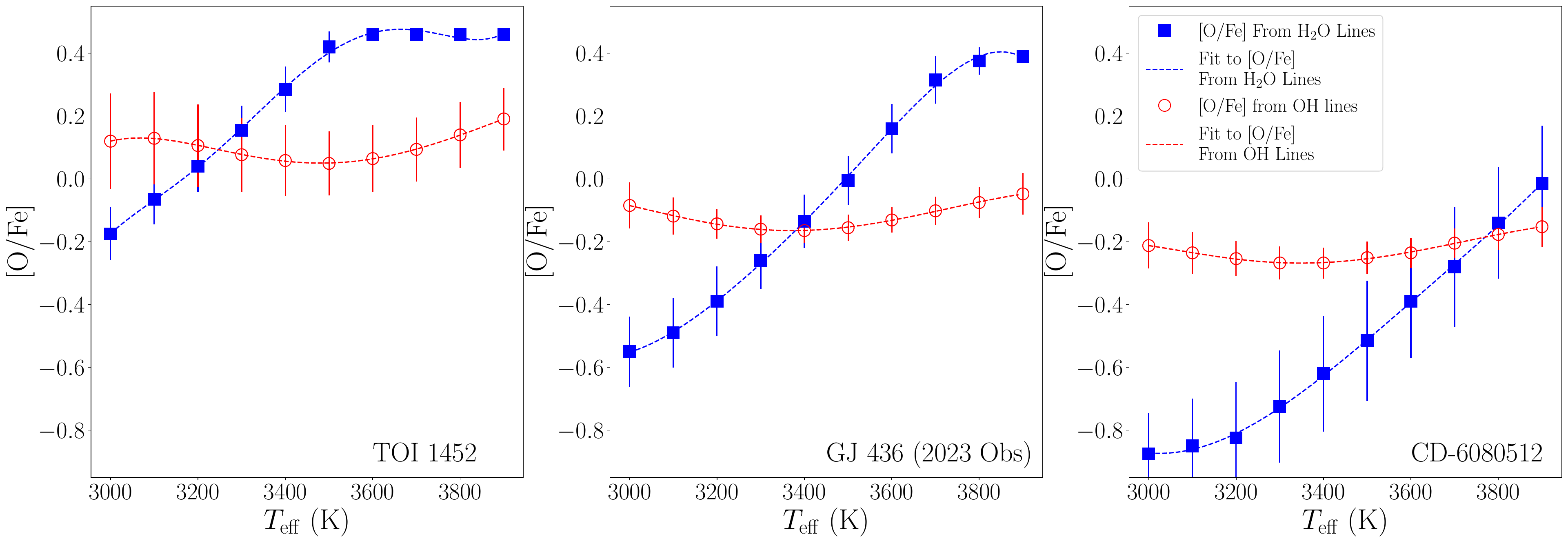}
    \caption{In this figure we illustrate the methodology to derive [O/H] abundances and $\teff{}$ for three stars (TOI 1452, GJ 436, and CD+608051). The O abundances derived using \water{} (blue squares) and OH lines (red circles) are plotted for different model atmospheres in steps of 100 K. A polynomial fit to each set of abundances is shown as solid red (OH) and blue (\water{}) lines. The $\teff$ that gives a consistent abundance is adopted as the final $\teff{}$ for each star.} 
    \label{fig:temp_oh}
\end{figure*}

Next, we created grids of synthetic spectra that varied in $\teff$ from 3000--4000 K in steps of 100 K, and varied in [O/H] abundance by 0.02 dex per model. Separate grids were made at log(g) values of 4, 4.5, 5, and 5.5. Other abundances were scaled with O and we initially used this element as a proxy for metallicity to ensure the molecular equilibrium was appropriate for our stars (e.g., in order to avoid improperly creating a carbon star for low [O/H] abundance models). We matched the log(g) to the photometrically derived value, given in Table \ref{tab:star_final_params}. Normalization followed three steps. First, the observed spectrum was normalized using a spline3 function within \texttt{iSpec} \citep{blancocuaresma14,blancocuaresma19} to remove the instrument profile, stellar continuum, and molecular line \water{} psuedocontinuum. Second, $\sim$8 \ang{} regions around each line were scaled to match the model spectrum and the generated synthetic spectrum was used to correct for the \water{} contributions to the spectrum. The narrow region was scaled by the average of the highest 5$\%$ quantile flux  from the synthesize spectra. Lastly, the OH and \water{} lines were visually inspected to ensure the fits were appropriate. 

The average OH and \water{} abundances derived for our sample of lines were compared at each $\teff$ step, as shown in Fig. \ref{fig:temp_oh}, where the plots illustrate that the OH lines give consistent abundances over a specific temperature range while \water{} lines behave differently. A fourth order polynomial was fit to each abundance distribution curve and we determined the $\teff{}$ where the OH line and the \water{} line abundance averages gave consistent [O/H] abundances. Effective temperature uncertainties were determined using a Monte Carlo simulation. We treated each O abundance for a given $\teff$ as a Gaussian distribution with a mean and standard deviation consistent with the same statistical values of our sample of OH and \water{} lines. We re-fit the abundances 5000 times using Gaussian random numbers. The 16$\%$ and 84$\%$ values from our Monte Carlo $\teff{}$ distribution as the 1$\sigma$ errors on the final measurements. Final $\teff{}$ and uncertainties are listed in Table \ref{tab:star_final_params}. For stars with [Fe/H] $<$ --0.2, we iterated the atmospheric parameter derivation and further abundance analysis with updated model atmospheres until the parameters matched the data. 

\begin{figure}
\centering
	\includegraphics[width=3.3in]{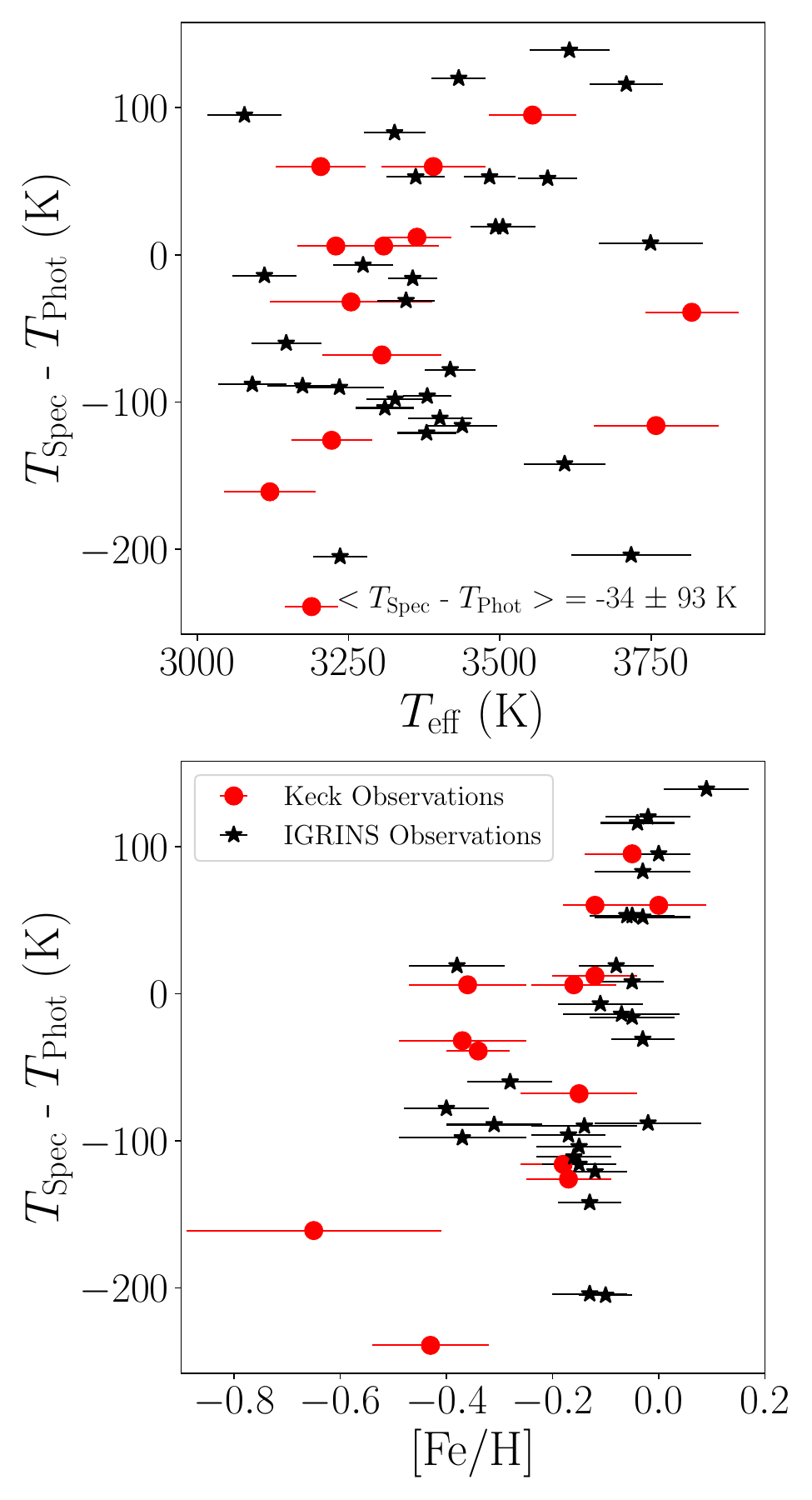}
    \caption{The difference between photometrically derived effective temperatures from \citet{mann15} and our abundances, as a function of $\teff{}$ (top panel) and [Fe/H] (bottom panel) are shown. Our sample observed by NIRSPEC on Keck is highlighted in red and stars with IGRINS spectra are black. } 
    \label{fig:teff_phot_comp}
\end{figure}

We compared our results to photometrically derived $\teff$ to our measurements from spectroscopy and found that the mean difference is $< \teff{}_{Spec} - \teff{}_{Phot} >$ = --34 $\pm$ 94 K. We examined whether the temperature differences increase or decrease with stellar $\teff{}$ in Fig. \ref{fig:teff_phot_comp} and find no large systematic offset in $\teff{}_{Spec}$ - $\teff{}_{Phot}$ for our cooler M-dwarfs compared to the hottest stars in our sample. We also find no large offset between $\teff{}$s derived from NIRSPEC spectra and IGRINS spectra. Typical uncertainties for \citet{mann15} are $\sim$60 K and our average uncertainty is  
54 K with a standard deviation of 13 K. The uncertainties added in quadrature are 81 K, similar to the total dispersion observed from our comparison sample. Our total scatter is also similar to \citet{souto20} which used a similar $\teff{}$ derivation methodology and found $< \teff{}_{Souto20} - \teff{}_{Mann15} >$ = 46 $\pm$ 90 K. We also examine our data to check for $\teff$ and metallicity dependent systematic differences with results shown in Fig. \ref{fig:teff_phot_comp}. No clear slope with $\teff$ is apparent and this suggests that there is no systematic error with with our $\teff{}$ derivation method exists. We do find that [Fe/H] = --0.25 model atmospheres lead to lower overall abundances and this effect is further discussed in section \ref{subsec::systematic_uncertainty}.

\begin{deluxetable*}{ l c c c c c c c c c c c c c c}
\tablewidth{0pt} 
\tabletypesize{\scriptsize}
\tablecaption{Stellar Parameters and Abundances \label{table::multiple_visits}} 
 \tablehead{\colhead{Name} &  \colhead{$\teff{}$} & \colhead{log(g)} & \colhead{Model} &  \colhead{Model} & \colhead{$\teff{}$} & \colhead{$\sigma \teff{}$} & \colhead{[Fe/H]} & \colhead{$\sigma$[Fe/H]} & \colhead{[O/H]} & \colhead{$\sigma$[O/H]} & \colhead{[Mg/H]} & \colhead{$\sigma$[Mg/H]} & \colhead{[Si/H]} & \colhead{$\sigma$[Si/H]} \\ \colhead{} & \colhead{Phot.} &  \colhead{Phot.} & \colhead{log(g)} & \colhead{[Fe/H]} & \colhead{Spec.} &  \colhead{Spec.} &  \colhead{} &  \colhead{} &  \colhead{} &  \colhead{} &  \colhead{} &  \colhead{} \\ \colhead{} & \colhead{(K)} &  \colhead{} & \colhead{} & \colhead{} & \colhead{(K)} &  \colhead{(K)} &  \colhead{} &  \colhead{} &  \colhead{} &  \colhead{} &  \colhead{} &  \colhead{}}
\startdata
BD-17588	&	3324	&	5.00	&	5	&	0	&	3235	&	73	&	--0.14	&	0.10	&	--0.07	&	0.05	&	--0.08	&	0.04	&	0.06	&	0.10 \\
CD-608051	&	3741	&	4.70	&	4.5	&	0	&	3749	&	86	&	--0.05	&	0.06	&	--0.19	&	0.05	&	--0.02	&	0.07	&	0.05	&	0.04 \\
G162-44	&	3312	&	4.88	&	5	&	0	&	3432	&	45	&	--0.02	&	0.08	&	0.13	&	0.03	&	--0.06	&	0.05	&	--0.09	&	0.09 \\
G50-16	&	3308	&	4.90	&	5	&	0	&	3361	&	48	&	--0.06	&	0.07	&	0.16	&	0.04	&	--0.07	&	0.06	&	--0.19	&	0.10 \\
GJ4102	&	3281	&	4.98	&	5	&	0	&	3274	&	50	&	--0.11	&	0.08	&	--0.08	&	0.05	&	--0.2	&	0.05	&	--0.09	&	0.09 \\ 
\enddata
\tablecomments{$\teff{}_{Phot}$ is derived from \citet{mann15} and solar abundances are from \citet{asplund09}}
\tablecomments{The full table is available in machine readable form.}
\label{tab:star_final_params}
\end{deluxetable*}

\subsection{Mg, Si, and Fe Abundances}

After [O/H] abundances were determined, grids of synthetic spectra were generated in 15 $\mbox{\AA}$ windows around selected Fe I and FeH features. Fe I and FeH lines were selected from \citet{souto17} and strong lines identified from initial synthesis were examined in a similar way to OH and \water{} lines discussed in section \ref{subsec::teff}. Each grid was varied in steps of $\delta$[Fe/H] = 0.02 dex, with spectroscopic $\teff{}$, photometric log(g) values, and other elements initially scaled to O. Once the grids were generated, the best fitting Fe abundances were derived using a $\chi ^{2}$ minimization, with the same routine as OH and \water{} abundance measurements. Each individual line was inspected for each star to ensure the continuum fit was appropriate, there were no unusual blends, and no wavelength shifts were needed. Individual line abundances were averaged to determine our final abundance, with the standard deviation from multiple lines used to determine the statistical uncertainty. 

\begin{figure}[h!]
\centering
	\includegraphics[width=3.3in]{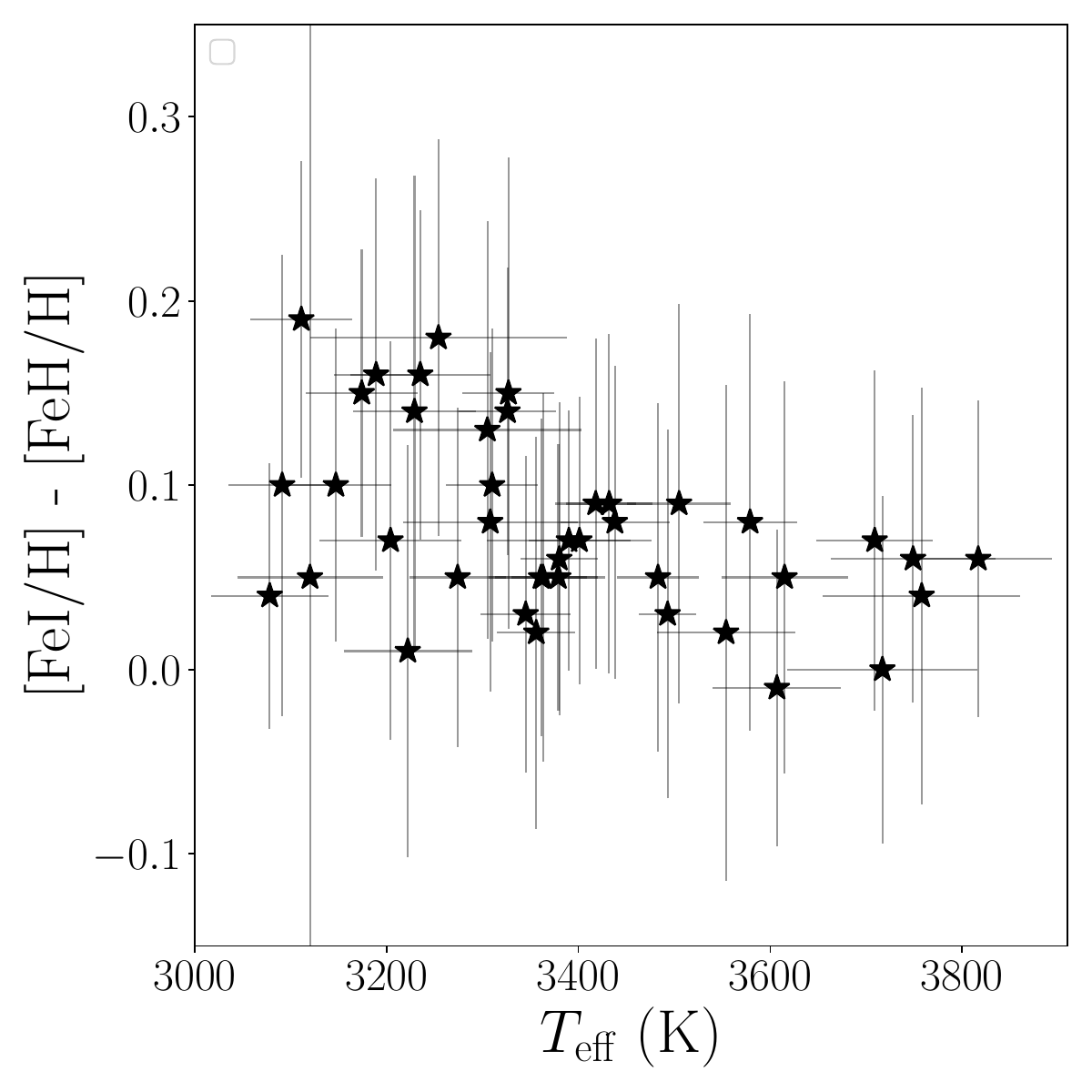}
    \caption{The difference between Fe abundances derived with Fe I lines and molecular FeH lines is shown as a function of $\teff{}$.} 
    \label{fig:iron}
\end{figure}

When looking at the abundances from Fe I versus FeH lines, we see a systematic difference that causes an average offset of 0.08 $\pm$ 0.05 dex. This abundance offset is similar to M-dwarf analysis from \citet{souto20}, who found an average difference of $< [\mathrm{Fe I/H}] - [\mathrm{FeH/H}]>$ = 0.12 $\pm$ 0.12 dex. However, we found the total offset increases with decreasing $\teff{}$, as shown in Fig. \ref{fig:iron}. Additionally, some FeH lines gave discrepant abundances on the order of -0.5 dex from the average ratio (primarily the FeH feature at 16935.09 \ang{}) for the coolest stars in our sample. In these instances, the FeH lines were not included in our final average abundances. The increasing offset may be due to our lack of model interpolation systematically affecting the coolest stars that fall between 5.0 $<$ log(g) $<$ 5.5, issues with computing each source of Fe (e.g., Fe depletion onto forming dust grains, \citealt{allard12}), or complications measuring Fe I for the coolest stars where only Fe I lines at 15207.53 \ang{} and 15490.339 \ang{} were consistently detectable. Abundance studies of M-dwarfs in wide-binaries with FGK stars and M-dwarfs in open clusters have found that FeH line parameters are likely not causing the abundance offsets \citep{silva-andrade26}. Some lines also fell into order gaps in the Keck spectra, such as the FeH line at 16050.3 \ang{}, and were not measured in stars observed by NIRSPEC. Caution may be warranted for future M-dwarf studies when interpreting FeH abundances for similarly cool M-dwarfs near $\teff{}$ $\sim$ 3000-3400 K. Some Fe I lines were also only detectable in the warmest M-dwarfs and were not suitable for abundance analysis in cooler stars. Line-by-line abundances for Fe, Mg, and Si can be found in Appendix A in Table 5. 

Mg and Si abundances followed the same methodology as Fe I lines, although only 3 Mg I lines and 3 Si I lines were found to be suitable for abundance analysis. For cooler stars, only 1-2 of the Mg I lines and 1 Si I line (16680.77 \ang{}) was reliably detectable (see Fig. \ref{fig:spectra_fits}). Due to the dearth of lines, statistical uncertainties for Mg, Si, and Fe were determined using a Monte Carlo simulation. Each flux value was treated as a random Gaussian number, with a standard deviation consistent with the SNR of the observation. For Keck spectra, we applied an SNR floor of 100 per pixel due to the difficulty estimating the SNR from the spectra itself. The IGRINS pipeline provided H-band SNR values for each pixel that were used for final abundance uncertainty calculations. Final abundances are listed in Table \ref{tab:star_final_params}. 

We used observations of our target stars across multiple dates to compare to the statistical uncertainties on our abundance measurements from our Monte Carlo fitting routine. Derived abundances for the nine program objects with multiple visits are shown in Table 6 in Appendix A. We find the $\teff{}$s and each abundance measurement are consistent within the uncertainties. We also note that [Si/H] has the largest star-to-star dispersion which increases with decreasing $\teff{}$. Large Si uncertainties are expected since atomic Si I lines are weakest in the coolest stars. Final uncertainties in Table \ref{tab:star_final_params} include the statistical uncertainty added in quadrature with uncertainties derived from varying the continuum by $\pm$ 0.005, as discussed in section \ref{subsec::systematic_uncertainty}.

\subsection{Additional Abundance Uncertainty Sources}
\label{subsec::systematic_uncertainty}

We varied the model parameters and re-measured abundances to quantify potential systematic uncertainties from atmospheric parameter errors. We chose five representative stars across our $\teff$ range (2MASSJ09442986-4546351, CD-608051, GJ 436, L375-2, LHS 1140), which varied from 3111 K - 3749 K. We re-created synthetic spectra grids for each element by varing the atmospheric model temperature by $\pm$ 100 K or log(g) by $\pm$ 0.5 for grids that varied [Fe/H], [Mg/Fe], and [Si/Fe]. Since O abundances and $\teff{}$ are derived simultaneously, we tested the impact when surface gravity is changed by 0.5 dex. We also tested the impact of altering the normalization level by $\pm$ 0.005 in normalized flux for the small number of [Fe/H], [Mg/H], and [Si/H]. The uncertainties were calculated for each line individually and the averages are shown in Table \ref{tab:sys_err}. 

We also quantify potential uncertainty on $\teff{}$ and the [O/H] ratio from our assumed [C/H] ratio. We created a grid of synthetic spectra at [C/O] = -0.1 dex, log(g) = 5.0, and [Fe/H] = 0.0 dex. The grid varied by $\delta$[O/H] = 0.02 dex for $\teff{}$s between  3000 – 3900 K (in steps of 100 K), following our previous grid creation methodology. We then refit $\teff{}$ and oxygen abundances for all stars with appropriate surface gravities and [Fe/H] ratios consistent with this grid. We found that measured [O/H] ratio were affected by an average of 0.09 $\pm$ 0.01 dex and $\teff{}$ was affected by –2 $\pm$ 20 K. We also inspected the line profiles for Fe I and Mg I lines between the two different grids and see minimal difference in the line profiles for each synthetic spectra with the best fitting [O/H]. We include the additional [O/H] uncertainty in Table \ref{tab:sys_err}.

\begin{deluxetable}{c c c c c}
\tablewidth{0pt} 
\tabletypesize{\footnotesize}
\tablecaption{Abundance Changes From Different Atmospheric Parameters\label{tab:sys_err}} 
 \tablehead{\colhead{Measured} & \colhead{$\teff{}$} & \colhead{log(g)} & \colhead{Normalization} &\colhead{[C/O] = -0.1 dex}  \\
 \colhead{Parameter} & \colhead{$\pm$ 100 K} & \colhead{$\pm$ 0.5 dex} & \colhead{$\pm$ 0.005} & \colhead{}}
\startdata
$\teff{}$ (K) & \nodata{} & 107 & \nodata{} & 2 \\
$\mathrm{[O/H]}$ (dex) & \nodata{} & 0.18 & \nodata{} & 0.09 \\
$\mathrm{[Fe I/H]}$ (dex) & 0.07 & 0.11 & 0.05 & \nodata{} \\
$\mathrm{[FeH/H]}$ (dex) & 0.12 & 0.03 & 0.04 & \nodata{}\\
$\mathrm{[Fe/H]}$ (dex) & 0.05 & 0.04 & 0.04 & \nodata{} \\
$\mathrm{[Mg/H]}$ (dex) & 0.12 & 0.09 & 0.04 & \nodata{}\\
$\mathrm{[Si/H]}$ (dex) & 0.10 & 0.12 & 0.06 & \nodata{}\\
\enddata
\end{deluxetable}

\begin{figure*}
\centering
	\includegraphics[width=7.2in]{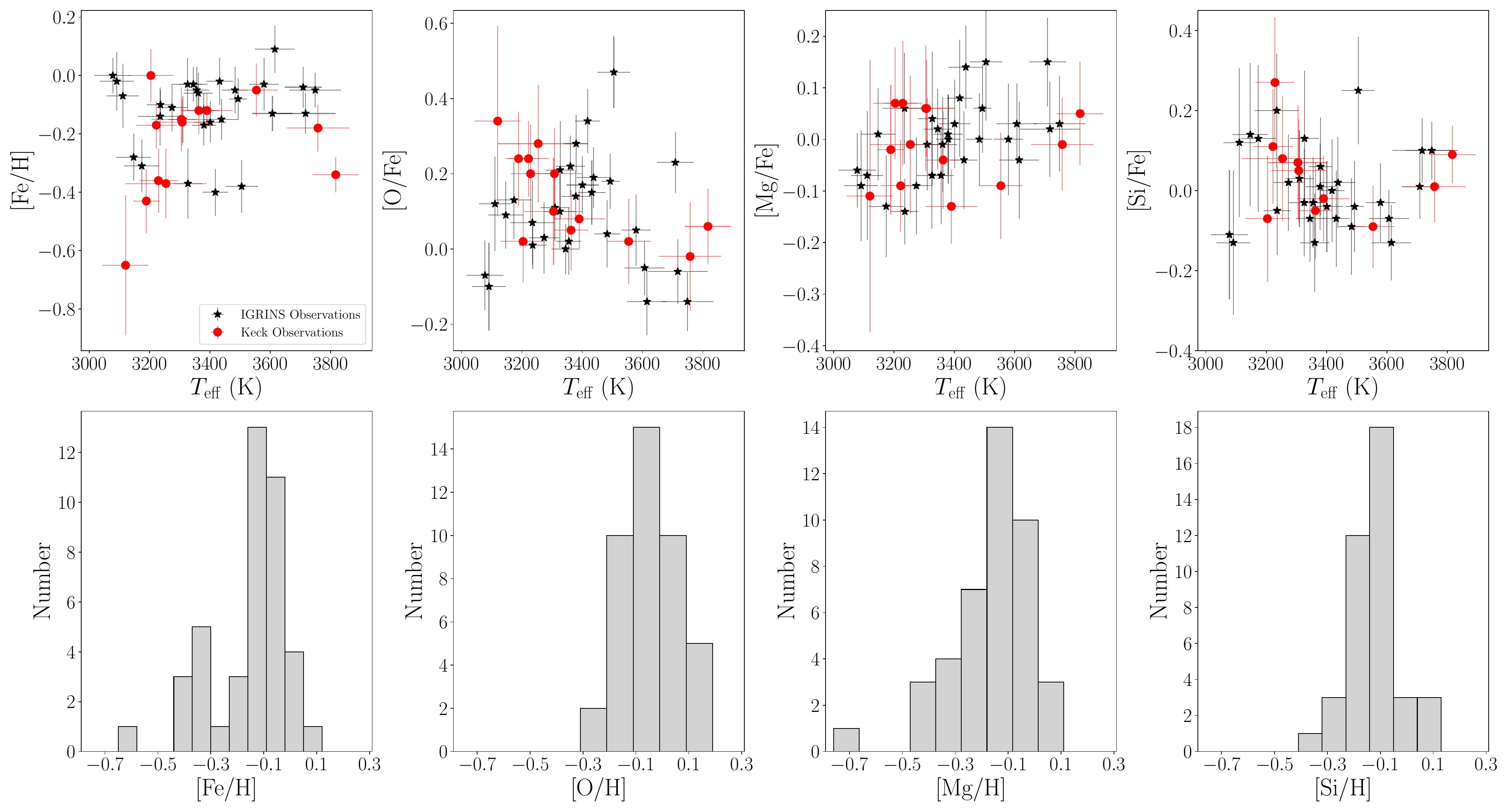}
    \caption{Top Panel: Measured abundance ratios are shown as a function of $\teff{}$ for our sample observed by NIRSPEC on Keck (in red) and stars with IGRINS spectra (black). Bottom Panel: Histograms for each [X/H] are shown for the full M-dwarfs sample. Bin sizes were chosen using the Knuth method \citep{knuth06} with the exception of [Si/H] where the average uncertainty of 0.09 dex was used.}
    \label{fig:teff_abundances}
\end{figure*}

Since we do not interpolate model atmospheres, the uncertainties are the maximum values given since our average $\teff$ uncertainty is 54 K and log(g) from photometry is not expected to differ by $\sim$0.5 dex. As a result, the uncertainties in Table \ref{tab:star_final_params} only include the statistical uncertainty. Each uncertainty value was also typically consistent for our five example stars, with the exception of the normalization related abundance uncertainty for Si I.  Since this line becomes extremely weak at low $\teff{}$s, the final uncertainty was $\teff$-dependent and we applied uncertainties of $\delta$[Si/H] = 0.04 for $\teff{}$s between 3500--3900 K, $\delta$[Si/H] = 0.08 dex for $\teff{}$ between 3200--3500 K, and anything cooler had a $\delta$[Si/H] = 0.13 dex added to the final uncertainty.  

Fig. \ref{fig:teff_phot_comp} and the bottom row of Fig. \ref{fig:teff_abundances}, also highlight potential offsets in Fe abundances for analyses that utilized atmospheric models at [Fe/H] = 0.0 dex compared to models with [Fe/H] = --0.25 dex. Other [X/H] elemental abundances do not show bimodal distributions based on the histograms in Fig. \ref{fig:teff_abundances} and only [Fe/H] abundances appear potentially impacted by a systematic effect. Bin sizes in the histogram where chosen based on the Knuth method \citep{knuth06} which led to sizes of 0.10 dex for [O/H], 0.07 dex for [Fe/H], and 0.10 dex for [Mg/H]. The methodology recommends a bin size of 0.05 dex for [Si/H]. However, since some stars had no measurable Si I lines, and there is a $\teff{}$-depending uncertainty when fitting Si I features for cooler M-dwarfs, we adopt a bin size of 0.09 dex to match the average uncertainty in Table \ref{tab:star_final_params}. We calculated the difference in [Fe/H] abundance for the nine relatively metal-poor stars that required an atmospheric model with [Fe/H] = --0.25 dex, as compared to the generated models at [Fe/H] = 0 dex, and found an average difference of [Fe/H] = --0.12 $\pm$ 0.03 dex. Since these are the most metal poor objects that are least appropriately fit using a MARCS model at [Fe/H] = 0.0 dex, then this gives a typical maximum systematic offset using a fixed MARCS model grid.

\begin{deluxetable*}{c c c c c c c c}
\tablewidth{0pt} 
\tabletypesize{\scriptsize}
\tablecaption{K2-72 Fe Abundance Changes For Different Atmospheric Models\label{tab:interp_test}} 
 \tablehead{\colhead{Model} & \colhead{Interpolated} & \colhead{[Fe/H]} & \colhead{$\sigma$[Fe/H]}  & \colhead{[Fe/H]} & \colhead{$\sigma$[Fe/H]} & \colhead{[Fe/H]} & \colhead{$\sigma$[Fe/H]} \\
 \colhead{} & \colhead{} & \colhead{(dex)} & \colhead{(dex)}  & \colhead{Only Fe I (dex)} & \colhead{Only Fe I (dex)} & \colhead{Only FeH (dex)} & \colhead{Only FeH (dex)}}
\startdata
$\mathrm{[Fe/H]}$ = 0	&	No	&	-0.17	&	0.06	&	-0.14	&	0.05	&	-0.21	&	0.06 \\
$\mathrm{[Fe/H]}$ = 0	&	Yes	&	-0.21	&	0.08	&	-0.16	&	0.07	&	-0.26	&	0.07	\\
$\mathrm{[Fe/H]}$ = -0.17	&	Yes	&	-0.29	&	0.09	&	-0.22	&	0.04	&	-0.36	&	0.06	\\
$\mathrm{[Fe/H]}$ = -0.25	&	No	&	-0.27	&	0.12	&	-0.18	&	0.07	&	-0.36	&	0.08	\\
$\mathrm{[Fe/H]}$ = -0.25	&	Yes	&	-0.33	&	0.14	&	-0.22	&	0.06	&	-0.44	&	0.08	\\
\enddata
\tablecomments{A $\teff{}$ = 3400 K and log(g) = 5 was adopted for each model.}
\end{deluxetable*}

The choice of discrete atmosphere models may contribute to a lack of abundances near [Fe/H] $\sim$ --0.2 dex, with additional small systematic uncertainties for stars near atmospheric parameter limits. To test whether this systematic uncertainty is larger than model interpolation uncertainties, we measured the abundance of K2-72 using multiple different atmospheric models. We chose K2-72 for the test because this object has an intermediate metallicity between models at [Fe/H] = --0.17 dex and a lower $\teff{}$ = 3380 K, where model interpolation becomes complicated due to rapid changes in atmospheric structure as molecules form (e.g., \citealt{wheeler24}). We measured abundances using MARCS models at a [Fe/H] = 0.0 dex and [Fe/H] = --0.25 dex, then created interpolated models at the same [Fe/H] using existing nearby MARCS models. Lastly, we interpolated a model at [Fe/H] = --0.17 dex as a direct comparison. Our results are shown in Table \ref{tab:interp_test}; we find the difference between the interpolated model and the unaltered model from the MARCS grid is --0.04 dex for [Fe/H] = 0.0 dex, and --0.06 dex for [Fe/H] = --0.25 dex. The molecular lines are also more impacted from changes in the model atmosphere, with iron abundances from [Fe/H] lines differing by --0.05 dex for the [Fe/H] = 0.0 dex model and --0.08 dex for the [Fe/H] = --0.25 model.  We also found that the interpolated model at [Fe/H] = --0.17 dex resulted in a more metal poor abundance than the MARCS grid model at [Fe/H] = --0.25 dex. 

By assessing the impact of whether or not to interpolate the model atmospheres, we have found that there is an impact to the measured abundances, but no clear way to minimize systematic uncertainties. Previous efforts have found that errors from model interpolation would likely become more significant for cooler stars \citep{wheeler24}. Either model interpolation or adopting the closest matching model from the MARCS grid likely introduces additional abundance errors of $\delta$[Fe/H] = $\sim$ 0.05 - 0.1 dex. Errors in atmospheric structure may also uniquely affect FeH lines as this molecule has a low dissociation energy of 1.54 eV \citep{hargreaves10} and other abundances showed no bimodality in Fig. \ref{fig:teff_abundances}.

We also tested how each average abundance varied with $\teff{}$ in Fig. \ref{fig:teff_abundances}. Unaccounted for blends, non-LTE affects, inaccuracies modeling a continuum dominated by molecular lines may all lead to $\teff{}$-dependent systematic errors. We fit a linear line to each [X/Fe] ratio, using the \texttt{emcee} package to estimate the uncertainty on the slope. We found slopes of $s_{\mathrm{[Fe/H]}}$ = 0.0013$^{+0.007}_{-0.008}$,  $s_{\mathrm{[O/Fe]}}$ = 0.011$^{+0.015}_{-0.014}$, $s_{\mathrm{[Mg/Fe]}}$ = 0.017 $\pm$ 0.008, $s_{\mathrm{[Si/Fe]}}$ = 0.004 $\pm$ 0.009, where $s$ is in units of dex per 100 K. Each element has a slope within 1$\sigma$  of 0.0 dex/100 K, suggesting there are no large $\teff$-dependent systematic errors, with the potential exception of Mg. The [Mg/Fe] ratio slope is $\sim$2$\sigma$ away from 0 and suggests that a star at $\sim$3100 K would differ by 0.14 dex from a star at 3900 K based on our linear fit. Additional high resolution abundances would be useful to determine which stars are more impacted by systematic effects. 

\section{Discussion}
\label{sec:discussion}

\subsection{M-Dwarf Abundance Comparisons}

\begin{figure*}[t!]
\centering
	\includegraphics[width=7.0in]{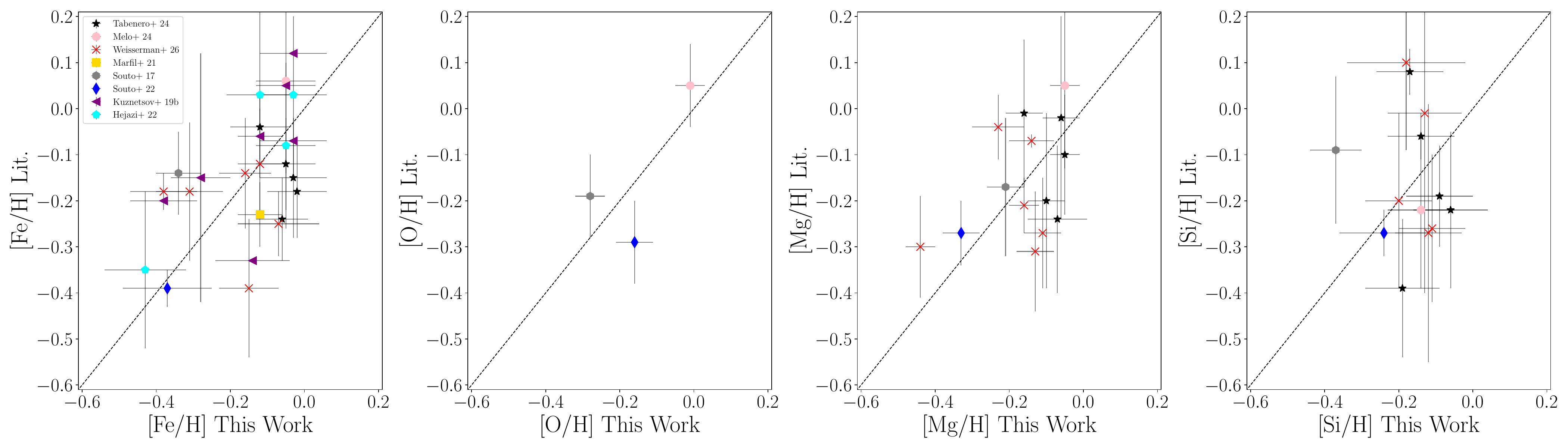}
    \caption{A comparison between our abundance measurements and literature values are shown for [Fe/H], [O/H], [Mg/H], and [Si/H]. Literature source abundances and citations are found in Table \ref{table::m_dwarf_lit_comp}. The black dashed line is a one-to-one line.}
    \label{fig:lit_comp}
\end{figure*}

Multiple stars from this work were analyzed in previous abundance studies that utilized low- to high-resolution spectroscopy. High-resolution infrared abundance studies are similar in methodological design to this work and so we compare our results to APOGEE analysis \citep{souto17,souto22,melo24} and CARMENES \citep{marfil21,tabernero24} for overlapping stars in Table \ref{tab:high_res_comp}. We have re-normalized the abundance ratios within Table \ref{tab:high_res_comp} to the \citet{asplund09} solar scale. In most cases, each abundance ratio and $\teff{}$ are within the 1$\sigma$ - 2$\sigma$ uncertainties from each study. The average differences are: $< \Delta \teff{} >$ = --73 $\pm$ 48 K, $< \Delta \mathrm{[O/H]} >$ = -0.02 $\pm$ 0.10, $< \Delta \mathrm{[Fe/H]} >$ = 0.01 $\pm$ 0.13, $< \Delta \mathrm{[Mg/H]} >$ = --0.01 $\pm$ 0.12, and $< \Delta \mathrm{[Si/H]} >$ = --0.02 $\pm$ 0.19. The averages do not include the LHS 1140 abundances since the Mg and Si abundances have large uncertainties in \citet{weisserman26}. The abundances from different literature sources are also displayed in Fig. \ref{fig:lit_comp}.

There is a systematic difference in the derived $\teff{}$ when compared to different data sets, such as analysis of CARMENES or APOGEE spectra, similar to what was observed when compared to photometric $\teff{}$ in section \ref{subsec::teff}. While the average offset between our abundance ratios and literature results is near 0.0 dex, literature sources vary systematically in different ways. For example, when compared to studies of CARMENES spectra \citep{marfil21,tabernero24}, we find an $< \Delta \mathrm{[Fe/H]} >$ = 0.13 $\pm$ 0.04 and when compared to APOGEE spectra $<\Delta \mathrm{[Fe/H]}>$ = --0.12 $\pm$ 0.10 \citep{souto17,souto22,melo24}. These systematic abundance differences come from analysis decisions, such as linelist differences (e.g. CARMENES uses J-band lines which are unavailable in the IGRINS and APOGEE bandpasses), synthetic synthesis tools, and model atmosphere choices. For example, this work uses MOOG while \citet{souto17,souto22} uses \texttt{Turbospectrum} \citep{plez12}. Additionally, both this work and \citet{souto17,souto22} use MARCS model atmospheres \citep{gustafsson08} while \citet{tabernero24} uses BT SETL model atmospheres \citep{allard12}. Different fitting and normalization routines can also lead to systematic differences. A more complex analysis of abundance measurement methodology is necessary to identify specific causes for systemic differences, similar to what has been done in FGK-stars \citep{hinkel16, jofre17}. We do note that while systematic differences exist, methodologies to study CARMENES and APOGEE spectra identified the relatively more metal-poor or metal-rich stars in the sample. For example, \citet{tabernero24} and our work find that G162-44 is 0.06 -- 0.08 dex more enhanced in [Fe/H] than G50-16, even if the results are offset. We do find agreement with \citet{weisserman26} abundances, which are based on the fitting routing of \citealt{gromek25}, and have  $< \Delta \mathrm{[Fe/H]} >$ = --0.01 $\pm$ 0.15, $< \Delta \mathrm{[Mg/H]} >$ = 0.01 $\pm$ 0.17, and $< \Delta \mathrm{[Si/H]} >$ = --0.02 $\pm$ 0.18. 

We also compare to \citet{kuznetsov19,hejazi22} which are both low-to-moderate-resolution spectroscopic abundance studies that have more than two stars in common with our sample. As shown in Table \ref{tab:high_res_comp}, the $< \Delta \teff{} >$ = -15 $\pm$ 151 K and $< \Delta \mathrm{[Fe/H]} >$ = --0.06 $\pm$ 0.11, are consistent with the uncertainties.  We overall find our abundances are broadly consistent with the literature results. 

We also obtained spectra from the IGRINS archive \citep{sawczynec25} of two M-dwarfs (not currently known to host planets) in a wide binary with an FGK companion to further test our abundance derivation methodology. Abundances were derived for LEP 16B and HD 125455B using the same methodology as the rest of our M-dwarf sample. These stars have $\teff{}$ = 3327 K and 3147 K, respectively, and test the reliability of the synthesis for our coolest stars. Median abundances from literature measurements for the primary stars, LEP 16 (or HD 24421) and HD 125455, were accessed from the Hypatia Catalog \citep{hinkel14}. Abundances were re-normalized to our solar scale of \citet{asplund09} and listed in Table \ref{table::m_dwarf_lit_comp} along with the literature sources of the abundances.

The warmer M-dwarf, LEP 16B, is consistent with the abundances with the primary star at levels $\lesssim$ 0.0 - 0.04 dex for [O/H], [Fe/H], [Mg/H], and [Si/H]. For HD 125455B, the abundances for [O/H], [Fe/H], and [Mg/H] have a larger offset ranging from -0.07 -- 0.12 dex, while [Si/H] is offset by --0.03 dex. In all cases but Mg for HD 125455B, the differences between the FGK primary and M-dwarf secondary are within 1$\sigma$. For [Mg/H], the Mg I lines weaken significantly at lower $\teff{}$s and the challenges with continuum placement may affect the final abundance ratios. The abundance offset is also similar to the difference between hot and cool M-dwarfs estimated from our [Mg/Fe] vs. $\teff{}$ analysis, shown in Fig. \ref{fig:teff_abundances}). We overall find minimal differences for LEP 16B, and slight offsets for HD 125455 B, suggesting our synthetic spectra grids and abundance measurement methodology is accurately deriving abundances for the warmer stars studied ($\teff{}$ $\gtrsim$ 3300 K), with potential small ($\sim$0.1 dex) offsets for the coolest stars in our sample. 

\subsection{Finding Older Stars with Enhanced $\alpha$-Elements from M-Dwarfs}
\label{subsec:thickdisk}

\begin{figure}
\centering
	\includegraphics[width=3.3in]{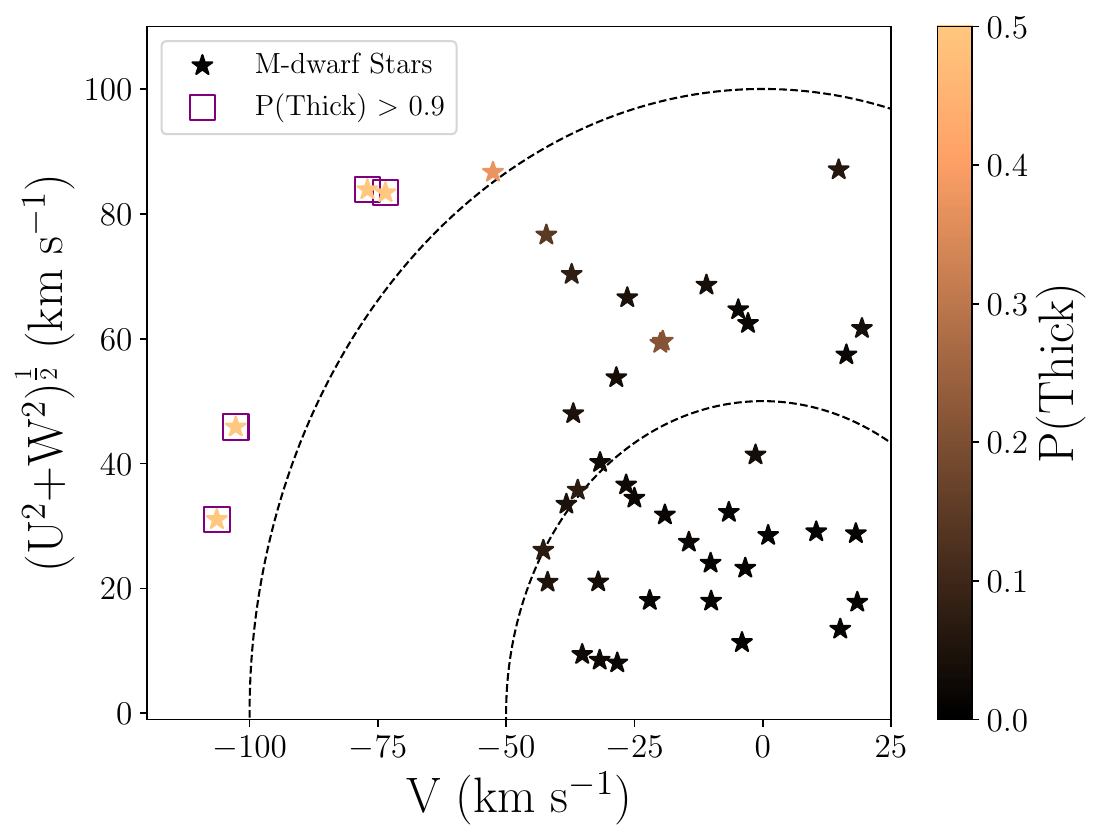}
    \caption{A Toomre diagram for our sample stars (star symbols) are shown with a color bar representing the probability that a star belongs in the thick disk stellar population per \citet{ramirez13}. Stars with a probability $>$90$\%$ are highlighted with a purple square.} 
    \label{fig:uvw}
\end{figure}

We classified our stellar sample as thin or thick disk stars using chemical and kinematic criteria in order to identify the stellar population membership of each M-dwarf planet-host. We followed the methodology of \citet{maas22} to compute our kinematic disk membership probabilities. We used Gaia astrometry \citep{gaia23} and coordinate transformation from \texttt{Astropy} \citep{astropy13}. The \texttt{pyia} code \citep{pyia18} was used to appropriately propagate uncertainties when computing the full UVW velocities. Disk membership probabilities were then computed using \citet{ramirez13} and plotted in Fig. \ref{fig:uvw}. Four stars have thick disk probabilities greater than 90$\%$: L181-1, Kepler 446, Kepler 42, and GJ 1252. 

\startlongtable
\begin{deluxetable*}{ l c c c c c c c c c c c}
\tablewidth{0pt} 
\tabletypesize{\footnotesize}
\tablecaption{M-Dwarf Abundance Comparisons \label{table::m_dwarf_lit_comp}} 
 \tablehead{\colhead{Star Name} & \colhead{$\teff{}$} & \colhead{$\sigma$ $\teff{}$} &  \colhead{[O/H]} &  \colhead{$\sigma$[O/H]} & \colhead{[Fe/H]} & \colhead{$\sigma$[Fe/H]} &\colhead{[Mg/H]} & \colhead{$\sigma$[Mg/H]} &\colhead{[Si/H]} & \colhead{$\sigma$[Si/H]} & \colhead{Source}\\ \colhead{} & \colhead{(K)} & \colhead{(K)} & \colhead{(dex)} & \colhead{(dex)} & \colhead{(dex)} & \colhead{(dex)} & \colhead{(dex)} & \colhead{(dex)}& \colhead{(dex)} & \colhead{(dex)}}
\startdata
  \multicolumn{12}{c}{M-Dwarf - FGK Binary Pairs}  \\ 
        \hline 
LEP 16B	&  3327	 & 48 &	--0.16	 &	0.05	 &	--0.37	 &	0.12	 &	--0.33	 &	0.05	 &	--0.24	 &	0.12 &	This Work \\ 
LEP 16  &	6108	& \nodata &	--0.12	 &	0.14	 &	--0.36	 &	0.12	 &	--0.31	 &	0.06	 &	--0.24	 &	0.06 &	(Hypatia) \\ 
\hline							
HD 125455B	& 3147	& 58 & --0.19	& 0.04	& --0.28& 	0.08	& --0.27	& 0.04	& --0.14	& 0.16&  This Work \\
HD 125455	&  5190	& \nodata & --0.08	& 0.13	& --0.21	& 0.03	& --0.15	& 0.06	& --0.11	& 0.04& (Hypatia) \\ 
        \hline
  \multicolumn{12}{c}{High-Resolution Abundance Analysis Comparison}  \\ 
        \hline 
G162-44	& 3432 & 45 & 0.13 & 0.03	& --0.02	& 0.08 & --0.06 &	0.05 & --0.09 &	0.09 & This Work \\
G162-44	&  3434	& 28 & \nodata{} & \nodata{} & --0.18 & 0.10 & --0.02 &	0.22 & --0.19 & 0.11 & (1) \\
\hline
G50-16 & 3361 & 48 & 0.16 & 0.04	& --0.06	& 0.07 & --0.07 &	0.08 & --0.19 &	0.10 & This Work \\
G50-16 &  3422	& 26 & \nodata{} & \nodata{} & --0.24 & 0.09 & --0.24 &	0.16 & --0.39 & 0.15 & (1) \\
\hline
GJ 436 & 3483 & 43 & -0.01 & 0.04	& --0.05	& 0.08 & --0.05 &	0.04 & --0.14 &	0.09 & This Work \\
GJ 436 &  3533	& 26 & \nodata{} & \nodata{} & --0.12 & 0.1 & --0.10 &	0.13 & --0.06 & 0.05 & (1) \\
GJ 436 &  3507	& 100 & 0.05 & 0.09 & 0.06 & 0.20 & 0.05 &	0.18 & 0.10 & 0.26 & (2) \\
\hline
Gliese486 & 3326 & 51 & 0.07 & 0.04	& --0.03	& 0.09 & --0.10 &	0.05 & --0.06 &	0.10 & This Work \\
Gliese486 &  3408	& 45 & \nodata{} & \nodata{} & --0.15 & 0.13 & --0.20 &	0.19 & --0.22 & 0.17 & (1) \\
\hline
L678-39 & 3379 & 49 & 0.02 & 0.03	& --0.12	& 0.06 & --0.11 &	0.05 & --0.11 &	0.09 & This Work \\
L678-39 &  3496	& 20 & \nodata{} & \nodata{} & --0.23 & 0.07 & \nodata{} & \nodata{} & \nodata{} & \nodata{} & (3) \\
L678-39 &  3505	& 51 & \nodata{} & \nodata{} & --0.12 & 0.12 & --0.27 & 0.12 & --0.26 & 0.16 & (4) \\
\hline
TOI-1468 & 3363 & 57 & -0.07 & 0.07	& --0.12	& 0.08 & --0.16 &	0.05 & --0.17 &	0.09 & This Work \\
TOI-1468 &  3496	& 25 & \nodata{} & \nodata{} & --0.04 & 0.07 & -0.01 & 0.16 & 0.08 & 0.05 & (1) \\
\hline
Kepler138 & 3817 & 77 & --0.28 & 0.08 & --0.34	& 0.06	& --0.31 & 0.05 &	--0.37 & 0.07 & This Work \\
Kepler138 &  3835	& 64 & --0.19 & 0.09 & --0.14 & 0.09 & -0.17 & 0.15 & --0.09 & 0.16 & (5) \\
\hline
TOI-2445 &  3254	& 134 & 0.09 & 0.10 & --0.37 & 0.12 & --0.26 & 0.04 & --0.44 & 0.10 & This Work \\
TOI-2445 & 3318 & 100 & 0.13 & 0.10 & --0.20 & 0.09	& --0.22 & 0.14 &	\nodata & \nodata{} & (2) \\
\hline
LEP 16B &  3327	& 48 & --0.16 & 0.05 & --0.37 & 0.12 & --0.33 & 0.05 & --0.24 & 0.12 & This Work \\
LEP 16B & 3455 & 100 & --0.29 & 0.09 & --0.39 & 0.04	& --0.27 & 0.07 & --0.27 & 0.05 &	 (6) \\
\hline
L320-124 &  3174	& 58 & --0.18 & 0.05 & --0.31 & 0.09 & --0.44 & 0.04 & --0.18 & 0.16 & This Work \\
L320-124  &  3229	& 75 & \nodata{} & \nodata{} & --0.18 & 0.15 & --0.30 & 0.11 & 0.10 & 0.19 & (4) \\
\hline
GJ 1252 & 3505 & 54 & 0.09 & 0.03	& --0.38	& 0.09 & --0.23 &	0.07 & --0.13 & 0.10 & This Work \\
GJ 1252  &  3458	& 137 & \nodata{} & \nodata{} & --0.18 & 0.04 & --0.04 & 0.07 & --0.01 & 0.39 & (4) \\
\hline
L98-59 &  3310	& 48 & --0.04 & 0.04 & --0.15 & 0.08 & --0.16 & 0.04 & --0.12 & 0.09 & This Work \\
L98-59  &  3415	& 135 & \nodata{} & \nodata{} & --0.29 & 0.15 & --0.21 & 0.06 & --0.27 & 0.28 & (4) \\
        \hline
LHS 1140 &  3125	&53 & 0.05 & 0.06 & --0.07 & 0.11 & --0.14 & 0.06 & 0.05 & 0.15 & This Work \\
LHS 1140  &  3096	& 48 & \nodata{} & \nodata{} & --0.25 & 0.07 & 0.07 & 0.14 & 0.57 & 0.05 & (4) \\
        \hline
L231-32 &  3401	& 53 & 0.01 & 0.03 & --0.16 & 0.07 & --0.13 & 0.05 & --0.20 & 0.09 & This Work \\
L231-32  &  3506	& 70 & \nodata{} & \nodata{} & --0.14 & 0.12 & --0.31 & 0.13 & --0.20 & 0.19 & (4) \\
        \hline
  \multicolumn{12}{c}{Low/Moderate-Resolution Abundance Analysis Comparison}  \\ 
        \hline 
GJ 367 & 3579 & 49 & 0.02 & 0.03	& --0.03	& 0.09 & --0.03 &	0.06 & --0.06 &	0.04 & This Work \\
GJ 367 &  3453	& \nodata{} & \nodata{} & \nodata{} & --0.07 &	\nodata{} & \nodata & \nodata{} & \nodata & \nodata{} & (7) \\
\hline
BD-17588 & 3235 & 73 & --0.07 & 0.05	& --0.14	& 0.10 & --0.08 &	0.04 & 0.06 &	0.10 & This Work \\
BD-17588 &  3339	& \nodata{} & \nodata{} & \nodata{} & --0.33 &	\nodata{} & \nodata & \nodata{} & \nodata & \nodata{} & (7) \\
\hline
GJ 436 & 3483 & 43 & -0.01 & 0.04	& --0.05	& 0.08 & --0.05 &	0.04 & --0.14 &	0.09 & This Work \\
GJ 436 &  3353	& \nodata{} & \nodata{} & \nodata{} & 0.05 &\nodata{} & \nodata & \nodata{} & \nodata & \nodata{} & (7) \\
GJ 436 &  3500	& 106 & \nodata{} & \nodata{} & --0.08 &	0.18 & \nodata & \nodata{} & \nodata & \nodata{} & (8) \\
\hline
Gliese486 & 3326 & 51 & 0.07 & 0.04	& --0.03	& 0.09 & --0.10 &	0.05 & --0.06 &	0.10 & This Work \\
Gliese486 &  3212	& \nodata{} & \nodata{} & \nodata{} & 0.12 &	\nodata{} & \nodata & \nodata{} & \nodata & \nodata{} & (7) \\
Gliese486 &  3300	& 91 & \nodata{} & \nodata{} & 0.03 &	0.17 & \nodata & \nodata{} & \nodata & \nodata{} & (8) \\
\hline
L181-1  & 3709 & 61 & 0.19 & 0.04	& --0.04	& 0.07 & 0.11 &	0.05 & --0.03 &	0.04 & This Work \\
L181-1 &  3569	& \nodata{} & \nodata{} & \nodata{} & 0.05 &	\nodata{} & \nodata & \nodata{} & \nodata & \nodata{} & (7) \\ \hline
L678-39 & 3379 & 49 & 0.02 & 0.03	& --0.12	& 0.06 & --0.11 &	0.05 & --0.11 &	0.09 & This Work \\
L678-39 &  3706	& \nodata{} & \nodata{} & \nodata{} & --0.06 & \nodata{} & \nodata{} & \nodata{} & \nodata{} & \nodata{} & (7) \\
\hline
TOI-1442 & 3390 & 86 & --0.04 & 0.05	& --0.12	& 0.09 & --0.25 &	0.04 & --0.14 &	0.09 & This Work \\
TOI-1442  &  3300	& 52 & \nodata{} & \nodata{} & 0.03 &	0.18 & \nodata & \nodata{} & \nodata & \nodata{} & (8)
\\
\hline
LHS 1478 & 3189 & 44 & --0.19 & 0.06	& --0.43	& 0.11 & --0.45 &	0.08 & \nodata{} &	\nodata{} & This Work \\
LHS 1478 &  3350	& 44 & \nodata{} & \nodata{} & --0.35 &	0.17 & \nodata & \nodata{} & \nodata & \nodata{} & (8) \\
\hline
GJ 1252 & 3505 & 54 & 0.09 & 0.03	& --0.38	& 0.09 & --0.23 &	0.07 & --0.13 & 0.10 & This Work \\
GJ 1252 &  3687	& \nodata{} & \nodata{} & \nodata{} & --0.20 &	\nodata{} & \nodata & \nodata{} & \nodata & \nodata{} & (7) \\
\hline
HD 125455B & 3147 & 44 & --0.19 & 0.04	& --0.28	& 0.08 & --0.27 &	0.04 & --0.14 &	0.16 & This Work \\
HD 125455B &  3163	& 46 & \nodata{} & \nodata{} & --0.15 &	0.27 & \nodata & \nodata{} & \nodata & \nodata{} & (7) \\
\enddata
\tablecomments{All abundances placed on \citet{asplund09} solar scale. }
\tablecomments{Sources are: (1) \citet{tabernero24}; (2) \citet{melo24}; (3) \citet{marfil21}; (4) \citet{weisserman26}; (5) \citet{souto17}; (6) \citet{souto22}; (7) \citet{kuznetsov19}; (8) \citet{hejazi22}; Hypatia abundances were derived from \citet{chen00,reddy03,valenti05,ramirez07,neves10,delgadomena10,petigura11,adibekyan12,ramirez13,wu15,montes18,luck18,souto20,costasilva20} }
\label{tab:high_res_comp}
\end{deluxetable*}

\begin{figure*}
\centering
	\includegraphics[width=7.in]{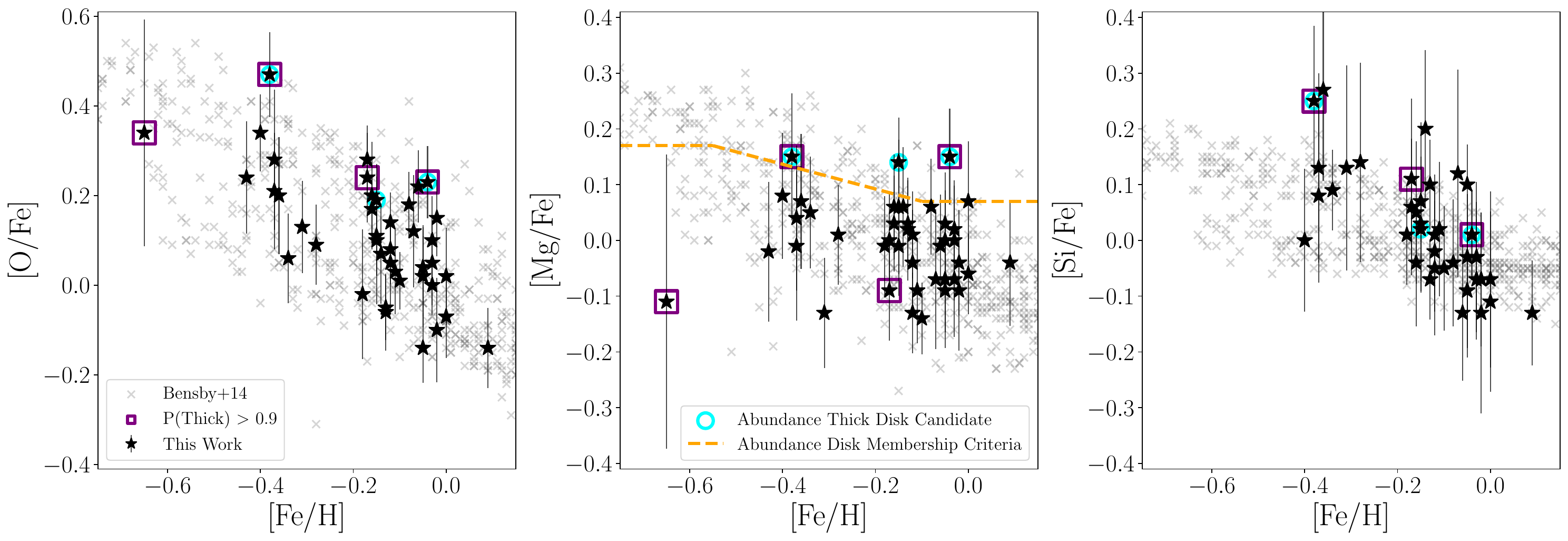}
    \caption{Abundance ratios as a function of metallicity are shown for our M-dwarf sample as black stars. Literature FGK abundances from \citet{bensby14} are shown as gray circles and a yellow dashed line indicates the dividing line between thin disk and thick disk stars from \citet{alinder25}. Kinematically and chemically identified thick disk stars are also highlighted in the plot.} 
    \label{fig:chemevol}
\end{figure*}

We applied chemical abundance cutoffs as well to distinguish thick disk stars. We focus on Mg as our $\alpha$-element tracer, since Si uncertainties for metal-poor stars were large, especially for weaker lines around metal-poor stars. We use the chemical criteria of \citet{alinder25}, to identify candidate thick disk stars. A piecewise linear function was used to separate high $\alpha$-abundances from thin disk stars and applies similar results to \citet{adibekyan12}. The abundance cutoffs were adjusted to the \citet{asplund09} scale and shown in Fig. \ref{fig:chemevol}. We identify three stars that have enhanced [Mg/Fe] ratios, TOI-269, L181-1, and GJ 1252.

Two stars meet both the kinematic and chemical definitions for thick disk membership: GJ 1252 and L181-1. L181-1 (or LHS 1815) was previously identified as a host of the first exoplanet discovered around a thick disk star \citep{gan20}. In contrast, GJ 1252 was found to have an age of 3.9 $\pm$ 0.4 Gyr as derived from gyrochronology \citep{crossfield22}, suggesting an age consistent with a thin disk star. Additional $\alpha$-elemental abundances could confirm the $\alpha$-enhanced nature of GJ 1252 and confirm that this star belongs to the thick disk. TOI-269 has a high [O/Fe] ratio, but has a kinematic thick disk probability of 0.02$\%$ and Mg abundances near the stellar population membership threshold. TOI-269 has a measured age from gyrochronology of 4.19 $\pm$ 0.97 Gyr using HARPS data and 2.16 $\pm$ 0.27 Gyr from WASP-South observations \citep{cointepas21}. Like GJ 1252, additional $\alpha$-elemental abundances could help confirm whether this object is truly a thick disk star, especially since the Mg abundance is well-within 1$\sigma$ threshold. 

Both Kepler 446 and Kepler 42 have kinematic criteria placing them as thick disk stars but the Mg abundances are not found to be $\alpha$-enhanced. Kepler 42 is the most metal-poor star in our sample ([Fe/H] = --0.65 $\pm$ 0.24) and has very large abundance [Mg/Fe] uncertainty of 0.26 dex. In contrast, Kepler 446 has relatively more precise [Mg/Fe] = --0.09 $\pm$ 0.09 which is more than 2$\sigma$ from the thin/thick disk criteria. Combining kinematics and abundance criteria is critical for identifying stars in different stellar populations, echoing similar results as FGK-star results (e.g., \citealt{adibekyan12,bensby14}).

\subsection{Molar Ratios of Planet Building-Blocks}

\begin{figure}[h!]
\centering
	\includegraphics[width=3.3in]{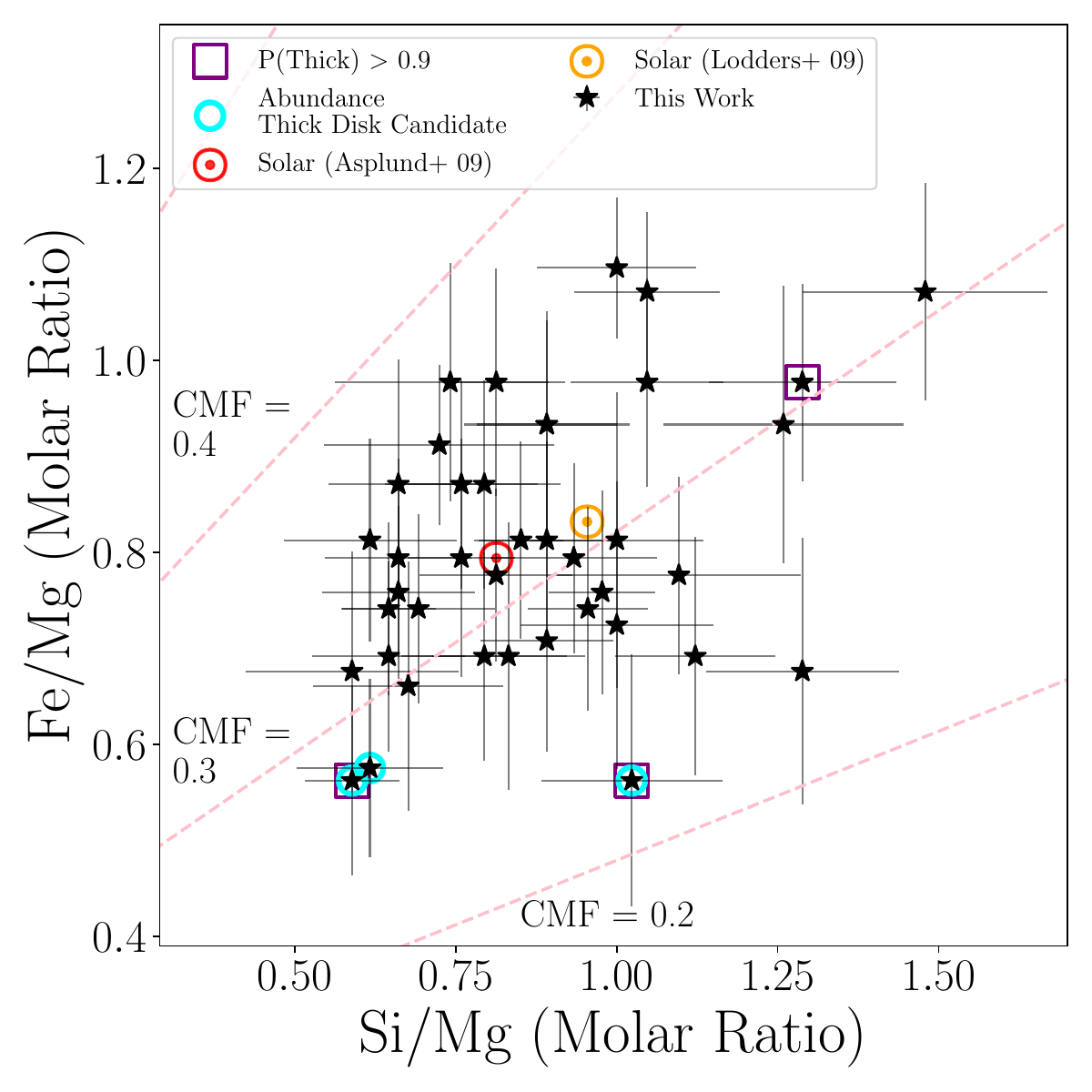}
    \caption{Molar ratios are plotted for our sample (black stars), in addition to solar values from \citet[][red solar symbol]{asplund09} and \citet[][gold solar symbol]{lodders09}. Thick disk stars identified with chemistry and kinematics are highlighted in the plot (classifications described in section \ref{subsec:thickdisk}). Lines of constant CMF values of 0.2, 0.3, and 0.4 are shown as pink dashed lines using formulae from \citet{unterborn23}.} 
    \label{fig:molar}
\end{figure}

We derived the molar ratios of the major rock-building elements, e.g., Fe/Mg and Mg/Si, and their respective uncertainties from the values in Table \ref{tab:star_final_params} using the formalism of \citet{hinkel22}. The results are shown in Fig. \ref{fig:molar} along with molar ratios from the solar abundance scale used in this work \citep{asplund09} and meteoritic value from \cite{lodders09}. Overall, we find that the solar values fall within the M-dwarf distribution of Si/Mg and Fe/Mg ratios for the majority of our sample. Additionally, our molar ratios are broadly consistent with those calculated from FGK-star abundances \citep{unterborn23,schulze26}, where the latter span in 0.14 $<$ Fe/Mg $<$ 1.36 and the most common Fe/Mg values range from $\sim$0.7 - 0.9 \citep[][using abundances from the Hypatia Catalog per \citealt{hinkel14}]{schulze26}. 

Molar ratios are commonly used in conjunction with exoplanet mass and radius measurements to characterize the interior structure of exoplanets. Models of the core mass fraction (CMF) aid in identifying super-Mercury worlds with large Fe cores, rocky super-Earths, or volatile-rich mini-Neptunes (e.g., \citealt{unterborn23, schulze26}). We computed which molar ratios led to constant CMF values, shown as contours in Fig. \ref{fig:molar}, as calculated using equations 9 and 10 from \citet{unterborn23}. We found our M-dwarf abundances are consistent with the CMFs of small planets found around FGK stars. Literature analysis of exoplanetary interiors based on M-dwarf abundances, including for some stars within our sample, have revealed four strong water world candidates and one super-Mercury \citep{souto18,weisserman26}. Our large sample will aid in statistically robust estimates of M-dwarf small planet interior compositions (see Paper II, Hinkel et al., in prep.). In addition, our chemically identified thick disk stars have the lowest Fe/Mg abundances, which are intriguing objects for follow-up interior structure analysis. 

\section{Conclusions }
\label{sec:conclusions}

We have measured abundances of 40 planet-hosting M-dwarfs and 2 M-dwarfs in a wide-binary pair with an FGK primary. We used high-resolution IR spectroscopy from IGRINS on Gemini South, the IGRINS archive, and using NIRSPEC on Keck to measure the abundances of O, Mg, Si, and Fe. Grids of synthetic spectra were generated using MOOG \citep{sneden73} with MARCS atmospheric models \citep{gustafsson08}, and a compiled linelist specific for M-dwarfs stars.  Our final results are given below:

\begin{itemize}
    \item{We derived $\teff{}$s and O abundances by finding the model atmosphere that gave consistent abundances from OH and \water{} lines. The methodology was based on \citet{souto17,souto20} and led to $\teff{}$ consistent with photometry \citep{mann15} with an average difference of --34 $\pm$ 93 K.}
    \item{Fe abundances were derived using Fe I and FeH lines. We found that stars with decreasing $\teff{}$ had a larger systematic offset between the [Fe/H] ratio derived by Fe I and FeH lines. The average difference for all stars was --0.08 $\pm$ 0.05 dex.}
    \item{No statistically significant slope was found for O, Si, or Fe as a function of $\teff{}$ for our derived stars. Only Mg was found to have a slope of $s_{\mathrm{[Mg/Fe]}}$ = 0.017 $\pm$ 0.008 (dex/100 K), suggesting that the coolest stars may have slight abundance offsets from the warmest M-dwarfs in our sample.}
    \item{We broadly find that our abundances are consistent with both high-resolution and low-resolution literature abundance measurements. However, offsets from $\pm$ 0.01 - 0.2 dex are found when abundances derived from different methodologies are compared to one another; such as comparisons to the APOGEE survey spectra \citep{souto22,melo24} or CARMENES spectra \citep{tabernero24}. Offsets between abundance analysis are likely due to different line selection criteria, radiative transfer code used, and model atmosphere choices.}
    \item{We use UVW velocities and chemistry to identify thick disk candidate planet hosts. Two stars, GJ 1252 and L181-1 have both kinematic and chemical abundances consistent with an older star population. GJ 1252 was previously identified as a younger star and further $\alpha$-elemental abundances would aid in confirming this object as a thick disk member.}
    \item{Molar ratios for our planet-hosting M-dwarfs were determined and are broadly consistent with FGK stars. }
\end{itemize}

\section{Acknowledgments}
All authors wish to acknowledge funding or support from the NASA Exoplanet Research Program (XRP) grant 20-XRP20\_2-0125 / NNH20ZDA001N. NRH would like to thank CHW3, Tatertot, and Lasagna. The research shown here acknowledges use of the Hypatia Catalog, an online compilation of stellar abundance data as described in Hinkel et al. (2014, AJ, 148, 54). This work used The Immersion Grating Infrared Spectrometer (IGRINS) which was developed under a collaboration between the University of Texas at Austin and the Korea Astronomy and Space Science Institute (KASI) with the financial support of the US National Science Foundation under grants AST-1229522, AST-1702267 and AST-1908892, McDonald Observatory of the University of Texas at Austin, the Korean GMT Project of KASI, the Mt. Cuba Astronomical Foundation and Gemini Observatory. The RRISA is maintained by the IGRINS Team with support from McDonald Observatory of the University of Texas at Austin and the US National Science Foundation under grant AST-1908892. This paper includes data taken at The McDonald Observatory of The University of Texas at Austin. Some of the data presented herein were obtained at Keck Observatory, which is a private 501(c)3 non-profit organization operated as a scientific partnership among the California Institute of Technology, the University of California, and the National Aeronautics and Space Administration. The Observatory was made possible by the generous financial support of the W. M. Keck Foundation. The authors wish to recognize and acknowledge the very significant cultural role and reverence that the summit of Maunakea has always had within the Native Hawaiian community. We are most fortunate to have the opportunity to conduct observations from this mountain. This work has made use of data from the European Space Agency (ESA) mission {\it Gaia} (\url{https://www.cosmos.esa.int/gaia}), processed by the {\it Gaia}
Data Processing and Analysis Consortium (DPAC, \url{https://www.cosmos.esa.int/web/gaia/dpac/consortium}). Funding for the DPAC has been provided by national institutions, in particular the institutions participating in the {\it Gaia} Multilateral Agreement. This research was supported in part by Lilly Endowment, Inc., through its support for the Indiana University Pervasive Technology Institute. We thank the referee for detailed, constructive feedback that led to an improved manuscript.

\facility{Keck:II (NIRSPEC), Gemini:South (IGRINS), McDonald Observatory:2.7m Harlan J. Smith Telescope (IGRINS)} 

\software{\texttt{scipy} \citep{scipy20}, \texttt{numpy} \citep{numpy}, \texttt{matplotlib} \citep{hunter07}, \texttt{astropy} \citep{astropy13,astropy18,astropy22}, \texttt{pyia} \citep{pyia18}, \texttt{iSpec} \citep{blancocuaresma14,blancocuaresma19}, \texttt{MOOG} \citep{sneden73}}

\bibliographystyle{aasjournalv7}
\bibliography{mdwarf_references}

@ARTICLE{schulze21,
       author = {{Schulze}, J.~G. and {Wang}, J. and {Johnson}, J.~A. and {Gaudi}, B.~S. and {Unterborn}, C.~T. and {Panero}, W.~R.},
        title = "{On the Probability That a Rocky Planet's Composition Reflects Its Host Star}",
      journal = {\psj},
         year = 2021,
        month = jun,
       volume = {2},
       number = {3},
          eid = {113},
        pages = {113},
          doi = {10.3847/PSJ/abcaa8},
archivePrefix = {arXiv},
       eprint = {2011.08893},
 primaryClass = {astro-ph.EP},
       adsurl = {https://ui.adsabs.harvard.edu/abs/2021PSJ.....2..113S}
}

@ARTICLE{mann15,
       author = {{Mann}, Andrew W. and {Feiden}, Gregory A. and {Gaidos}, Eric and {Boyajian}, Tabetha and {von Braun}, Kaspar},
        title = "{How to Constrain Your M Dwarf: Measuring Effective Temperature, Bolometric Luminosity, Mass, and Radius}",
      journal = {\apj},
         year = 2015,
        month = may,
       volume = {804},
       number = {1},
          eid = {64},
        pages = {64},
          doi = {10.1088/0004-637X/804/1/64},
archivePrefix = {arXiv},
       eprint = {1501.01635},
 primaryClass = {astro-ph.SR},
       adsurl = {https://ui.adsabs.harvard.edu/abs/2015ApJ...804...64M}
}

@ARTICLE{gaia23,
       author = {{Gaia Collaboration} and {Vallenari}, A. and {Brown}, A.~G.~A. and {Prusti}, T. and {de Bruijne}, J.~H.~J. and {Arenou}, F. and {Babusiaux}, C. and {Biermann}, M. and {Creevey}, O.~L. and {Ducourant}, C. and {Evans}, D.~W. and {Eyer}, L. and {Guerra}, R. and {Hutton}, A. and {Jordi}, C. and {Klioner}, S.~A. and {Lammers}, U.~L. and {Lindegren}, L. and {Luri}, X. and {Mignard}, F. and {Panem}, C. and {Pourbaix}, D. and {Randich}, S. and {Sartoretti}, P. and {Soubiran}, C. and {Tanga}, P. and {Walton}, N.~A. and {Bailer-Jones}, C.~A.~L. and {Bastian}, U. and {Drimmel}, R. and {Jansen}, F. and {Katz}, D. and {Lattanzi}, M.~G. and {van Leeuwen}, F. and {Bakker}, J. and {Cacciari}, C. and {Casta{\~n}eda}, J. and {De Angeli}, F. and {Fabricius}, C. and {Fouesneau}, M. and {Fr{\'e}mat}, Y. and {Galluccio}, L. and {Guerrier}, A. and {Heiter}, U. and {Masana}, E. and {Messineo}, R. and {Mowlavi}, N. and {Nicolas}, C. and {Nienartowicz}, K. and {Pailler}, F. and {Panuzzo}, P. and {Riclet}, F. and {Roux}, W. and {Seabroke}, G.~M. and {Sordo}, R. and {Th{\'e}venin}, F. and {Gracia-Abril}, G. and {Portell}, J. and {Teyssier}, D. and {Altmann}, M. and {Andrae}, R. and {Audard}, M. and {Bellas-Velidis}, I. and {Benson}, K. and {Berthier}, J. and {Blomme}, R. and {Burgess}, P.~W. and {Busonero}, D. and {Busso}, G. and {C{\'a}novas}, H. and {Carry}, B. and {Cellino}, A. and {Cheek}, N. and {Clementini}, G. and {Damerdji}, Y. and {Davidson}, M. and {de Teodoro}, P. and {Nu{\~n}ez Campos}, M. and {Delchambre}, L. and {Dell'Oro}, A. and {Esquej}, P. and {Fern{\'a}ndez-Hern{\'a}ndez}, J. and {Fraile}, E. and {Garabato}, D. and {Garc{\'\i}a-Lario}, P. and {Gosset}, E. and {Haigron}, R. and {Halbwachs}, J. -L. and {Hambly}, N.~C. and {Harrison}, D.~L. and {Hern{\'a}ndez}, J. and {Hestroffer}, D. and {Hodgkin}, S.~T. and {Holl}, B. and {Jan{\ss}en}, K. and {Jevardat de Fombelle}, G. and {Jordan}, S. and {Krone-Martins}, A. and {Lanzafame}, A.~C. and {L{\"o}ffler}, W. and {Marchal}, O. and {Marrese}, P.~M. and {Moitinho}, A. and {Muinonen}, K. and {Osborne}, P. and {Pancino}, E. and {Pauwels}, T. and {Recio-Blanco}, A. and {Reyl{\'e}}, C. and {Riello}, M. and {Rimoldini}, L. and {Roegiers}, T. and {Rybizki}, J. and {Sarro}, L.~M. and {Siopis}, C. and {Smith}, M. and {Sozzetti}, A. and {Utrilla}, E. and {van Leeuwen}, M. and {Abbas}, U. and {{\'A}brah{\'a}m}, P. and {Abreu Aramburu}, A. and {Aerts}, C. and {Aguado}, J.~J. and {Ajaj}, M. and {Aldea-Montero}, F. and {Altavilla}, G. and {{\'A}lvarez}, M.~A. and {Alves}, J. and {Anders}, F. and {Anderson}, R.~I. and {Anglada Varela}, E. and {Antoja}, T. and {Baines}, D. and {Baker}, S.~G. and {Balaguer-N{\'u}{\~n}ez}, L. and {Balbinot}, E. and {Balog}, Z. and {Barache}, C. and {Barbato}, D. and {Barros}, M. and {Barstow}, M.~A. and {Bartolom{\'e}}, S. and {Bassilana}, J. -L. and {Bauchet}, N. and {Becciani}, U. and {Bellazzini}, M. and {Berihuete}, A. and {Bernet}, M. and {Bertone}, S. and {Bianchi}, L. and {Binnenfeld}, A. and {Blanco-Cuaresma}, S. and {Blazere}, A. and {Boch}, T. and {Bombrun}, A. and {Bossini}, D. and {Bouquillon}, S. and {Bragaglia}, A. and {Bramante}, L. and {Breedt}, E. and {Bressan}, A. and {Brouillet}, N. and {Brugaletta}, E. and {Bucciarelli}, B. and {Burlacu}, A. and {Butkevich}, A.~G. and {Buzzi}, R. and {Caffau}, E. and {Cancelliere}, R. and {Cantat-Gaudin}, T. and {Carballo}, R. and {Carlucci}, T. and {Carnerero}, M.~I. and {Carrasco}, J.~M. and {Casamiquela}, L. and {Castellani}, M. and {Castro-Ginard}, A. and {Chaoul}, L. and {Charlot}, P. and {Chemin}, L. and {Chiaramida}, V. and {Chiavassa}, A. and {Chornay}, N. and {Comoretto}, G. and {Contursi}, G. and {Cooper}, W.~J. and {Cornez}, T. and {Cowell}, S. and {Crifo}, F. and {Cropper}, M. and {Crosta}, M. and {Crowley}, C. and {Dafonte}, C. and {Dapergolas}, A. and {David}, M. and {David}, P. and {de Laverny}, P. and {De Luise}, F. and {De March}, R. and {De Ridder}, J. and {de Souza}, R. and {de Torres}, A. and {del Peloso}, E.~F. and {del Pozo}, E. and {Delbo}, M. and {Delgado}, A. and {Delisle}, J. -B. and {Demouchy}, C. and {Dharmawardena}, T.~E. and {Di Matteo}, P. and {Diakite}, S. and {Diener}, C. and {Distefano}, E. and {Dolding}, C. and {Edvardsson}, B. and {Enke}, H. and {Fabre}, C. and {Fabrizio}, M. and {Faigler}, S. and {Fedorets}, G. and {Fernique}, P. and {Fienga}, A. and {Figueras}, F. and {Fournier}, Y. and {Fouron}, C. and {Fragkoudi}, F. and {Gai}, M. and {Garcia-Gutierrez}, A. and {Garcia-Reinaldos}, M. and {Garc{\'\i}a-Torres}, M. and {Garofalo}, A. and {Gavel}, A. and {Gavras}, P. and {Gerlach}, E. and {Geyer}, R. and {Giacobbe}, P. and {Gilmore}, G. and {Girona}, S. and {Giuffrida}, G. and {Gomel}, R. and {Gomez}, A. and {Gonz{\'a}lez-N{\'u}{\~n}ez}, J. and {Gonz{\'a}lez-Santamar{\'\i}a}, I. and {Gonz{\'a}lez-Vidal}, J.~J. and {Granvik}, M. and {Guillout}, P. and {Guiraud}, J. and {Guti{\'e}rrez-S{\'a}nchez}, R. and {Guy}, L.~P. and {Hatzidimitriou}, D. and {Hauser}, M. and {Haywood}, M. and {Helmer}, A. and {Helmi}, A. and {Sarmiento}, M.~H. and {Hidalgo}, S.~L. and {Hilger}, T. and {H{\l}adczuk}, N. and {Hobbs}, D. and {Holland}, G. and {Huckle}, H.~E. and {Jardine}, K. and {Jasniewicz}, G. and {Jean-Antoine Piccolo}, A. and {Jim{\'e}nez-Arranz}, {\'O}. and {Jorissen}, A. and {Juaristi Campillo}, J. and {Julbe}, F. and {Karbevska}, L. and {Kervella}, P. and {Khanna}, S. and {Kontizas}, M. and {Kordopatis}, G. and {Korn}, A.~J. and {K{\'o}sp{\'a}l}, {\'A}. and {Kostrzewa-Rutkowska}, Z. and {Kruszy{\'n}ska}, K. and {Kun}, M. and {Laizeau}, P. and {Lambert}, S. and {Lanza}, A.~F. and {Lasne}, Y. and {Le Campion}, J. -F. and {Lebreton}, Y. and {Lebzelter}, T. and {Leccia}, S. and {Leclerc}, N. and {Lecoeur-Taibi}, I. and {Liao}, S. and {Licata}, E.~L. and {Lindstr{\o}m}, H.~E.~P. and {Lister}, T.~A. and {Livanou}, E. and {Lobel}, A. and {Lorca}, A. and {Loup}, C. and {Madrero Pardo}, P. and {Magdaleno Romeo}, A. and {Managau}, S. and {Mann}, R.~G. and {Manteiga}, M. and {Marchant}, J.~M. and {Marconi}, M. and {Marcos}, J. and {Marcos Santos}, M.~M.~S. and {Mar{\'\i}n Pina}, D. and {Marinoni}, S. and {Marocco}, F. and {Marshall}, D.~J. and {Martin Polo}, L. and {Mart{\'\i}n-Fleitas}, J.~M. and {Marton}, G. and {Mary}, N. and {Masip}, A. and {Massari}, D. and {Mastrobuono-Battisti}, A. and {Mazeh}, T. and {McMillan}, P.~J. and {Messina}, S. and {Michalik}, D. and {Millar}, N.~R. and {Mints}, A. and {Molina}, D. and {Molinaro}, R. and {Moln{\'a}r}, L. and {Monari}, G. and {Mongui{\'o}}, M. and {Montegriffo}, P. and {Montero}, A. and {Mor}, R. and {Mora}, A. and {Morbidelli}, R. and {Morel}, T. and {Morris}, D. and {Muraveva}, T. and {Murphy}, C.~P. and {Musella}, I. and {Nagy}, Z. and {Noval}, L. and {Oca{\~n}a}, F. and {Ogden}, A. and {Ordenovic}, C. and {Osinde}, J.~O. and {Pagani}, C. and {Pagano}, I. and {Palaversa}, L. and {Palicio}, P.~A. and {Pallas-Quintela}, L. and {Panahi}, A. and {Payne-Wardenaar}, S. and {Pe{\~n}alosa Esteller}, X. and {Penttil{\"a}}, A. and {Pichon}, B. and {Piersimoni}, A.~M. and {Pineau}, F. -X. and {Plachy}, E. and {Plum}, G. and {Poggio}, E. and {Pr{\v{s}}a}, A. and {Pulone}, L. and {Racero}, E. and {Ragaini}, S. and {Rainer}, M. and {Raiteri}, C.~M. and {Rambaux}, N. and {Ramos}, P. and {Ramos-Lerate}, M. and {Re Fiorentin}, P. and {Regibo}, S. and {Richards}, P.~J. and {Rios Diaz}, C. and {Ripepi}, V. and {Riva}, A. and {Rix}, H. -W. and {Rixon}, G. and {Robichon}, N. and {Robin}, A.~C. and {Robin}, C. and {Roelens}, M. and {Rogues}, H.~R.~O. and {Rohrbasser}, L. and {Romero-G{\'o}mez}, M. and {Rowell}, N. and {Royer}, F. and {Ruz Mieres}, D. and {Rybicki}, K.~A. and {Sadowski}, G. and {S{\'a}ez N{\'u}{\~n}ez}, A. and {Sagrist{\`a} Sell{\'e}s}, A. and {Sahlmann}, J. and {Salguero}, E. and {Samaras}, N. and {Sanchez Gimenez}, V. and {Sanna}, N. and {Santove{\~n}a}, R. and {Sarasso}, M. and {Schultheis}, M. and {Sciacca}, E. and {Segol}, M. and {Segovia}, J.~C. and {S{\'e}gransan}, D. and {Semeux}, D. and {Shahaf}, S. and {Siddiqui}, H.~I. and {Siebert}, A. and {Siltala}, L. and {Silvelo}, A. and {Slezak}, E. and {Slezak}, I. and {Smart}, R.~L. and {Snaith}, O.~N. and {Solano}, E. and {Solitro}, F. and {Souami}, D. and {Souchay}, J. and {Spagna}, A. and {Spina}, L. and {Spoto}, F. and {Steele}, I.~A. and {Steidelm{\"u}ller}, H. and {Stephenson}, C.~A. and {S{\"u}veges}, M. and {Surdej}, J. and {Szabados}, L. and {Szegedi-Elek}, E. and {Taris}, F. and {Taylor}, M.~B. and {Teixeira}, R. and {Tolomei}, L. and {Tonello}, N. and {Torra}, F. and {Torra}, J. and {Torralba Elipe}, G. and {Trabucchi}, M. and {Tsounis}, A.~T. and {Turon}, C. and {Ulla}, A. and {Unger}, N. and {Vaillant}, M.~V. and {van Dillen}, E. and {van Reeven}, W. and {Vanel}, O. and {Vecchiato}, A. and {Viala}, Y. and {Vicente}, D. and {Voutsinas}, S. and {Weiler}, M. and {Wevers}, T. and {Wyrzykowski}, {\L}. and {Yoldas}, A. and {Yvard}, P. and {Zhao}, H. and {Zorec}, J. and {Zucker}, S. and {Zwitter}, T.},
        title = "{Gaia Data Release 3. Summary of the content and survey properties}",
      journal = {\aap},
         year = 2023,
        month = jun,
       volume = {674},
          eid = {A1},
        pages = {A1},
          doi = {10.1051/0004-6361/202243940},
archivePrefix = {arXiv},
       eprint = {2208.00211},
 primaryClass = {astro-ph.GA},
       adsurl = {https://ui.adsabs.harvard.edu/abs/2023A&A...674A...1G}
}

@ARTICLE{gaia16,
       author = {{Gaia Collaboration} and {Prusti}, T. and {de Bruijne}, J.~H.~J. and {Brown}, A.~G.~A. and {Vallenari}, A. and {Babusiaux}, C. and {Bailer-Jones}, C.~A.~L. and {Bastian}, U. and {Biermann}, M. and {Evans}, D.~W. and {Eyer}, L. and {Jansen}, F. and {Jordi}, C. and {Klioner}, S.~A. and {Lammers}, U. and {Lindegren}, L. and {Luri}, X. and {Mignard}, F. and {Milligan}, D.~J. and {Panem}, C. and {Poinsignon}, V. and {Pourbaix}, D. and {Randich}, S. and {Sarri}, G. and {Sartoretti}, P. and {Siddiqui}, H.~I. and {Soubiran}, C. and {Valette}, V. and {van Leeuwen}, F. and {Walton}, N.~A. and {Aerts}, C. and {Arenou}, F. and {Cropper}, M. and {Drimmel}, R. and {H{\o}g}, E. and {Katz}, D. and {Lattanzi}, M.~G. and {O'Mullane}, W. and {Grebel}, E.~K. and {Holland}, A.~D. and {Huc}, C. and {Passot}, X. and {Bramante}, L. and {Cacciari}, C. and {Casta{\~n}eda}, J. and {Chaoul}, L. and {Cheek}, N. and {De Angeli}, F. and {Fabricius}, C. and {Guerra}, R. and {Hern{\'a}ndez}, J. and {Jean-Antoine-Piccolo}, A. and {Masana}, E. and {Messineo}, R. and {Mowlavi}, N. and {Nienartowicz}, K. and {Ord{\'o}{\~n}ez-Blanco}, D. and {Panuzzo}, P. and {Portell}, J. and {Richards}, P.~J. and {Riello}, M. and {Seabroke}, G.~M. and {Tanga}, P. and {Th{\'e}venin}, F. and {Torra}, J. and {Els}, S.~G. and {Gracia-Abril}, G. and {Comoretto}, G. and {Garcia-Reinaldos}, M. and {Lock}, T. and {Mercier}, E. and {Altmann}, M. and {Andrae}, R. and {Astraatmadja}, T.~L. and {Bellas-Velidis}, I. and {Benson}, K. and {Berthier}, J. and {Blomme}, R. and {Busso}, G. and {Carry}, B. and {Cellino}, A. and {Clementini}, G. and {Cowell}, S. and {Creevey}, O. and {Cuypers}, J. and {Davidson}, M. and {De Ridder}, J. and {de Torres}, A. and {Delchambre}, L. and {Dell'Oro}, A. and {Ducourant}, C. and {Fr{\'e}mat}, Y. and {Garc{\'\i}a-Torres}, M. and {Gosset}, E. and {Halbwachs}, J. -L. and {Hambly}, N.~C. and {Harrison}, D.~L. and {Hauser}, M. and {Hestroffer}, D. and {Hodgkin}, S.~T. and {Huckle}, H.~E. and {Hutton}, A. and {Jasniewicz}, G. and {Jordan}, S. and {Kontizas}, M. and {Korn}, A.~J. and {Lanzafame}, A.~C. and {Manteiga}, M. and {Moitinho}, A. and {Muinonen}, K. and {Osinde}, J. and {Pancino}, E. and {Pauwels}, T. and {Petit}, J. -M. and {Recio-Blanco}, A. and {Robin}, A.~C. and {Sarro}, L.~M. and {Siopis}, C. and {Smith}, M. and {Smith}, K.~W. and {Sozzetti}, A. and {Thuillot}, W. and {van Reeven}, W. and {Viala}, Y. and {Abbas}, U. and {Abreu Aramburu}, A. and {Accart}, S. and {Aguado}, J.~J. and {Allan}, P.~M. and {Allasia}, W. and {Altavilla}, G. and {{\'A}lvarez}, M.~A. and {Alves}, J. and {Anderson}, R.~I. and {Andrei}, A.~H. and {Anglada Varela}, E. and {Antiche}, E. and {Antoja}, T. and {Ant{\'o}n}, S. and {Arcay}, B. and {Atzei}, A. and {Ayache}, L. and {Bach}, N. and {Baker}, S.~G. and {Balaguer-N{\'u}{\~n}ez}, L. and {Barache}, C. and {Barata}, C. and {Barbier}, A. and {Barblan}, F. and {Baroni}, M. and {Barrado y Navascu{\'e}s}, D. and {Barros}, M. and {Barstow}, M.~A. and {Becciani}, U. and {Bellazzini}, M. and {Bellei}, G. and {Bello Garc{\'\i}a}, A. and {Belokurov}, V. and {Bendjoya}, P. and {Berihuete}, A. and {Bianchi}, L. and {Bienaym{\'e}}, O. and {Billebaud}, F. and {Blagorodnova}, N. and {Blanco-Cuaresma}, S. and {Boch}, T. and {Bombrun}, A. and {Borrachero}, R. and {Bouquillon}, S. and {Bourda}, G. and {Bouy}, H. and {Bragaglia}, A. and {Breddels}, M.~A. and {Brouillet}, N. and {Br{\"u}semeister}, T. and {Bucciarelli}, B. and {Budnik}, F. and {Burgess}, P. and {Burgon}, R. and {Burlacu}, A. and {Busonero}, D. and {Buzzi}, R. and {Caffau}, E. and {Cambras}, J. and {Campbell}, H. and {Cancelliere}, R. and {Cantat-Gaudin}, T. and {Carlucci}, T. and {Carrasco}, J.~M. and {Castellani}, M. and {Charlot}, P. and {Charnas}, J. and {Charvet}, P. and {Chassat}, F. and {Chiavassa}, A. and {Clotet}, M. and {Cocozza}, G. and {Collins}, R.~S. and {Collins}, P. and {Costigan}, G. and {Crifo}, F. and {Cross}, N.~J.~G. and {Crosta}, M. and {Crowley}, C. and {Dafonte}, C. and {Damerdji}, Y. and {Dapergolas}, A. and {David}, P. and {David}, M. and {De Cat}, P. and {de Felice}, F. and {de Laverny}, P. and {De Luise}, F. and {De March}, R. and {de Martino}, D. and {de Souza}, R. and {Debosscher}, J. and {del Pozo}, E. and {Delbo}, M. and {Delgado}, A. and {Delgado}, H.~E. and {di Marco}, F. and {Di Matteo}, P. and {Diakite}, S. and {Distefano}, E. and {Dolding}, C. and {Dos Anjos}, S. and {Drazinos}, P. and {Dur{\'a}n}, J. and {Dzigan}, Y. and {Ecale}, E. and {Edvardsson}, B. and {Enke}, H. and {Erdmann}, M. and {Escolar}, D. and {Espina}, M. and {Evans}, N.~W. and {Eynard Bontemps}, G. and {Fabre}, C. and {Fabrizio}, M. and {Faigler}, S. and {Falc{\~a}o}, A.~J. and {Farr{\`a}s Casas}, M. and {Faye}, F. and {Federici}, L. and {Fedorets}, G. and {Fern{\'a}ndez-Hern{\'a}ndez}, J. and {Fernique}, P. and {Fienga}, A. and {Figueras}, F. and {Filippi}, F. and {Findeisen}, K. and {Fonti}, A. and {Fouesneau}, M. and {Fraile}, E. and {Fraser}, M. and {Fuchs}, J. and {Furnell}, R. and {Gai}, M. and {Galleti}, S. and {Galluccio}, L. and {Garabato}, D. and {Garc{\'\i}a-Sedano}, F. and {Gar{\'e}}, P. and {Garofalo}, A. and {Garralda}, N. and {Gavras}, P. and {Gerssen}, J. and {Geyer}, R. and {Gilmore}, G. and {Girona}, S. and {Giuffrida}, G. and {Gomes}, M. and {Gonz{\'a}lez-Marcos}, A. and {Gonz{\'a}lez-N{\'u}{\~n}ez}, J. and {Gonz{\'a}lez-Vidal}, J.~J. and {Granvik}, M. and {Guerrier}, A. and {Guillout}, P. and {Guiraud}, J. and {G{\'u}rpide}, A. and {Guti{\'e}rrez-S{\'a}nchez}, R. and {Guy}, L.~P. and {Haigron}, R. and {Hatzidimitriou}, D. and {Haywood}, M. and {Heiter}, U. and {Helmi}, A. and {Hobbs}, D. and {Hofmann}, W. and {Holl}, B. and {Holland}, G. and {Hunt}, J.~A.~S. and {Hypki}, A. and {Icardi}, V. and {Irwin}, M. and {Jevardat de Fombelle}, G. and {Jofr{\'e}}, P. and {Jonker}, P.~G. and {Jorissen}, A. and {Julbe}, F. and {Karampelas}, A. and {Kochoska}, A. and {Kohley}, R. and {Kolenberg}, K. and {Kontizas}, E. and {Koposov}, S.~E. and {Kordopatis}, G. and {Koubsky}, P. and {Kowalczyk}, A. and {Krone-Martins}, A. and {Kudryashova}, M. and {Kull}, I. and {Bachchan}, R.~K. and {Lacoste-Seris}, F. and {Lanza}, A.~F. and {Lavigne}, J. -B. and {Le Poncin-Lafitte}, C. and {Lebreton}, Y. and {Lebzelter}, T. and {Leccia}, S. and {Leclerc}, N. and {Lecoeur-Taibi}, I. and {Lemaitre}, V. and {Lenhardt}, H. and {Leroux}, F. and {Liao}, S. and {Licata}, E. and {Lindstr{\o}m}, H.~E.~P. and {Lister}, T.~A. and {Livanou}, E. and {Lobel}, A. and {L{\"o}ffler}, W. and {L{\'o}pez}, M. and {Lopez-Lozano}, A. and {Lorenz}, D. and {Loureiro}, T. and {MacDonald}, I. and {Magalh{\~a}es Fernandes}, T. and {Managau}, S. and {Mann}, R.~G. and {Mantelet}, G. and {Marchal}, O. and {Marchant}, J.~M. and {Marconi}, M. and {Marie}, J. and {Marinoni}, S. and {Marrese}, P.~M. and {Marschalk{\'o}}, G. and {Marshall}, D.~J. and {Mart{\'\i}n-Fleitas}, J.~M. and {Martino}, M. and {Mary}, N. and {Matijevi{\v{c}}}, G. and {Mazeh}, T. and {McMillan}, P.~J. and {Messina}, S. and {Mestre}, A. and {Michalik}, D. and {Millar}, N.~R. and {Miranda}, B.~M.~H. and {Molina}, D. and {Molinaro}, R. and {Molinaro}, M. and {Moln{\'a}r}, L. and {Moniez}, M. and {Montegriffo}, P. and {Monteiro}, D. and {Mor}, R. and {Mora}, A. and {Morbidelli}, R. and {Morel}, T. and {Morgenthaler}, S. and {Morley}, T. and {Morris}, D. and {Mulone}, A.~F. and {Muraveva}, T. and {Musella}, I. and {Narbonne}, J. and {Nelemans}, G. and {Nicastro}, L. and {Noval}, L. and {Ord{\'e}novic}, C. and {Ordieres-Mer{\'e}}, J. and {Osborne}, P. and {Pagani}, C. and {Pagano}, I. and {Pailler}, F. and {Palacin}, H. and {Palaversa}, L. and {Parsons}, P. and {Paulsen}, T. and {Pecoraro}, M. and {Pedrosa}, R. and {Pentik{\"a}inen}, H. and {Pereira}, J. and {Pichon}, B. and {Piersimoni}, A.~M. and {Pineau}, F. -X. and {Plachy}, E. and {Plum}, G. and {Poujoulet}, E. and {Pr{\v{s}}a}, A. and {Pulone}, L. and {Ragaini}, S. and {Rago}, S. and {Rambaux}, N. and {Ramos-Lerate}, M. and {Ranalli}, P. and {Rauw}, G. and {Read}, A. and {Regibo}, S. and {Renk}, F. and {Reyl{\'e}}, C. and {Ribeiro}, R.~A. and {Rimoldini}, L. and {Ripepi}, V. and {Riva}, A. and {Rixon}, G. and {Roelens}, M. and {Romero-G{\'o}mez}, M. and {Rowell}, N. and {Royer}, F. and {Rudolph}, A. and {Ruiz-Dern}, L. and {Sadowski}, G. and {Sagrist{\`a} Sell{\'e}s}, T. and {Sahlmann}, J. and {Salgado}, J. and {Salguero}, E. and {Sarasso}, M. and {Savietto}, H. and {Schnorhk}, A. and {Schultheis}, M. and {Sciacca}, E. and {Segol}, M. and {Segovia}, J.~C. and {Segransan}, D. and {Serpell}, E. and {Shih}, I. -C. and {Smareglia}, R. and {Smart}, R.~L. and {Smith}, C. and {Solano}, E. and {Solitro}, F. and {Sordo}, R. and {Soria Nieto}, S. and {Souchay}, J. and {Spagna}, A. and {Spoto}, F. and {Stampa}, U. and {Steele}, I.~A. and {Steidelm{\"u}ller}, H. and {Stephenson}, C.~A. and {Stoev}, H. and {Suess}, F.~F. and {S{\"u}veges}, M. and {Surdej}, J. and {Szabados}, L. and {Szegedi-Elek}, E. and {Tapiador}, D. and {Taris}, F. and {Tauran}, G. and {Taylor}, M.~B. and {Teixeira}, R. and {Terrett}, D. and {Tingley}, B. and {Trager}, S.~C. and {Turon}, C. and {Ulla}, A. and {Utrilla}, E. and {Valentini}, G. and {van Elteren}, A. and {Van Hemelryck}, E. and {van Leeuwen}, M. and {Varadi}, M. and {Vecchiato}, A. and {Veljanoski}, J. and {Via}, T. and {Vicente}, D. and {Vogt}, S. and {Voss}, H. and {Votruba}, V. and {Voutsinas}, S. and {Walmsley}, G. and {Weiler}, M. and {Weingrill}, K. and {Werner}, D. and {Wevers}, T. and {Whitehead}, G. and {Wyrzykowski}, {\L}. and {Yoldas}, A. and {{\v{Z}}erjal}, M. and {Zucker}, S. and {Zurbach}, C. and {Zwitter}, T. and {Alecu}, A. and {Allen}, M. and {Allende Prieto}, C. and {Amorim}, A. and {Anglada-Escud{\'e}}, G. and {Arsenijevic}, V. and {Azaz}, S. and {Balm}, P. and {Beck}, M. and {Bernstein}, H. -H. and {Bigot}, L. and {Bijaoui}, A. and {Blasco}, C. and {Bonfigli}, M. and {Bono}, G. and {Boudreault}, S. and {Bressan}, A. and {Brown}, S. and {Brunet}, P. -M. and {Bunclark}, P. and {Buonanno}, R. and {Butkevich}, A.~G. and {Carret}, C. and {Carrion}, C. and {Chemin}, L. and {Ch{\'e}reau}, F. and {Corcione}, L. and {Darmigny}, E. and {de Boer}, K.~S. and {de Teodoro}, P. and {de Zeeuw}, P.~T. and {Delle Luche}, C. and {Domingues}, C.~D. and {Dubath}, P. and {Fodor}, F. and {Fr{\'e}zouls}, B. and {Fries}, A. and {Fustes}, D. and {Fyfe}, D. and {Gallardo}, E. and {Gallegos}, J. and {Gardiol}, D. and {Gebran}, M. and {Gomboc}, A. and {G{\'o}mez}, A. and {Grux}, E. and {Gueguen}, A. and {Heyrovsky}, A. and {Hoar}, J. and {Iannicola}, G. and {Isasi Parache}, Y. and {Janotto}, A. -M. and {Joliet}, E. and {Jonckheere}, A. and {Keil}, R. and {Kim}, D. -W. and {Klagyivik}, P. and {Klar}, J. and {Knude}, J. and {Kochukhov}, O. and {Kolka}, I. and {Kos}, J. and {Kutka}, A. and {Lainey}, V. and {LeBouquin}, D. and {Liu}, C. and {Loreggia}, D. and {Makarov}, V.~V. and {Marseille}, M.~G. and {Martayan}, C. and {Martinez-Rubi}, O. and {Massart}, B. and {Meynadier}, F. and {Mignot}, S. and {Munari}, U. and {Nguyen}, A. -T. and {Nordlander}, T. and {Ocvirk}, P. and {O'Flaherty}, K.~S. and {Olias Sanz}, A. and {Ortiz}, P. and {Osorio}, J. and {Oszkiewicz}, D. and {Ouzounis}, A. and {Palmer}, M. and {Park}, P. and {Pasquato}, E. and {Peltzer}, C. and {Peralta}, J. and {P{\'e}turaud}, F. and {Pieniluoma}, T. and {Pigozzi}, E. and {Poels}, J. and {Prat}, G. and {Prod'homme}, T. and {Raison}, F. and {Rebordao}, J.~M. and {Risquez}, D. and {Rocca-Volmerange}, B. and {Rosen}, S. and {Ruiz-Fuertes}, M.~I. and {Russo}, F. and {Sembay}, S. and {Serraller Vizcaino}, I. and {Short}, A. and {Siebert}, A. and {Silva}, H. and {Sinachopoulos}, D. and {Slezak}, E. and {Soffel}, M. and {Sosnowska}, D. and {Strai{\v{z}}ys}, V. and {ter Linden}, M. and {Terrell}, D. and {Theil}, S. and {Tiede}, C. and {Troisi}, L. and {Tsalmantza}, P. and {Tur}, D. and {Vaccari}, M. and {Vachier}, F. and {Valles}, P. and {Van Hamme}, W. and {Veltz}, L. and {Virtanen}, J. and {Wallut}, J. -M. and {Wichmann}, R. and {Wilkinson}, M.~I. and {Ziaeepour}, H. and {Zschocke}, S.},
        title = "{The Gaia mission}",
      journal = {\aap},
         year = 2016,
        month = nov,
       volume = {595},
          eid = {A1},
        pages = {A1},
          doi = {10.1051/0004-6361/201629272},
archivePrefix = {arXiv},
       eprint = {1609.04153},
 primaryClass = {astro-ph.IM},
       adsurl = {https://ui.adsabs.harvard.edu/abs/2016A&A...595A...1G}
}

@ARTICLE{skrutskie06,
       author = {{Skrutskie}, M.~F. and {Cutri}, R.~M. and {Stiening}, R. and {Weinberg}, M.~D. and {Schneider}, S. and {Carpenter}, J.~M. and {Beichman}, C. and {Capps}, R. and {Chester}, T. and {Elias}, J. and {Huchra}, J. and {Liebert}, J. and {Lonsdale}, C. and {Monet}, D.~G. and {Price}, S. and {Seitzer}, P. and {Jarrett}, T. and {Kirkpatrick}, J.~D. and {Gizis}, J.~E. and {Howard}, E. and {Evans}, T. and {Fowler}, J. and {Fullmer}, L. and {Hurt}, R. and {Light}, R. and {Kopan}, E.~L. and {Marsh}, K.~A. and {McCallon}, H.~L. and {Tam}, R. and {Van Dyk}, S. and {Wheelock}, S.},
        title = "{The Two Micron All Sky Survey (2MASS)}",
      journal = {\aj},
         year = 2006,
        month = feb,
       volume = {131},
       number = {2},
        pages = {1163-1183},
          doi = {10.1086/498708},
       adsurl = {https://ui.adsabs.harvard.edu/abs/2006AJ....131.1163S}
}

@ARTICLE{delfosse00,
       author = {{Delfosse}, X. and {Forveille}, T. and {S{\'e}gransan}, D. and {Beuzit}, J. -L. and {Udry}, S. and {Perrier}, C. and {Mayor}, M.},
        title = "{Accurate masses of very low mass stars. IV. Improved mass-luminosity relations}",
      journal = {\aap},
         year = 2000,
        month = dec,
       volume = {364},
        pages = {217-224},
          doi = {10.48550/arXiv.astro-ph/0010586},
archivePrefix = {arXiv},
       eprint = {astro-ph/0010586},
 primaryClass = {astro-ph},
       adsurl = {https://ui.adsabs.harvard.edu/abs/2000A&A...364..217D}
}

@PHDTHESIS{sneden73,
       author = {{Sneden}, Christopher Alan},
        title = "{Carbon and Nitrogen Abundances in Metal-Poor Stars.}",
       school = {University of Texas, Austin},
         year = 1973,
        month = jan,
       adsurl = {https://ui.adsabs.harvard.edu/abs/1973PhDT.......180S}
}

@INPROCEEDINGS{mclean98,
       author = {{McLean}, Ian S. and {Becklin}, Eric E. and {Bendiksen}, Oddvar and {Brims}, George and {Canfield}, John and {Figer}, Donald F. and {Graham}, James R. and {Hare}, Jonah and {Lacayanga}, Fred and {Larkin}, James E. and {Larson}, Samuel B. and {Levenson}, Nancy and {Magnone}, Nick and {Teplitz}, Harry and {Wong}, Woon},
        title = "{Design and development of NIRSPEC: a near-infrared echelle spectrograph for the Keck II telescope}",
    booktitle = {Infrared Astronomical Instrumentation},
         year = 1998,
       editor = {{Fowler}, Albert M.},
       series = {Society of Photo-Optical Instrumentation Engineers (SPIE) Conference Series},
       volume = {3354},
        month = aug,
        pages = {566-578},
          doi = {10.1117/12.317283},
       adsurl = {https://ui.adsabs.harvard.edu/abs/1998SPIE.3354..566M}
}

@INPROCEEDINGS{tody86,
       author = {{Tody}, Doug},
        title = "{The IRAF Data Reduction and Analysis System}",
    booktitle = {Instrumentation in astronomy VI},
         year = 1986,
       editor = {{Crawford}, David L.},
       series = {Society of Photo-Optical Instrumentation Engineers (SPIE) Conference Series},
       volume = {627},
        month = jan,
        pages = {733},
          doi = {10.1117/12.968154},
       adsurl = {https://ui.adsabs.harvard.edu/abs/1986SPIE..627..733T}
}

@INPROCEEDINGS{tody93,
       author = {{Tody}, Doug},
        title = "{IRAF in the Nineties}",
    booktitle = {Astronomical Data Analysis Software and Systems II},
         year = 1993,
       editor = {{Hanisch}, R.~J. and {Brissenden}, R.~J.~V. and {Barnes}, J.},
       series = {Astronomical Society of the Pacific Conference Series},
       volume = {52},
        month = jan,
        pages = {173},
       adsurl = {https://ui.adsabs.harvard.edu/abs/1993ASPC...52..173T}
}

@ARTICLE{brooke16,
       author = {{Brooke}, James S.~A. and {Bernath}, Peter F. and {Western}, Colin M. and {Sneden}, Christopher and {Af{\c{s}}ar}, Melike and {Li}, Gang and {Gordon}, Iouli E.},
        title = "{Line strengths of rovibrational and rotational transitions in the X$^{2}$ {\ensuremath{\Pi}} ground state of OH}",
      journal = {\jqsrt},
         year = 2016,
        month = jan,
       volume = {168},
        pages = {142-157},
          doi = {10.1016/j.jqsrt.2015.07.021},
       adsurl = {https://ui.adsabs.harvard.edu/abs/2016JQSRT.168..142B}
}

@ARTICLE{oliva15,
       author = {{Oliva}, E. and {Origlia}, L. and {Scuderi}, S. and {Benatti}, S. and {Carleo}, I. and {Lapenna}, E. and {Mucciarelli}, A. and {Baffa}, C. and {Biliotti}, V. and {Carbonaro}, L. and {Falcini}, G. and {Giani}, E. and {Iuzzolino}, M. and {Massi}, F. and {Sanna}, N. and {Sozzi}, M. and {Tozzi}, A. and {Ghedina}, A. and {Ghinassi}, F. and {Lodi}, M. and {Harutyunyan}, A. and {Pedani}, M.},
        title = "{Lines and continuum sky emission in the near infrared: observational constraints from deep high spectral resolution spectra with GIANO-TNG}",
      journal = {\aap},
         year = 2015,
        month = sep,
       volume = {581},
          eid = {A47},
        pages = {A47},
          doi = {10.1051/0004-6361/201526291},
archivePrefix = {arXiv},
       eprint = {1506.09004},
 primaryClass = {astro-ph.IM},
       adsurl = {https://ui.adsabs.harvard.edu/abs/2015A&A...581A..47O}
}

@ARTICLE{rousselot00,
       author = {{Rousselot}, P. and {Lidman}, C. and {Cuby}, J.-G. and {Moreels}, G. and {Monnet}, G.},
        title = "{Night-sky spectral atlas of OH emission lines in the near-infrared}",
      journal = {\aap},
         year = 2000,
        month = feb,
       volume = {354},
        pages = {1134-1150},
       adsurl = {https://ui.adsabs.harvard.edu/abs/2000A&A...354.1134R}
}

@ARTICLE{blancocuaresma14,
       author = {{Blanco-Cuaresma}, S. and {Soubiran}, C. and {Heiter}, U. and {Jofr{\'e}}, P.},
        title = "{Determining stellar atmospheric parameters and chemical abundances of FGK stars with iSpec}",
      journal = {\aap},
         year = 2014,
        month = sep,
       volume = {569},
          eid = {A111},
        pages = {A111},
          doi = {10.1051/0004-6361/201423945},
archivePrefix = {arXiv},
       eprint = {1407.2608},
 primaryClass = {astro-ph.IM},
       adsurl = {https://ui.adsabs.harvard.edu/abs/2014A&A...569A.111B}
}

@ARTICLE{blancocuaresma19,
       author = {{Blanco-Cuaresma}, Sergi},
        title = "{Modern stellar spectroscopy caveats}",
      journal = {\mnras},
         year = 2019,
        month = jun,
       volume = {486},
       number = {2},
        pages = {2075-2101},
          doi = {10.1093/mnras/stz549},
archivePrefix = {arXiv},
       eprint = {1902.09558},
 primaryClass = {astro-ph.SR},
       adsurl = {https://ui.adsabs.harvard.edu/abs/2019MNRAS.486.2075B}
}

@ARTICLE{hejazi24,
       author = {{Hejazi}, Neda and {Crossfield}, Ian J.~M. and {Souto}, Diogo and {Brande}, Jonathan and {Nordlander}, Thomas and {Marfil}, Emilio and {Cunha}, Katia and {Coria}, David R. and {Maas}, Zachary G. and {Polanski}, Alex S. and {Hinkel}, Natalie R. and {Hand}, Joseph E.},
        title = "{High-resolution Elemental Abundance Measurements of Cool JWST Planet Hosts Using AutoSpecFit: An Application to the Sub-Neptune K2-18b's Host M Dwarf}",
      journal = {\apj},
         year = 2024,
        month = sep,
       volume = {973},
       number = {1},
          eid = {31},
        pages = {31},
          doi = {10.3847/1538-4357/ad61dc},
archivePrefix = {arXiv},
       eprint = {2407.07869},
 primaryClass = {astro-ph.EP},
       adsurl = {https://ui.adsabs.harvard.edu/abs/2024ApJ...973...31H}
}

@ARTICLE{smith21,
       author = {{Smith}, Verne V. and {Bizyaev}, Dmitry and {Cunha}, Katia and {Shetrone}, Matthew D. and {Souto}, Diogo and {Allende Prieto}, Carlos and {Masseron}, Thomas and {M{\'e}sz{\'a}ros}, Szabolcs and {J{\"o}nsson}, Henrik and {Hasselquist}, Sten and {Osorio}, Yeisson and {Garc{\'\i}a-Hern{\'a}ndez}, D.~A. and {Plez}, Bertrand and {Beaton}, Rachael L. and {Holtzman}, Jon and {Majewski}, Steven R. and {Stringfellow}, Guy S. and {Sobeck}, Jennifer},
        title = "{The APOGEE Data Release 16 Spectral Line List}",
      journal = {\aj},
         year = 2021,
        month = jun,
       volume = {161},
       number = {6},
          eid = {254},
        pages = {254},
          doi = {10.3847/1538-3881/abefdc},
archivePrefix = {arXiv},
       eprint = {2103.10112},
 primaryClass = {astro-ph.SR},
       adsurl = {https://ui.adsabs.harvard.edu/abs/2021AJ....161..254S}
}

@ARTICLE{shetrone15,
       author = {{Shetrone}, M. and {Bizyaev}, D. and {Lawler}, J.~E. and {Allende Prieto}, C. and {Johnson}, J.~A. and {Smith}, V.~V. and {Cunha}, K. and {Holtzman}, J. and {Garc{\'\i}a P{\'e}rez}, A.~E. and {M{\'e}sz{\'a}ros}, Sz. and {Sobeck}, J. and {Zamora}, O. and {Garc{\'\i}a-Hern{\'a}ndez}, D.~A. and {Souto}, D. and {Chojnowski}, D. and {Koesterke}, L. and {Majewski}, S. and {Zasowski}, G.},
        title = "{The SDSS-III APOGEE Spectral Line List for H-band Spectroscopy}",
      journal = {\apjs},
         year = 2015,
        month = dec,
       volume = {221},
       number = {2},
          eid = {24},
        pages = {24},
          doi = {10.1088/0067-0049/221/2/24},
archivePrefix = {arXiv},
       eprint = {1502.04080},
 primaryClass = {astro-ph.IM},
       adsurl = {https://ui.adsabs.harvard.edu/abs/2015ApJS..221...24S}
}

@ARTICLE{majewski17,
       author = {{Majewski}, Steven R. and {Schiavon}, Ricardo P. and {Frinchaboy}, Peter M. and {Allende Prieto}, Carlos and {Barkhouser}, Robert and {Bizyaev}, Dmitry and {Blank}, Basil and {Brunner}, Sophia and {Burton}, Adam and {Carrera}, Ricardo and {Chojnowski}, S. Drew and {Cunha}, K{\'a}tia and {Epstein}, Courtney and {Fitzgerald}, Greg and {Garc{\'\i}a P{\'e}rez}, Ana E. and {Hearty}, Fred R. and {Henderson}, Chuck and {Holtzman}, Jon A. and {Johnson}, Jennifer A. and {Lam}, Charles R. and {Lawler}, James E. and {Maseman}, Paul and {M{\'e}sz{\'a}ros}, Szabolcs and {Nelson}, Matthew and {Nguyen}, Duy Coung and {Nidever}, David L. and {Pinsonneault}, Marc and {Shetrone}, Matthew and {Smee}, Stephen and {Smith}, Verne V. and {Stolberg}, Todd and {Skrutskie}, Michael F. and {Walker}, Eric and {Wilson}, John C. and {Zasowski}, Gail and {Anders}, Friedrich and {Basu}, Sarbani and {Beland}, Stephane and {Blanton}, Michael R. and {Bovy}, Jo and {Brownstein}, Joel R. and {Carlberg}, Joleen and {Chaplin}, William and {Chiappini}, Cristina and {Eisenstein}, Daniel J. and {Elsworth}, Yvonne and {Feuillet}, Diane and {Fleming}, Scott W. and {Galbraith-Frew}, Jessica and {Garc{\'\i}a}, Rafael A. and {Garc{\'\i}a-Hern{\'a}ndez}, D. An{\'\i}bal and {Gillespie}, Bruce A. and {Girardi}, L{\'e}o and {Gunn}, James E. and {Hasselquist}, Sten and {Hayden}, Michael R. and {Hekker}, Saskia and {Ivans}, Inese and {Kinemuchi}, Karen and {Klaene}, Mark and {Mahadevan}, Suvrath and {Mathur}, Savita and {Mosser}, Beno{\^\i}t and {Muna}, Demitri and {Munn}, Jeffrey A. and {Nichol}, Robert C. and {O'Connell}, Robert W. and {Parejko}, John K. and {Robin}, A.~C. and {Rocha-Pinto}, Helio and {Schultheis}, Matthias and {Serenelli}, Aldo M. and {Shane}, Neville and {Silva Aguirre}, Victor and {Sobeck}, Jennifer S. and {Thompson}, Benjamin and {Troup}, Nicholas W. and {Weinberg}, David H. and {Zamora}, Olga},
        title = "{The Apache Point Observatory Galactic Evolution Experiment (APOGEE)}",
      journal = {\aj},
         year = 2017,
        month = sep,
       volume = {154},
       number = {3},
          eid = {94},
        pages = {94},
          doi = {10.3847/1538-3881/aa784d},
archivePrefix = {arXiv},
       eprint = {1509.05420},
 primaryClass = {astro-ph.IM},
       adsurl = {https://ui.adsabs.harvard.edu/abs/2017AJ....154...94M}
}

@ARTICLE{polyansky18,
       author = {{Polyansky}, Oleg L. and {Kyuberis}, Aleksandra A. and {Zobov}, Nikolai F. and {Tennyson}, Jonathan and {Yurchenko}, Sergei N. and {Lodi}, Lorenzo},
        title = "{ExoMol molecular line lists XXX: a complete high-accuracy line list for water}",
      journal = {\mnras},
         year = 2018,
        month = oct,
       volume = {480},
       number = {2},
        pages = {2597-2608},
          doi = {10.1093/mnras/sty1877},
archivePrefix = {arXiv},
       eprint = {1807.04529},
 primaryClass = {astro-ph.EP},
       adsurl = {https://ui.adsabs.harvard.edu/abs/2018MNRAS.480.2597P}
}

@ARTICLE{li15,
       author = {{Li}, Gang and {Gordon}, Iouli E. and {Rothman}, Laurence S. and {Tan}, Yan and {Hu}, Shui-Ming and {Kassi}, Samir and {Campargue}, Alain and {Medvedev}, Emile S.},
        title = "{Rovibrational Line Lists for Nine Isotopologues of the CO Molecule in the X $^{1}${\ensuremath{\Sigma}}$^{+}$ Ground Electronic State}",
      journal = {\apjs},
         year = 2015,
        month = jan,
       volume = {216},
       number = {1},
          eid = {15},
        pages = {15},
          doi = {10.1088/0067-0049/216/1/15},
       adsurl = {https://ui.adsabs.harvard.edu/abs/2015ApJS..216...15L}
}

@ARTICLE{mckemmish16,
       author = {{McKemmish}, Laura K. and {Yurchenko}, Sergei N. and {Tennyson}, Jonathan},
        title = "{ExoMol line lists - XVIII. The high-temperature spectrum of VO}",
      journal = {\mnras},
         year = 2016,
        month = nov,
       volume = {463},
       number = {1},
        pages = {771-793},
          doi = {10.1093/mnras/stw1969},
archivePrefix = {arXiv},
       eprint = {1609.06120},
 primaryClass = {astro-ph.SR},
       adsurl = {https://ui.adsabs.harvard.edu/abs/2016MNRAS.463..771M}
}

@ARTICLE{mckemmish17,
       author = {{McKemmish}, Laura K. and {Masseron}, Thomas and {Sheppard}, Samuel and {Sandeman}, Elizabeth and {Schofield}, Zak and {Furtenbacher}, Tibor and {Cs{\'a}sz{\'a}r}, Attila G. and {Tennyson}, Jonathan and {Sousa-Silva}, Clara},
        title = "{Marvel Analysis of the Measured High-resolution Rovibronic Spectra of TiO}",
      journal = {\apjs},
         year = 2017,
        month = feb,
       volume = {228},
       number = {2},
          eid = {15},
        pages = {15},
          doi = {10.3847/1538-4365/228/2/15},
archivePrefix = {arXiv},
       eprint = {1703.03477},
 primaryClass = {astro-ph.SR},
       adsurl = {https://ui.adsabs.harvard.edu/abs/2017ApJS..228...15M}
}

@ARTICLE{dulick03,
       author = {{Dulick}, M. and {Bauschlicher}, Jr., C.~W. and {Burrows}, Adam and {Sharp}, C.~M. and {Ram}, R.~S. and {Bernath}, Peter},
        title = "{Line Intensities and Molecular Opacities of the FeH F $^{4}${\ensuremath{\Delta}}$_{i}$-X $^{4}${\ensuremath{\Delta}}$_{i}$ Transition}",
      journal = {\apj},
         year = 2003,
        month = sep,
       volume = {594},
       number = {1},
        pages = {651-663},
          doi = {10.1086/376791},
archivePrefix = {arXiv},
       eprint = {astro-ph/0305162},
 primaryClass = {astro-ph},
       adsurl = {https://ui.adsabs.harvard.edu/abs/2003ApJ...594..651D}
}

@ARTICLE{hargreaves10,
       author = {{Hargreaves}, Robert J. and {Hinkle}, Kenneth H. and {Bauschlicher}, Jr., Charles W. and {Wende}, Sebastian and {Seifahrt}, Andreas and {Bernath}, Peter F.},
        title = "{High-resolution 1.6 {\ensuremath{\mu}}m Spectra of FeH in M and L Dwarfs}",
      journal = {\aj},
         year = 2010,
        month = oct,
       volume = {140},
       number = {4},
        pages = {919-924},
          doi = {10.1088/0004-6256/140/4/919},
       adsurl = {https://ui.adsabs.harvard.edu/abs/2010AJ....140..919H}
}

@ARTICLE{brooke14,
       author = {{Brooke}, James S.~A. and {Ram}, Ram S. and {Western}, Colin M. and {Li}, Gang and {Schwenke}, David W. and {Bernath}, Peter F.},
        title = "{Einstein A Coefficients and Oscillator Strengths for the A $^{2}${\ensuremath{\Pi}}-X $^{2}${\ensuremath{\Sigma}}$^{+}$ (Red) and B$^{2}${\ensuremath{\Sigma}}$^{+}$-X$^{2}${\ensuremath{\Sigma}}$^{+}$ (Violet) Systems and Rovibrational Transitions in the X $^{2}${\ensuremath{\Sigma}}$^{+}$ State of CN}",
      journal = {\apjs},
         year = 2014,
        month = feb,
       volume = {210},
       number = {2},
          eid = {23},
        pages = {23},
          doi = {10.1088/0067-0049/210/2/23},
       adsurl = {https://ui.adsabs.harvard.edu/abs/2014ApJS..210...23B}
}

@ARTICLE{flores24,
       author = {{Flores}, Isabella and {Maas}, Zachary G. and {Hinkel}, Natalie R.},
        title = "{An Exploration of Molecular Lines for Abundance Analysis in M-dwarf IR Spectra}",
      journal = {Research Notes of the American Astronomical Society},
         year = 2024,
        month = dec,
       volume = {8},
       number = {12},
          eid = {315},
        pages = {315},
          doi = {10.3847/2515-5172/ad9f60},
       adsurl = {https://ui.adsabs.harvard.edu/abs/2024RNAAS...8..315F}
}

@ARTICLE{souto20,
       author = {{Souto}, Diogo and {Cunha}, Katia and {Smith}, Verne V. and {Allende Prieto}, C. and {Burgasser}, Adam and {Covey}, Kevin and {Garc{\'\i}a-Hern{\'a}ndez}, D.~A. and {Holtzman}, Jon A. and {Johnson}, Jennifer A. and {J{\"o}nsson}, Henrik and {Mahadevan}, Suvrath and {Majewski}, Steven R. and {Masseron}, Thomas and {Shetrone}, Matthew and {Rojas-Ayala}, B{\'a}rbara and {Sobeck}, Jennifer and {Stassun}, Keivan G. and {Terrien}, Ryan and {Teske}, Johanna and {Wanderley}, F{\'a}bio and {Zamora}, Olga},
        title = "{Stellar Characterization of M Dwarfs from the APOGEE Survey: A Calibrator Sample for M-dwarf Metallicities}",
      journal = {\apj},
         year = 2020,
        month = feb,
       volume = {890},
       number = {2},
          eid = {133},
        pages = {133},
          doi = {10.3847/1538-4357/ab6d07},
archivePrefix = {arXiv},
       eprint = {2001.05597},
 primaryClass = {astro-ph.SR},
       adsurl = {https://ui.adsabs.harvard.edu/abs/2020ApJ...890..133S}
}

@ARTICLE{souto17,
       author = {{Souto}, D. and {Cunha}, K. and {Garc{\'\i}a-Hern{\'a}ndez}, D.~A. and {Zamora}, O. and {Allende Prieto}, C. and {Smith}, V.~V. and {Mahadevan}, S. and {Blake}, C. and {Johnson}, J.~A. and {J{\"o}nsson}, H. and {Pinsonneault}, M. and {Holtzman}, J. and {Majewski}, S.~R. and {Shetrone}, M. and {Teske}, J. and {Nidever}, D. and {Schiavon}, R. and {Sobeck}, J. and {Garc{\'\i}a P{\'e}rez}, A.~E. and {G{\'o}mez Maqueo Chew}, Y. and {Stassun}, K.},
        title = "{Chemical Abundances of M-dwarfs from the APOGEE Survey. I. The Exoplanet Hosting Stars Kepler-138 and Kepler-186}",
      journal = {\apj},
         year = 2017,
        month = feb,
       volume = {835},
       number = {2},
          eid = {239},
        pages = {239},
          doi = {10.3847/1538-4357/835/2/239},
archivePrefix = {arXiv},
       eprint = {1612.01598},
 primaryClass = {astro-ph.SR},
       adsurl = {https://ui.adsabs.harvard.edu/abs/2017ApJ...835..239S}
}

@ARTICLE{asplund09,
       author = {{Asplund}, Martin and {Grevesse}, Nicolas and {Sauval}, A. Jacques and {Scott}, Pat},
        title = "{The Chemical Composition of the Sun}",
      journal = {\araa},
         year = 2009,
        month = sep,
       volume = {47},
       number = {1},
        pages = {481-522},
          doi = {10.1146/annurev.astro.46.060407.145222},
archivePrefix = {arXiv},
       eprint = {0909.0948},
 primaryClass = {astro-ph.SR},
       adsurl = {https://ui.adsabs.harvard.edu/abs/2009ARA&A..47..481A}
}

@ARTICLE{souto22,
       author = {{Souto}, Diogo and {Cunha}, Katia and {Smith}, Verne V. and {Allende Prieto}, C. and {Covey}, Kevin and {Garc{\'\i}a-Hern{\'a}ndez}, D.~A. and {Holtzman}, Jon A. and {J{\"o}nsson}, Henrik and {Mahadevan}, Suvrath and {Majewski}, Steven R. and {Masseron}, Thomas and {Pinsonneault}, Marc and {Schneider}, Donald P. and {Shetrone}, Matthew and {Stassun}, Keivan G. and {Terrien}, Ryan and {Zamora}, Olga and {Stringfellow}, Guy S. and {Lane}, Richard R. and {Nitschelm}, Christian and {Rojas-Ayala}, B{\'a}rbara},
        title = "{Detailed Chemical Abundances for a Benchmark Sample of M Dwarfs from the APOGEE Survey}",
      journal = {\apj},
         year = 2022,
        month = mar,
       volume = {927},
       number = {1},
          eid = {123},
        pages = {123},
          doi = {10.3847/1538-4357/ac4891},
archivePrefix = {arXiv},
       eprint = {2201.00891},
 primaryClass = {astro-ph.SR},
       adsurl = {https://ui.adsabs.harvard.edu/abs/2022ApJ...927..123S}
}

@ARTICLE{tabernero24,
       author = {{Tabernero}, H.~M. and {Shan}, Y. and {Caballero}, J.~A. and {Duque-Arribas}, C. and {Montes}, D. and {Gonz{\'a}lez Hern{\'a}ndez}, J.~I. and {Zapatero Osorio}, M.~R. and {Schweitzer}, A. and {Henning}, Th. and {Cort{\'e}s-Contreras}, M. and {Quirrenbach}, A. and {Amado}, P.~J. and {Reiners}, A. and {Ribas}, I. and {Bergond}, G. and {Morales}, J.~C.},
        title = "{The CARMENES search for exoplanets around M dwarfs: Magnesium and silicon abundances of K7{\textendash}M5.5 stars}",
      journal = {\aap},
         year = 2024,
        month = sep,
       volume = {689},
          eid = {A223},
        pages = {A223},
          doi = {10.1051/0004-6361/202450054},
archivePrefix = {arXiv},
       eprint = {2407.19969},
 primaryClass = {astro-ph.SR},
       adsurl = {https://ui.adsabs.harvard.edu/abs/2024A&A...689A.223T}
}

@ARTICLE{melo24,
       author = {{Melo}, Edypo and {Souto}, Diogo and {Cunha}, Katia and {Smith}, Verne V. and {Wanderley}, F{\'a}bio and {Grilo}, Vinicius and {Camara}, Deusalete and {Murta}, Kely and {Hejazi}, Neda and {Crossfield}, Ian J.~M. and {Teske}, Johanna and {Luque}, Rafael and {Zhang}, Michael and {Bean}, Jacob},
        title = "{Stellar Characterization and Chemical Abundances of Exoplanet-hosting M Dwarfs from APOGEE Spectra: Future JWST Targets}",
      journal = {\apj},
         year = 2024,
        month = oct,
       volume = {973},
       number = {2},
          eid = {90},
        pages = {90},
          doi = {10.3847/1538-4357/ad5004},
archivePrefix = {arXiv},
       eprint = {2406.00111},
 primaryClass = {astro-ph.SR},
       adsurl = {https://ui.adsabs.harvard.edu/abs/2024ApJ...973...90M}
}

@ARTICLE{marfil21,
       author = {{Marfil}, E. and {Tabernero}, H.~M. and {Montes}, D. and {Caballero}, J.~A. and {L{\'a}zaro}, F.~J. and {Gonz{\'a}lez Hern{\'a}ndez}, J.~I. and {Nagel}, E. and {Passegger}, V.~M. and {Schweitzer}, A. and {Ribas}, I. and {Reiners}, A. and {Quirrenbach}, A. and {Amado}, P.~J. and {Cifuentes}, C. and {Cort{\'e}s-Contreras}, M. and {Dreizler}, S. and {Duque-Arribas}, C. and {Galad{\'\i}-Enr{\'\i}quez}, D. and {Henning}, Th. and {Jeffers}, S.~V. and {Kaminski}, A. and {K{\"u}rster}, M. and {Lafarga}, M. and {L{\'o}pez-Gallifa}, {\'A}. and {Morales}, J.~C. and {Shan}, Y. and {Zechmeister}, M.},
        title = "{The CARMENES search for exoplanets around M dwarfs. Stellar atmospheric parameters of target stars with SteParSyn}",
      journal = {\aap},
         year = 2021,
        month = dec,
       volume = {656},
          eid = {A162},
        pages = {A162},
          doi = {10.1051/0004-6361/202141980},
archivePrefix = {arXiv},
       eprint = {2110.07329},
 primaryClass = {astro-ph.SR},
       adsurl = {https://ui.adsabs.harvard.edu/abs/2021A&A...656A.162M}
}

@ARTICLE{kuznetsov19,
       author = {{Kuznetsov}, M.~K. and {del Burgo}, C. and {Pavlenko}, Ya. V. and {Frith}, J.},
        title = "{Characterization of a Sample of Southern M Dwarfs Using Harps and X-shooter Spectra}",
      journal = {\apj},
         year = 2019,
        month = jun,
       volume = {878},
       number = {2},
          eid = {134},
        pages = {134},
          doi = {10.3847/1538-4357/ab1fe9},
       adsurl = {https://ui.adsabs.harvard.edu/abs/2019ApJ...878..134K}
}

@ARTICLE{hejazi22,
       author = {{Hejazi}, Neda and {L{\'e}pine}, S{\'e}bastien and {Nordlander}, Thomas},
        title = "{Chemical Properties of the Local Disk and Halo. II. Abundances of 3745 M Dwarfs and Subdwarfs from Improved Model Fitting of Low-resolution Spectra}",
      journal = {\apj},
         year = 2022,
        month = mar,
       volume = {927},
       number = {1},
          eid = {122},
        pages = {122},
          doi = {10.3847/1538-4357/ac4e16},
archivePrefix = {arXiv},
       eprint = {2201.07460},
 primaryClass = {astro-ph.SR},
       adsurl = {https://ui.adsabs.harvard.edu/abs/2022ApJ...927..122H}
}

@software{plez12,
       author = {{Plez}, B.},
        title = "{Turbospectrum: Code for spectral synthesis}",
 howpublished = {Astrophysics Source Code Library, record ascl:1205.004},
         year = 2012,
        month = may,
          eid = {ascl:1205.004},
archivePrefix = {ascl},
       eprint = {1205.004},
       adsurl = {https://ui.adsabs.harvard.edu/abs/2012ascl.soft05004P}
}

@ARTICLE{gustafsson08,
       author = {{Gustafsson}, B. and {Edvardsson}, B. and {Eriksson}, K. and {J{\o}rgensen}, U.~G. and {Nordlund}, {\r{A}}. and {Plez}, B.},
        title = "{A grid of MARCS model atmospheres for late-type stars. I. Methods and general properties}",
      journal = {\aap},
         year = 2008,
        month = aug,
       volume = {486},
       number = {3},
        pages = {951-970},
          doi = {10.1051/0004-6361:200809724},
archivePrefix = {arXiv},
       eprint = {0805.0554},
 primaryClass = {astro-ph},
       adsurl = {https://ui.adsabs.harvard.edu/abs/2008A&A...486..951G}
}

@ARTICLE{allard12,
       author = {{Allard}, F. and {Homeier}, D. and {Freytag}, B.},
        title = "{Models of very-low-mass stars, brown dwarfs and exoplanets}",
      journal = {Philosophical Transactions of the Royal Society of London Series A},
         year = 2012,
        month = jun,
       volume = {370},
       number = {1968},
        pages = {2765-2777},
          doi = {10.1098/rsta.2011.0269},
archivePrefix = {arXiv},
       eprint = {1112.3591},
 primaryClass = {astro-ph.SR},
       adsurl = {https://ui.adsabs.harvard.edu/abs/2012RSPTA.370.2765A}
}

@ARTICLE{dressing15,
       author = {{Dressing}, Courtney D. and {Charbonneau}, David},
        title = "{The Occurrence of Potentially Habitable Planets Orbiting M Dwarfs Estimated from the Full Kepler Dataset and an Empirical Measurement of the Detection Sensitivity}",
      journal = {\apj},
         year = 2015,
        month = jul,
       volume = {807},
       number = {1},
          eid = {45},
        pages = {45},
          doi = {10.1088/0004-637X/807/1/45},
archivePrefix = {arXiv},
       eprint = {1501.01623},
 primaryClass = {astro-ph.EP},
       adsurl = {https://ui.adsabs.harvard.edu/abs/2015ApJ...807...45D}
}

@ARTICLE{apogeedr17,
       author = {{Abdurro'uf} and {Accetta}, Katherine and {Aerts}, Conny and {Silva Aguirre}, V{\'\i}ctor and {Ahumada}, Romina and {Ajgaonkar}, Nikhil and {Filiz Ak}, N. and {Alam}, Shadab and {Allende Prieto}, Carlos and {Almeida}, Andr{\'e}s and {Anders}, Friedrich and {Anderson}, Scott F. and {Andrews}, Brett H. and {Anguiano}, Borja and {Aquino-Ort{\'\i}z}, Erik and {Arag{\'o}n-Salamanca}, Alfonso and {Argudo-Fern{\'a}ndez}, Maria and {Ata}, Metin and {Aubert}, Marie and {Avila-Reese}, Vladimir and {Badenes}, Carles and {Barb{\'a}}, Rodolfo H. and {Barger}, Kat and {Barrera-Ballesteros}, Jorge K. and {Beaton}, Rachael L. and {Beers}, Timothy C. and {Belfiore}, Francesco and {Bender}, Chad F. and {Bernardi}, Mariangela and {Bershady}, Matthew A. and {Beutler}, Florian and {Bidin}, Christian Moni and {Bird}, Jonathan C. and {Bizyaev}, Dmitry and {Blanc}, Guillermo A. and {Blanton}, Michael R. and {Boardman}, Nicholas Fraser and {Bolton}, Adam S. and {Boquien}, M{\'e}d{\'e}ric and {Borissova}, Jura and {Bovy}, Jo and {Brandt}, W.~N. and {Brown}, Jordan and {Brownstein}, Joel R. and {Brusa}, Marcella and {Buchner}, Johannes and {Bundy}, Kevin and {Burchett}, Joseph N. and {Bureau}, Martin and {Burgasser}, Adam and {Cabang}, Tuesday K. and {Campbell}, Stephanie and {Cappellari}, Michele and {Carlberg}, Joleen K. and {Wanderley}, F{\'a}bio Carneiro and {Carrera}, Ricardo and {Cash}, Jennifer and {Chen}, Yan-Ping and {Chen}, Wei-Huai and {Cherinka}, Brian and {Chiappini}, Cristina and {Choi}, Peter Doohyun and {Chojnowski}, S. Drew and {Chung}, Haeun and {Clerc}, Nicolas and {Cohen}, Roger E. and {Comerford}, Julia M. and {Comparat}, Johan and {da Costa}, Luiz and {Covey}, Kevin and {Crane}, Jeffrey D. and {Cruz-Gonzalez}, Irene and {Culhane}, Connor and {Cunha}, Katia and {Dai}, Y. Sophia and {Damke}, Guillermo and {Darling}, Jeremy and {Davidson}, Jr., James W. and {Davies}, Roger and {Dawson}, Kyle and {De Lee}, Nathan and {Diamond-Stanic}, Aleksandar M. and {Cano-D{\'\i}az}, Mariana and {S{\'a}nchez}, Helena Dom{\'\i}nguez and {Donor}, John and {Duckworth}, Chris and {Dwelly}, Tom and {Eisenstein}, Daniel J. and {Elsworth}, Yvonne P. and {Emsellem}, Eric and {Eracleous}, Mike and {Escoffier}, Stephanie and {Fan}, Xiaohui and {Farr}, Emily and {Feng}, Shuai and {Fern{\'a}ndez-Trincado}, Jos{\'e} G. and {Feuillet}, Diane and {Filipp}, Andreas and {Fillingham}, Sean P. and {Frinchaboy}, Peter M. and {Fromenteau}, Sebastien and {Galbany}, Llu{\'\i}s and {Garc{\'\i}a}, Rafael A. and {Garc{\'\i}a-Hern{\'a}ndez}, D.~A. and {Ge}, Junqiang and {Geisler}, Doug and {Gelfand}, Joseph and {G{\'e}ron}, Tobias and {Gibson}, Benjamin J. and {Goddy}, Julian and {Godoy-Rivera}, Diego and {Grabowski}, Kathleen and {Green}, Paul J. and {Greener}, Michael and {Grier}, Catherine J. and {Griffith}, Emily and {Guo}, Hong and {Guy}, Julien and {Hadjara}, Massinissa and {Harding}, Paul and {Hasselquist}, Sten and {Hayes}, Christian R. and {Hearty}, Fred and {Hern{\'a}ndez}, Jes{\'u}s and {Hill}, Lewis and {Hogg}, David W. and {Holtzman}, Jon A. and {Horta}, Danny and {Hsieh}, Bau-Ching and {Hsu}, Chin-Hao and {Hsu}, Yun-Hsin and {Huber}, Daniel and {Huertas-Company}, Marc and {Hutchinson}, Brian and {Hwang}, Ho Seong and {Ibarra-Medel}, H{\'e}ctor J. and {Chitham}, Jacob Ider and {Ilha}, Gabriele S. and {Imig}, Julie and {Jaekle}, Will and {Jayasinghe}, Tharindu and {Ji}, Xihan and {Johnson}, Jennifer A. and {Jones}, Amy and {J{\"o}nsson}, Henrik and {Katkov}, Ivan and {Khalatyan}, Dr., Arman and {Kinemuchi}, Karen and {Kisku}, Shobhit and {Knapen}, Johan H. and {Kneib}, Jean-Paul and {Kollmeier}, Juna A. and {Kong}, Miranda and {Kounkel}, Marina and {Kreckel}, Kathryn and {Krishnarao}, Dhanesh and {Lacerna}, Ivan and {Lane}, Richard R. and {Langgin}, Rachel and {Lavender}, Ramon and {Law}, David R. and {Lazarz}, Daniel and {Leung}, Henry W. and {Leung}, Ho-Hin and {Lewis}, Hannah M. and {Li}, Cheng and {Li}, Ran and {Lian}, Jianhui and {Liang}, Fu-Heng and {Lin}, Lihwai and {Lin}, Yen-Ting and {Lin}, Sicheng and {Lintott}, Chris and {Long}, Dan and {Longa-Pe{\~n}a}, Pen{\'e}lope and {L{\'o}pez-Cob{\'a}}, Carlos and {Lu}, Shengdong and {Lundgren}, Britt F. and {Luo}, Yuanze and {Mackereth}, J. Ted and {de la Macorra}, Axel and {Mahadevan}, Suvrath and {Majewski}, Steven R. and {Manchado}, Arturo and {Mandeville}, Travis and {Maraston}, Claudia and {Margalef-Bentabol}, Berta and {Masseron}, Thomas and {Masters}, Karen L. and {Mathur}, Savita and {McDermid}, Richard M. and {Mckay}, Myles and {Merloni}, Andrea and {Merrifield}, Michael and {Meszaros}, Szabolcs and {Miglio}, Andrea and {Di Mille}, Francesco and {Minniti}, Dante and {Minsley}, Rebecca and {Monachesi}, Antonela},
        title = "{The Seventeenth Data Release of the Sloan Digital Sky Surveys: Complete Release of MaNGA, MaStar, and APOGEE-2 Data}",
      journal = {\apjs},
         year = 2022,
        month = apr,
       volume = {259},
       number = {2},
          eid = {35},
        pages = {35},
          doi = {10.3847/1538-4365/ac4414},
archivePrefix = {arXiv},
       eprint = {2112.02026},
 primaryClass = {astro-ph.GA},
       adsurl = {https://ui.adsabs.harvard.edu/abs/2022ApJS..259...35A}
}

@ARTICLE{reddy03,
       author = {{Reddy}, Bacham E. and {Tomkin}, Jocelyn and {Lambert}, David L. and {Allende Prieto}, Carlos},
        title = "{The chemical compositions of Galactic disc F and G dwarfs}",
      journal = {\mnras},
         year = 2003,
        month = mar,
       volume = {340},
       number = {1},
        pages = {304-340},
          doi = {10.1046/j.1365-8711.2003.06305.x},
archivePrefix = {arXiv},
       eprint = {astro-ph/0211551},
 primaryClass = {astro-ph},
       adsurl = {https://ui.adsabs.harvard.edu/abs/2003MNRAS.340..304R}
}

@ARTICLE{ramirez07,
       author = {{Ram{\'\i}rez}, I. and {Allende Prieto}, C. and {Lambert}, D.~L.},
        title = "{Oxygen abundances in nearby stars. Clues to the formation and evolution of the Galactic disk}",
      journal = {\aap},
         year = 2007,
        month = apr,
       volume = {465},
       number = {1},
        pages = {271-289},
          doi = {10.1051/0004-6361:20066619},
archivePrefix = {arXiv},
       eprint = {astro-ph/0701362},
 primaryClass = {astro-ph},
       adsurl = {https://ui.adsabs.harvard.edu/abs/2007A&A...465..271R}
}

@ARTICLE{ramirez13,
       author = {{Ram{\'\i}rez}, I. and {Allende Prieto}, C. and {Lambert}, D.~L.},
        title = "{Oxygen Abundances in Nearby FGK Stars and the Galactic Chemical Evolution of the Local Disk and Halo}",
      journal = {\apj},
         year = 2013,
        month = feb,
       volume = {764},
       number = {1},
          eid = {78},
        pages = {78},
          doi = {10.1088/0004-637X/764/1/78},
archivePrefix = {arXiv},
       eprint = {1301.1582},
 primaryClass = {astro-ph.SR},
       adsurl = {https://ui.adsabs.harvard.edu/abs/2013ApJ...764...78R}
}

@ARTICLE{wu15,
       author = {{Wu}, Xiaoshu and {Wang}, Liang and {Shi}, Jianrong and {Zhao}, Gang and {Grupp}, Frank},
        title = "{Palladium and silver abundances in stars with [Fe/H] > -2.6}",
      journal = {\aap},
         year = 2015,
        month = jul,
       volume = {579},
          eid = {A8},
        pages = {A8},
          doi = {10.1051/0004-6361/201525679},
archivePrefix = {arXiv},
       eprint = {1505.04356},
 primaryClass = {astro-ph.SR},
       adsurl = {https://ui.adsabs.harvard.edu/abs/2015A&A...579A...8W}
}

@ARTICLE{chen00,
       author = {{Chen}, Y.~Q. and {Nissen}, P.~E. and {Zhao}, G. and {Zhang}, H.~W. and {Benoni}, T.},
        title = "{Chemical composition of 90 F and G disk dwarfs}",
      journal = {\aaps},
         year = 2000,
        month = feb,
       volume = {141},
        pages = {491-506},
          doi = {10.1051/aas:2000124},
archivePrefix = {arXiv},
       eprint = {astro-ph/9912342},
 primaryClass = {astro-ph},
       adsurl = {https://ui.adsabs.harvard.edu/abs/2000A&AS..141..491C}
}

@ARTICLE{hinkel16,
       author = {{Hinkel}, Natalie R. and {Young}, Patrick A. and {Pagano}, Michael D. and {Desch}, Steven J. and {Anbar}, Ariel D. and {Adibekyan}, Vardan and {Blanco-Cuaresma}, Sergi and {Carlberg}, Joleen K. and {Delgado Mena}, Elisa and {Liu}, Fan and {Nordlander}, Thomas and {Sousa}, Sergio G. and {Korn}, Andreas and {Gruyters}, Pieter and {Heiter}, Ulrike and {Jofr{\'e}}, Paula and {Santos}, Nuno C. and {Soubiran}, Caroline},
        title = "{A Comparison of Stellar Elemental Abundance Techniques and Measurements}",
      journal = {\apjs},
         year = 2016,
        month = sep,
       volume = {226},
       number = {1},
          eid = {4},
        pages = {4},
          doi = {10.3847/0067-0049/226/1/4},
archivePrefix = {arXiv},
       eprint = {1607.03130},
 primaryClass = {astro-ph.SR},
       adsurl = {https://ui.adsabs.harvard.edu/abs/2016ApJS..226....4H}
}

@ARTICLE{montes18,
       author = {{Montes}, D. and {Gonz{\'a}lez-Peinado}, R. and {Tabernero}, H.~M. and {Caballero}, J.~A. and {Marfil}, E. and {Alonso-Floriano}, F.~J. and {Cort{\'e}s-Contreras}, M. and {Gonz{\'a}lez Hern{\'a}ndez}, J.~I. and {Klutsch}, A. and {Moreno-J{\'o}dar}, C.},
        title = "{Calibrating the metallicity of M dwarfs in wide physical binaries with F-, G-, and K-primaries - I: High-resolution spectroscopy with HERMES: stellar parameters, abundances, and kinematics}",
      journal = {\mnras},
         year = 2018,
        month = sep,
       volume = {479},
       number = {1},
        pages = {1332-1382},
          doi = {10.1093/mnras/sty1295},
archivePrefix = {arXiv},
       eprint = {1805.05394},
 primaryClass = {astro-ph.SR},
       adsurl = {https://ui.adsabs.harvard.edu/abs/2018MNRAS.479.1332M}
}

@ARTICLE{adibekyan12,
       author = {{Adibekyan}, V. Zh. and {Sousa}, S.~G. and {Santos}, N.~C. and {Delgado Mena}, E. and {Gonz{\'a}lez Hern{\'a}ndez}, J.~I. and {Israelian}, G. and {Mayor}, M. and {Khachatryan}, G.},
        title = "{Chemical abundances of 1111 FGK stars from the HARPS GTO planet search program. Galactic stellar populations and planets}",
      journal = {\aap},
         year = 2012,
        month = sep,
       volume = {545},
          eid = {A32},
        pages = {A32},
          doi = {10.1051/0004-6361/201219401},
archivePrefix = {arXiv},
       eprint = {1207.2388},
 primaryClass = {astro-ph.EP},
       adsurl = {https://ui.adsabs.harvard.edu/abs/2012A&A...545A..32A}
}

@ARTICLE{costasilva20,
       author = {{Costa Silva}, A.~R. and {Delgado Mena}, E. and {Tsantaki}, M.},
        title = "{Chemical abundances of 1111 FGK stars from the HARPS-GTO planet search sample. III. Sulfur}",
      journal = {\aap},
         year = 2020,
        month = feb,
       volume = {634},
          eid = {A136},
        pages = {A136},
          doi = {10.1051/0004-6361/201936523},
archivePrefix = {arXiv},
       eprint = {1912.08659},
 primaryClass = {astro-ph.SR},
       adsurl = {https://ui.adsabs.harvard.edu/abs/2020A&A...634A.136C}
}

@ARTICLE{delgadomena10,
       author = {{Delgado Mena}, E. and {Israelian}, G. and {Gonz{\'a}lez Hern{\'a}ndez}, J.~I. and {Bond}, J.~C. and {Santos}, N.~C. and {Udry}, S. and {Mayor}, M.},
        title = "{Chemical Clues on the Formation of Planetary Systems: C/O Versus Mg/Si for HARPS GTO Sample}",
      journal = {\apj},
         year = 2010,
        month = dec,
       volume = {725},
       number = {2},
        pages = {2349-2358},
          doi = {10.1088/0004-637X/725/2/2349},
archivePrefix = {arXiv},
       eprint = {1009.5224},
 primaryClass = {astro-ph.SR},
       adsurl = {https://ui.adsabs.harvard.edu/abs/2010ApJ...725.2349D}
}

@ARTICLE{neves10,
       author = {{Gonz{\'a}lez Hern{\'a}ndez}, J.~I. and {Israelian}, G. and {Santos}, N.~C. and {Sousa}, S. and {Delgado-Mena}, E. and {Neves}, V. and {Udry}, S.},
        title = "{Searching for the Signatures of Terrestrial Planets in Solar Analogs}",
      journal = {\apj},
         year = 2010,
        month = sep,
       volume = {720},
       number = {2},
        pages = {1592-1602},
          doi = {10.1088/0004-637X/720/2/1592},
archivePrefix = {arXiv},
       eprint = {1007.0580},
 primaryClass = {astro-ph.SR},
       adsurl = {https://ui.adsabs.harvard.edu/abs/2010ApJ...720.1592G}
}

@ARTICLE{luck18,
       author = {{Luck}, R. Earle},
        title = "{Abundances in the Local Region. III. Southern F, G, and K Dwarfs}",
      journal = {\aj},
         year = 2018,
        month = mar,
       volume = {155},
       number = {3},
          eid = {111},
        pages = {111},
          doi = {10.3847/1538-3881/aaa9b5},
       adsurl = {https://ui.adsabs.harvard.edu/abs/2018AJ....155..111L}
}

@ARTICLE{petigura11,
       author = {{Petigura}, Erik A. and {Marcy}, Geoffrey W.},
        title = "{Carbon and Oxygen in Nearby Stars: Keys to Protoplanetary Disk Chemistry}",
      journal = {\apj},
         year = 2011,
        month = jul,
       volume = {735},
       number = {1},
          eid = {41},
        pages = {41},
          doi = {10.1088/0004-637X/735/1/41},
archivePrefix = {arXiv},
       eprint = {1106.5449},
 primaryClass = {astro-ph.SR},
       adsurl = {https://ui.adsabs.harvard.edu/abs/2011ApJ...735...41P}
}

@ARTICLE{valenti05,
       author = {{Valenti}, Jeff A. and {Fischer}, Debra A.},
        title = "{Spectroscopic Properties of Cool Stars (SPOCS). I. 1040 F, G, and K Dwarfs from Keck, Lick, and AAT Planet Search Programs}",
      journal = {\apjs},
         year = 2005,
        month = jul,
       volume = {159},
       number = {1},
        pages = {141-166},
          doi = {10.1086/430500},
       adsurl = {https://ui.adsabs.harvard.edu/abs/2005ApJS..159..141V}
}

@ARTICLE{wheeler24,
       author = {{Wheeler}, Adam J. and {Casey}, Andrew R. and {Abruzzo}, Matthew W.},
        title = "{Korg: Fitting, Model Atmosphere Interpolation, and Brackett Lines}",
      journal = {\aj},
         year = 2024,
        month = feb,
       volume = {167},
       number = {2},
          eid = {83},
        pages = {83},
          doi = {10.3847/1538-3881/ad19cc},
archivePrefix = {arXiv},
       eprint = {2310.19823},
 primaryClass = {astro-ph.SR},
       adsurl = {https://ui.adsabs.harvard.edu/abs/2024AJ....167...83W}
}

@ARTICLE{abia20,
       author = {{Abia}, C. and {Tabernero}, H.~M. and {Korotin}, S.~A. and {Montes}, D. and {Marfil}, E. and {Caballero}, J.~A. and {Straniero}, O. and {Prantzos}, N. and {Ribas}, I. and {Reiners}, A. and {Quirrenbach}, A. and {Amado}, P.~J. and {B{\'e}jar}, V.~J.~S. and {Cort{\'e}s-Contreras}, M. and {Dreizler}, S. and {Henning}, Th. and {Jeffers}, S.~V. and {Kaminski}, A. and {K{\"u}rster}, M. and {Lafarga}, M. and {L{\'o}pez-Gallifa}, {\'A}. and {Morales}, J.~C. and {Nagel}, E. and {Passegger}, V.~M. and {Pedraz}, S. and {Rodr{\'\i}guez L{\'o}pez}, C. and {Schweitzer}, A. and {Zechmeister}, M.},
        title = "{The CARMENES search for exoplanets around M dwarfs. Rubidium abundances in nearby cool stars}",
      journal = {\aap},
         year = 2020,
        month = oct,
       volume = {642},
          eid = {A227},
        pages = {A227},
          doi = {10.1051/0004-6361/202039032},
archivePrefix = {arXiv},
       eprint = {2009.00876},
 primaryClass = {astro-ph.SR},
       adsurl = {https://ui.adsabs.harvard.edu/abs/2020A&A...642A.227A}
}

@ARTICLE{sawczynec25,
       author = {{Sawczynec}, Erica and {Kaplan}, Kyle F. and {Mace}, Gregory N. and {Lee}, Jae-Joon and {Jaffe}, Daniel T. and {Park}, Chan and {Yuk}, In-Soo and {Chun}, Moo-Young and {Pak}, Soojong and {Hwang}, Narae and {Jeong}, Ueejeong and {Kim}, Hwihyun and {Kim}, Hyun-Jeong and {Kim}, Kang-Min and {Kim}, Sanghyuk and {Le}, Huynh Anh N. and {Lee}, Hye-In and {Lee}, Sungho and {Oh}, Heeyoung and {Oh}, Jae Sok and {Park}, Byeong-Gon and {Park}, Woojin and {Yu}, Young-Sam},
        title = "{10 Years of Archival High-resolution NIR Spectra: The Raw and Reduced IGRINS Spectral Archive (RRISA)}",
      journal = {\pasp},
         year = 2025,
        month = mar,
       volume = {137},
       number = {3},
          eid = {034505},
        pages = {034505},
          doi = {10.1088/1538-3873/adba89},
archivePrefix = {arXiv},
       eprint = {2503.05867},
 primaryClass = {astro-ph.IM},
       adsurl = {https://ui.adsabs.harvard.edu/abs/2025PASP..137c4505S}
}

@ARTICLE{hinkel14,
       author = {{Hinkel}, Natalie R. and {Timmes}, F.~X. and {Young}, Patrick A. and {Pagano}, Michael D. and {Turnbull}, Margaret C.},
        title = "{Stellar Abundances in the Solar Neighborhood: The Hypatia Catalog}",
      journal = {\aj},
         year = 2014,
        month = sep,
       volume = {148},
       number = {3},
          eid = {54},
        pages = {54},
          doi = {10.1088/0004-6256/148/3/54},
archivePrefix = {arXiv},
       eprint = {1405.6719},
 primaryClass = {astro-ph.SR},
       adsurl = {https://ui.adsabs.harvard.edu/abs/2014AJ....148...54H}
}

@ARTICLE{gan20,
       author = {{Gan}, Tianjun and {Shporer}, Avi and {Livingston}, John H. and {Collins}, Karen A. and {Mao}, Shude and {Trani}, Alessandro A. and {Gandolfi}, Davide and {Hirano}, Teruyuki and {Luque}, Rafael and {Stassun}, Keivan G. and {Ziegler}, Carl and {Howell}, Steve B. and {Hellier}, Coel and {Irwin}, Jonathan M. and {Winters}, Jennifer G. and {Anderson}, David R. and {Brice{\~n}o}, C{\'e}sar and {Law}, Nicholas and {Mann}, Andrew W. and {Bonfils}, Xavier and {Astudillo-Defru}, Nicola and {Jensen}, Eric L.~N. and {Anglada-Escud{\'e}}, Guillem and {Ricker}, George R. and {Vanderspek}, Roland and {Latham}, David W. and {Seager}, Sara and {Winn}, Joshua N. and {Jenkins}, Jon M. and {Furesz}, Gabor and {Guerrero}, Natalia M. and {Quintana}, Elisa and {Twicken}, Joseph D. and {Caldwell}, Douglas A. and {Tenenbaum}, Peter and {Huang}, Chelsea X. and {Rowden}, Pamela and {Rojas-Ayala}, B{\'a}rbara},
        title = "{LHS 1815b: The First Thick-disk Planet Detected by TESS}",
      journal = {\aj},
         year = 2020,
        month = apr,
       volume = {159},
       number = {4},
          eid = {160},
        pages = {160},
          doi = {10.3847/1538-3881/ab775a},
archivePrefix = {arXiv},
       eprint = {2003.04525},
 primaryClass = {astro-ph.EP},
       adsurl = {https://ui.adsabs.harvard.edu/abs/2020AJ....159..160G}
}

@ARTICLE{alinder25,
       author = {{Alinder}, Simon and {Bensby}, Thomas and {McMillan}, Paul},
        title = "{Impact of selection criteria on the structural parameters of the Galactic thin and thick discs}",
      journal = {arXiv e-prints},
         year = 2025,
        month = nov,
          eid = {arXiv:2511.10092},
        pages = {arXiv:2511.10092},
          doi = {10.48550/arXiv.2511.10092},
archivePrefix = {arXiv},
       eprint = {2511.10092},
 primaryClass = {astro-ph.GA},
       adsurl = {https://ui.adsabs.harvard.edu/abs/2025arXiv251110092A}
}

@ARTICLE{maas22,
       author = {{Maas}, Zachary G. and {Hawkins}, Keith and {Hinkel}, Natalie R. and {Cargile}, Phillip and {Janowiecki}, Steven and {Nelson}, Tyler},
        title = "{The Galactic Distribution of Phosphorus: A Survey of 163 Disk and Halo Stars}",
      journal = {\aj},
         year = 2022,
        month = aug,
       volume = {164},
       number = {2},
          eid = {61},
        pages = {61},
          doi = {10.3847/1538-3881/ac77f8},
archivePrefix = {arXiv},
       eprint = {2206.03528},
 primaryClass = {astro-ph.SR},
       adsurl = {https://ui.adsabs.harvard.edu/abs/2022AJ....164...61M}
}

@software{pyia18,
       author = {{Price-Whelan}, Adrian},
        title = "{adrn/pyia: v0.2}",
         year = 2018,
        month = apr,
          eid = {10.5281/zenodo.1228136},
          doi = {10.5281/zenodo.1228136},
      version = {v0.2},
    publisher = {Zenodo},
       adsurl = {https://ui.adsabs.harvard.edu/abs/2018zndo...1228136P}
}

@ARTICLE{crossfield22,
       author = {{Crossfield}, Ian J.~M. and {Malik}, Matej and {Hill}, Michelle L. and {Kane}, Stephen R. and {Foley}, Bradford and {Polanski}, Alex S. and {Coria}, David and {Brande}, Jonathan and {Zhang}, Yanzhe and {Wienke}, Katherine and {Kreidberg}, Laura and {Cowan}, Nicolas B. and {Dragomir}, Diana and {Gorjian}, Varoujan and {Mikal-Evans}, Thomas and {Benneke}, Bj{\"o}rn and {Christiansen}, Jessie L. and {Deming}, Drake and {Morales}, Farisa Y.},
        title = "{GJ 1252b: A Hot Terrestrial Super-Earth with No Atmosphere}",
      journal = {\apjl},
         year = 2022,
        month = sep,
       volume = {937},
       number = {1},
          eid = {L17},
        pages = {L17},
          doi = {10.3847/2041-8213/ac886b},
archivePrefix = {arXiv},
       eprint = {2208.09479},
 primaryClass = {astro-ph.EP},
       adsurl = {https://ui.adsabs.harvard.edu/abs/2022ApJ...937L..17C}
}

@ARTICLE{foley16,
       author = {{Foley}, Bradford J. and {Driscoll}, Peter E.},
        title = "{Whole planet coupling between climate, mantle, and core: Implications for rocky planet evolution}",
      journal = {Geochemistry, Geophysics, Geosystems},
         year = 2016,
        month = may,
       volume = {17},
       number = {5},
        pages = {1885-1914},
          doi = {10.1002/2015GC006210},
archivePrefix = {arXiv},
       eprint = {1711.06801},
 primaryClass = {astro-ph.EP},
       adsurl = {https://ui.adsabs.harvard.edu/abs/2016GGG....17.1885F}
}

@ARTICLE{lisse20,
       author = {{Lisse}, Carey M. and {Desch}, Steven J. and {Unterborn}, Cayman T. and {Kane}, Stephen R. and {Young}, Patrick R. and {Hartnett}, Hilairy E. and {Hinkel}, Natalie R. and {Shim}, Sang-Heon and {Mamajek}, Eric E. and {Izenberg}, Noam R.},
        title = "{A Geologically Robust Procedure for Observing Rocky Exoplanets to Ensure that Detection of Atmospheric Oxygen Is a Modern Earth-like Biosignature}",
      journal = {\apjl},
         year = 2020,
        month = jul,
       volume = {898},
       number = {1},
          eid = {L17},
        pages = {L17},
          doi = {10.3847/2041-8213/ab9b91},
archivePrefix = {arXiv},
       eprint = {2006.07403},
 primaryClass = {astro-ph.EP},
       adsurl = {https://ui.adsabs.harvard.edu/abs/2020ApJ...898L..17L}
}

@ARTICLE{cointepas21,
       author = {{Cointepas}, M. and {Almenara}, J.~M. and {Bonfils}, X. and {Bouchy}, F. and {Astudillo-Defru}, N. and {Murgas}, F. and {Otegi}, J.~F. and {Wyttenbach}, A. and {Anderson}, D.~R. and {Artigau}, {\'E}. and {Canto Martins}, B.~L. and {Charbonneau}, D. and {Collins}, K.~A. and {Collins}, K.~I. and {Correia}, J.-J. and {Curaba}, S. and {Delboulb{\'e}}, A. and {Delfosse}, X. and {D{\'\i}az}, R.~F. and {Dorn}, C. and {Doyon}, R. and {Feautrier}, P. and {Figueira}, P. and {Forveille}, T. and {Gaisne}, G. and {Gan}, T. and {Gluck}, L. and {Helled}, R. and {Hellier}, C. and {Jocou}, L. and {Kern}, P. and {Lafrasse}, S. and {Law}, N. and {Le{\~a}o}, I.~C. and {Lovis}, C. and {Magnard}, Y. and {Mann}, A.~W. and {Maurel}, D. and {de Medeiros}, J.~R. and {Melo}, C. and {Moulin}, T. and {Pepe}, F. and {Rabou}, P. and {Rochat}, S. and {Rodriguez}, D.~R. and {Roux}, A. and {Santos}, N.~C. and {S{\'e}gransan}, D. and {Stadler}, E. and {Ting}, E.~B. and {Twicken}, J.~D. and {Udry}, S. and {Waalkes}, W.~C. and {West}, R.~G. and {W{\"u}nsche}, A. and {Ziegler}, C. and {Ricker}, G. and {Vanderspek}, R. and {Latham}, D.~W. and {Seager}, S. and {Winn}, J. and {Jenkins}, J.~M.},
        title = "{TOI-269 b: an eccentric sub-Neptune transiting a M2 dwarf revisited with ExTrA}",
      journal = {\aap},
         year = 2021,
        month = jun,
       volume = {650},
          eid = {A145},
        pages = {A145},
          doi = {10.1051/0004-6361/202140328},
archivePrefix = {arXiv},
       eprint = {2104.14782},
 primaryClass = {astro-ph.EP},
       adsurl = {https://ui.adsabs.harvard.edu/abs/2021A&A...650A.145C}
}

@ARTICLE{astropy13,
       author = {{Astropy Collaboration} and {Robitaille}, Thomas P. and {Tollerud}, Erik J. and {Greenfield}, Perry and {Droettboom}, Michael and {Bray}, Erik and {Aldcroft}, Tom and {Davis}, Matt and {Ginsburg}, Adam and {Price-Whelan}, Adrian M. and {Kerzendorf}, Wolfgang E. and {Conley}, Alexander and {Crighton}, Neil and {Barbary}, Kyle and {Muna}, Demitri and {Ferguson}, Henry and {Grollier}, Fr{\'e}d{\'e}ric and {Parikh}, Madhura M. and {Nair}, Prasanth H. and {Unther}, Hans M. and {Deil}, Christoph and {Woillez}, Julien and {Conseil}, Simon and {Kramer}, Roban and {Turner}, James E.~H. and {Singer}, Leo and {Fox}, Ryan and {Weaver}, Benjamin A. and {Zabalza}, Victor and {Edwards}, Zachary I. and {Azalee Bostroem}, K. and {Burke}, D.~J. and {Casey}, Andrew R. and {Crawford}, Steven M. and {Dencheva}, Nadia and {Ely}, Justin and {Jenness}, Tim and {Labrie}, Kathleen and {Lim}, Pey Lian and {Pierfederici}, Francesco and {Pontzen}, Andrew and {Ptak}, Andy and {Refsdal}, Brian and {Servillat}, Mathieu and {Streicher}, Ole},
        title = "{Astropy: A community Python package for astronomy}",
      journal = {\aap},
         year = 2013,
        month = oct,
       volume = {558},
          eid = {A33},
        pages = {A33},
          doi = {10.1051/0004-6361/201322068},
archivePrefix = {arXiv},
       eprint = {1307.6212},
 primaryClass = {astro-ph.IM},
       adsurl = {https://ui.adsabs.harvard.edu/abs/2013A&A...558A..33A}
}

@ARTICLE{bensby14,
       author = {{Bensby}, T. and {Feltzing}, S. and {Oey}, M.~S.},
        title = "{Exploring the Milky Way stellar disk. A detailed elemental abundance study of 714 F and G dwarf stars in the solar neighbourhood}",
      journal = {\aap},
         year = 2014,
        month = feb,
       volume = {562},
          eid = {A71},
        pages = {A71},
          doi = {10.1051/0004-6361/201322631},
archivePrefix = {arXiv},
       eprint = {1309.2631},
 primaryClass = {astro-ph.GA},
       adsurl = {https://ui.adsabs.harvard.edu/abs/2014A&A...562A..71B}
}

@ARTICLE{reyle21,
       author = {{Reyl{\'e}}, C. and {Jardine}, K. and {Fouqu{\'e}}, P. and {Caballero}, J.~A. and {Smart}, R.~L. and {Sozzetti}, A.},
        title = "{The 10 parsec sample in the Gaia era}",
      journal = {\aap},
         year = 2021,
        month = jun,
       volume = {650},
          eid = {A201},
        pages = {A201},
          doi = {10.1051/0004-6361/202140985},
archivePrefix = {arXiv},
       eprint = {2104.14972},
 primaryClass = {astro-ph.SR},
       adsurl = {https://ui.adsabs.harvard.edu/abs/2021A&A...650A.201R}
}

@ARTICLE{casagrande08,
       author = {{Casagrande}, Luca and {Flynn}, Chris and {Bessell}, Michael},
        title = "{M dwarfs: effective temperatures, radii and metallicities}",
      journal = {\mnras},
         year = 2008,
        month = sep,
       volume = {389},
       number = {2},
        pages = {585-607},
          doi = {10.1111/j.1365-2966.2008.13573.x},
archivePrefix = {arXiv},
       eprint = {0806.2471},
 primaryClass = {astro-ph},
       adsurl = {https://ui.adsabs.harvard.edu/abs/2008MNRAS.389..585C}
}

@ARTICLE{allard00,
       author = {{Allard}, France and {Hauschildt}, Peter H. and {Schwenke}, David},
        title = "{TiO and H$_{2}$O Absorption Lines in Cool Stellar Atmospheres}",
      journal = {\apj},
         year = 2000,
        month = sep,
       volume = {540},
       number = {2},
        pages = {1005-1015},
          doi = {10.1086/309366},
archivePrefix = {arXiv},
       eprint = {astro-ph/0008465},
 primaryClass = {astro-ph},
       adsurl = {https://ui.adsabs.harvard.edu/abs/2000ApJ...540.1005A}
}

@ARTICLE{hejazi15,
       author = {{Hejazi}, N. and {De Robertis}, M.~M. and {Dawson}, P.~C.},
        title = "{Optical-Near Infrared Photometric Calibration of M Dwarf Metallicity and Its Application}",
      journal = {\aj},
         year = 2015,
        month = apr,
       volume = {149},
       number = {4},
          eid = {140},
        pages = {140},
          doi = {10.1088/0004-6256/149/4/140},
archivePrefix = {arXiv},
       eprint = {1502.07460},
 primaryClass = {astro-ph.SR},
       adsurl = {https://ui.adsabs.harvard.edu/abs/2015AJ....149..140H}
}

@ARTICLE{rojasayala12,
       author = {{Rojas-Ayala}, B{\'a}rbara and {Covey}, Kevin R. and {Muirhead}, Philip S. and {Lloyd}, James P.},
        title = "{Metallicity and Temperature Indicators in M Dwarf K-band Spectra: Testing New and Updated Calibrations with Observations of 133 Solar Neighborhood M Dwarfs}",
      journal = {\apj},
         year = 2012,
        month = apr,
       volume = {748},
       number = {2},
          eid = {93},
        pages = {93},
          doi = {10.1088/0004-637X/748/2/93},
archivePrefix = {arXiv},
       eprint = {1112.4567},
 primaryClass = {astro-ph.SR},
       adsurl = {https://ui.adsabs.harvard.edu/abs/2012ApJ...748...93R}
}

@ARTICLE{newton15,
       author = {{Newton}, Elisabeth R. and {Charbonneau}, David and {Irwin}, Jonathan and {Mann}, Andrew W.},
        title = "{An Empirical Calibration to Estimate Cool Dwarf Fundamental Parameters from H-band Spectra}",
      journal = {\apj},
         year = 2015,
        month = feb,
       volume = {800},
       number = {2},
          eid = {85},
        pages = {85},
          doi = {10.1088/0004-637X/800/2/85},
archivePrefix = {arXiv},
       eprint = {1412.2758},
 primaryClass = {astro-ph.SR},
       adsurl = {https://ui.adsabs.harvard.edu/abs/2015ApJ...800...85N}
}

@ARTICLE{mann13,
       author = {{Mann}, Andrew W. and {Brewer}, John M. and {Gaidos}, Eric and {L{\'e}pine}, S{\'e}bastien and {Hilton}, Eric J.},
        title = "{Prospecting in Late-type Dwarfs: A Calibration of Infrared and Visible Spectroscopic Metallicities of Late K and M Dwarfs Spanning 1.5 dex}",
      journal = {\aj},
         year = 2013,
        month = feb,
       volume = {145},
       number = {2},
          eid = {52},
        pages = {52},
          doi = {10.1088/0004-6256/145/2/52},
archivePrefix = {arXiv},
       eprint = {1211.4630},
 primaryClass = {astro-ph.SR},
       adsurl = {https://ui.adsabs.harvard.edu/abs/2013AJ....145...52M}
}

@ARTICLE{neves14,
       author = {{Neves}, V. and {Bonfils}, X. and {Santos}, N.~C. and {Delfosse}, X. and {Forveille}, T. and {Allard}, F. and {Udry}, S.},
        title = "{Metallicity of M dwarfs. IV. A high-precision [Fe/H] and T$_{eff}$ technique from high-resolution optical spectra for M dwarfs}",
      journal = {\aap},
         year = 2014,
        month = aug,
       volume = {568},
          eid = {A121},
        pages = {A121},
          doi = {10.1051/0004-6361/201424139},
archivePrefix = {arXiv},
       eprint = {1406.6127},
 primaryClass = {astro-ph.SR},
       adsurl = {https://ui.adsabs.harvard.edu/abs/2014A&A...568A.121N}
}

@ARTICLE{lindgren17,
       author = {{Lindgren}, Sara and {Heiter}, Ulrike},
        title = "{Metallicity determination of M dwarfs. Expanded parameter range in metallicity and effective temperature}",
      journal = {\aap},
         year = 2017,
        month = aug,
       volume = {604},
          eid = {A97},
        pages = {A97},
          doi = {10.1051/0004-6361/201730715},
archivePrefix = {arXiv},
       eprint = {1705.08785},
 primaryClass = {astro-ph.SR},
       adsurl = {https://ui.adsabs.harvard.edu/abs/2017A&A...604A..97L}
}

@ARTICLE{lindgren16,
       author = {{Lindgren}, Sara and {Heiter}, Ulrike and {Seifahrt}, Andreas},
        title = "{Metallicity determination of M dwarfs. High-resolution infrared spectroscopy}",
      journal = {\aap},
         year = 2016,
        month = feb,
       volume = {586},
          eid = {A100},
        pages = {A100},
          doi = {10.1051/0004-6361/201526602},
archivePrefix = {arXiv},
       eprint = {1510.06642},
 primaryClass = {astro-ph.SR},
       adsurl = {https://ui.adsabs.harvard.edu/abs/2016A&A...586A.100L}
}

@ARTICLE{henry24,
       author = {{Henry}, Todd J. and {Jao}, Wei-Chun},
        title = "{The Character of M Dwarfs}",
      journal = {\araa},
         year = 2024,
        month = sep,
       volume = {62},
       number = {1},
        pages = {593-633},
          doi = {10.1146/annurev-astro-052722-102740},
       adsurl = {https://ui.adsabs.harvard.edu/abs/2024ARA&A..62..593H}
}

\section{Appendix A}
\label{sec:appendixA}

\FloatBarrier

Table of line-by-line abundances and abundances for stars with multiple visits for stars in our sample are shown in this appendix. We also give OH and \water{} line information used to derive [O/H] and $\teff{}$s for our target stars in Table \ref{tab:oxy_linelist}. We note that \water{} lines are blended and a representative strong \water{} line was chosen to denote the region fit.

\begin{splitdeluxetable*}{c c c c c c c c B c c c c c c c c c  B c c c  c c c c c c c}
\tablewidth{0pt} 
\tabletypesize{\scriptsize}
\tablecaption{Line-By-Line Abundances} 
 \tablehead{\colhead{Names} & \colhead{Date Observed} & \colhead{Fe I}  & \colhead{Fe I}  & \colhead{Fe I}  & \colhead{Fe I} & \colhead{Fe I}  & \colhead{Fe I}  & \colhead{Fe I}  & \colhead{Fe I} & \colhead{Fe I} & \colhead{FeH} & \colhead{FeH} & \colhead{FeH} & \colhead{FeH} & \colhead{FeH} & \colhead{FeH} & \colhead{FeH} & \colhead{FeH} & \colhead{Mg I} & \colhead{Mg I} & \colhead{Mg I}  & \colhead{Si I} & \colhead{Si I} & \colhead{Si I} \\ \colhead{} & \colhead{UT}  & \colhead{15194.492 $\ang{}$}  & \colhead{15207.53 $\ang{}$}  & \colhead{15294.562 $\ang{}$}  & \colhead{15381.99 $\ang{}$} & \colhead{15490.339 $\ang{}$}  & \colhead{15611.14 $\ang{}$}  & \colhead{15648.515 $\ang{}$}  & \colhead{15692.751 $\ang{}$} & \colhead{16009.61 $\ang{}$} & \colhead{16050.38 $\ang{}$} & \colhead{16057.34 $\ang{}$} & \colhead{16245.746 $\ang{}$} & \colhead{16249.71 $\ang{}$} & \colhead{16284.665 $\ang{}$} & \colhead{16814.063 $\ang{}$} & \colhead{16889.575 $\ang{}$} & \colhead{16935.09} & \colhead{15740.666 $\ang{}$} & \colhead{15748.938 $\ang{}$} & \colhead{15765.792 $\ang{}$}  & \colhead{15960.063 $\ang{}$} & \colhead{16094.79 $\ang{}$} & \colhead{16680.77 $\ang{}$} }
\startdata
BD-17588	&	2020-11-01T02:19:38.960		&	\nodata{}	&	-0.06	&	\nodata{}	&	\nodata{}	&	-0.02	&	-0.18	&	\nodata{}	&	\nodata{}	&	\nodata{}	&	-0.22	&	-0.26	&	\nodata{}	&	\nodata{}	&	\nodata{}	&	\nodata{}	&	\nodata{}	&	\nodata{}	&	\nodata{}	&	-0.1	&	-0.07	&	\nodata{}	&	\nodata{}	&	0.06	\\
CD-608051	&	2021-04-26T10:04:53.947		&	0	&	\nodata{}	&	\nodata{}	&	\nodata{}	&	0.02	&	\nodata{}	&	\nodata{}	&	-0.06	&	\nodata{}	&	-0.08	&	-0.08	&	-0.14	&	0	&	\nodata{}	&	-0.04	&	-0.1	&	\nodata{}	&	0.03	&	-0.01	&	-0.1	&	0.02	&	0.09	&	0.08	\\
G162-44	&	2021-02-24T03:50:12.213		&	\nodata{}	&	0.08	&	\nodata{}	&	\nodata{}	&	0.1	&	-0.02	&	\nodata{}	&	\nodata{}	&	\nodata{}	&	-0.1	&	0.08	&	-0.08	&	\nodata{}	&	-0.12	&	\nodata{}	&	\nodata{}	&	\nodata{}	&	-0.04	&	-0.06	&	-0.1	&	-0.2	&	\nodata{}	&	-0.05	\\
G162-44	&	2022-01-24T07:44:44.999		&	\nodata{}	&	0.02	&	\nodata{}	&	\nodata{}	&	0.06	&	0	&	\nodata{}	&	\nodata{}	&	\nodata{}	&	-0.08	&	0.02	&	-0.16	&	-0.06	&	-0.08	&	-0.08	&	\nodata{}	&	\nodata{}	&	-0.09	&	-0.09	&	-0.08	&	-0.1	&	\nodata{}	&	-0.05	\\
G162-44	&	2021-01-06T03:38:06.981		&	\nodata{}	&	0.02	&	\nodata{}	&	\nodata{}	&	0.06	&	-0.08	&	\nodata{}	&	\nodata{}	&	\nodata{}	&	-0.06	&	-0.02	&	-0.14	&	\nodata{}	&	-0.06	&	-0.02	&	\nodata{}	&	\nodata{}	&	-0.06	&	-0.08	&	0.04	&	-0.11	&	\nodata{}	&	0.01	\\
\enddata
\tablecomments{The full table is available in machine readable form.}
\end{splitdeluxetable*}

\begin{deluxetable}{ l c c c}
\tablewidth{0pt} 
\tabletypesize{\scriptsize}
\tablecaption{OH and \water{} Linelist \label{tab:oxy_linelist}} 
 \tablehead{\colhead{$\lambda_{\mathrm{air}}$} & \colhead{Molecule} &  \colhead{$\chi$ } &  \colhead{log(gf)} \\  \colhead{\ang} & \colhead{} & \colhead{(eV)} & \colhead{(dex)} }
\startdata
15145.78	&	OH	&	0.16	&	-5.52	\\
15266.17	&	OH	&	0.21	&	-5.50	\\
15383.33	&	OH	&	1.02	&	-5.11	\\
15407.31	&	OH	&	0.25	&	-5.44	\\
15409.18	&	OH	&	0.25	&	-5.44	\\
15422.37	&	OH	&	0.25	&	-5.39	\\
15457.60	&	OH	&	0.93	&	-5.19	\\
15458.23	&	OH	&	0.93	&	-5.19	\\
15463.90	&	OH	&	1.37	&	-5.20	\\
15496.35	&	OH	&	0.91	&	-5.24	\\
15496.93	&	OH	&	0.91	&	-5.24	\\
15535.47	&	OH	&	0.51	&	-5.30	\\
15558.04	&	OH	&	0.30	&	-5.38	\\
15560.25	&	OH	&	0.30	&	-5.38	\\
15572.09	&	OH	&	0.30	&	-5.34	\\
15626.71	&	OH	&	0.54	&	-5.27	\\
15651.91	&	OH	&	0.80	&	-5.56	\\
15653.49	&	OH	&	0.53	&	-5.20	\\
15897.71	&	OH	&	0.41	&	-5.24	\\
\hline						
15139.01	&	\water{}	&	0.38	&	-5.86	\\
15152.65	&	\water{}	&	0.38	&	-5.92	\\
15339.99	&	\water{}	&	0.75	&	-6.91	\\
15342.16	&	\water{}	&	0.49	&	-5.72	\\
15392.64	&	\water{}	&	0.53	&	-5.65	\\
\enddata
\end{deluxetable}

\begin{deluxetable*}{ l c c c c c c c}
\tablewidth{0pt} 
\tabletypesize{\scriptsize}
\tablecaption{M-Dwarfs with Multiple Visits \label{tab:starmultiplevisits}} 
 \tablehead{\colhead{Name} & \colhead{Obs Date} &  \colhead{$\teff{}$ (K) } &  \colhead{[O/H]} &  \colhead{[Fe/H]} & \colhead{[Mg/H]} & \colhead{[Si/H]} }
\startdata
GJ 367   & 2022-03-13T01:35:37.705 & 3567 & 0.02 & -0.02 & -0.04 & -0.05  \\
GJ 367   & 2022-05-07T01:17:39.107 & 3590 & 0.01 & -0.03 & -0.03 & -0.07 \\
\hline
G162-44	& 2021-02-24T03:50:12.213 & 3435 & 0.13	& 0.00	& -0.07	& -0.13	\\
G162-44	& 2022-01-24T07:44:44.999 & 3427	& 0.14	& -0.04	& -0.09	& -0.06	\\
G162-44	& 2021-01-06T03:38:06.981 & 3434	& 0.13 & -0.03	& -0.03	& -0.08	\\
\hline
GJ4102	&	2023-03-26T09:46:13.080	&	3277	&	-0.08	&	-0.12	&	-0.19	&	-0.12	\\
GJ4102	&	2023-03-26T09:30:45.239	&	3270	&	-0.08	&	-0.11	&	-0.22	&	-0.06	\\
\hline
GJ436	&	2023-04-01T02:16:46.909	&	3496	&	0.02	&	-0.05	&	-0.05	&	-0.12	\\
GJ436	&	2016-02-29-10:23:22.932	&	3469	&	-0.05	&	-0.04	&	-0.04	&	-0.16	\\
\hline
Gliese486	&	2021-05-25T00:13:09.871	&	3340	&	0.08	&	-0.01	&	-0.08	&	-0.04\\
Gliese486	&	2021-05-02T23:42:06.584	&	3326	&	0.08	&	-0.01	&	-0.11	&	-0.06\\
Gliese486	&	2022-02-28T07:08:49.895	&	3312	&	0.06	&	-0.08	&	-0.11	&	-0.08\\
\hline
L231-32	&	2021-10-17T02:48:17.344	&	3403	&	0.01	&	-0.15	&	-0.14	&	-0.20	\\
L231-32	&	2020-10-31T08:25:43.470	&	3398	&	0.00	&	-0.16	&	-0.12	&	-0.21	\\
\hline
L98-59	&	2021-03-13T02:28:50.907	&	3299	&	-0.03	&	-0.13	&	-0.21	&	-0.13	\\
L98-59	&	2021-02-11T07:16:29.358	&	3323	&	-0.03	&	-0.16	&	-0.15	&	-0.14	\\
L98-59	&	2020-11-02T08:36:30.060	&	3308	&	-0.05	&	-0.15	&	-0.13	&	-0.10	\\
\hline
LP961-53	&	2022-02-27T05:40:18.792	&	3519	&	-0.13	&	-0.14	&	-0.09	&	-0.22	\\
LP961-53	&	2023-12-10T08:01:39.269	&	3695	&	-0.14	&	-0.14	&	-0.12	&	-0.17	\\
\hline
TOI-269	&	2023-09-28T03:58:24.081	&	3413	&	0.05	&	-0.13	&	0.00	&	\nodata{} \\
TOI-269	&	2023-12-08T08:11:23.488	&	3462	&	0.02	&	-0.17	&	-0.02	&	-0.13	\\
\enddata
\end{deluxetable*}

\end{document}